\documentclass{SciPost}

\hypersetup{
  colorlinks,
  linkcolor={red!50!black},
  citecolor={blue!50!black},
  urlcolor={blue!80!black}
}

\usepackage[bitstream-charter]{mathdesign}
\DeclareSymbolFont{usualmathcal}{OMS}{cmsy}{m}{n}
\DeclareSymbolFontAlphabet{\mathcal}{usualmathcal}

\fancypagestyle{SPstyle}{
  \fancyhf{}
  \lhead{\colorbox{scipostblue}{\bfseries\color{white}~SciPost Phys. Rev. }}
  \rhead{\bfseries\color{scipostdeepblue}~Submission }
  
  \fancyfoot[C]{\textbf{\thepage}}
}

\usepackage[T1]{fontenc}
\usepackage{nicematrix}
\usepackage{longtable}
\usepackage{siunitx}
\usepackage{mathtools}
\usepackage{dsfont}
\usepackage{pifont}
\usepackage{bm}

\newcommand{\ket}[1]{\left|#1\right>}
\newcommand{\braket}[1]{\left\langle#1\right\rangle}

\newcommand{\re}{\mathrm{e}}
\newcommand{\ri}{\mathrm{i}}

\DeclareMathOperator{\Tr}{Tr}

\newcommand{\conductance}{\mathcal{G}}
\newcommand{\ldos}{\rho_n^{\rm ldos}(E)}

\newcommand\bovermat[2]{%
  \makebox[0pt][l]{$\smash{\overbrace{\phantom{%
    \begin{matrix}#2\end{matrix}}}^{#1}}$}#2}

\graphicspath{{figures/}}

\begin{document}

\pagestyle{SPstyle}

\begin{center}
  {\Large\textbf{\color{scipostdeepblue}{Computational quantum transport: a scattering approach perspective}}}
\end{center}

\begin{center}
  \textbf{Xavier Waintal\textsuperscript{1},
    Michael Wimmer\textsuperscript{2,3},
    Anton Akhmerov\textsuperscript{3},\\
    Christoph Groth\textsuperscript{1$\star$},
    Branislav K. Nikoli\'{c}\textsuperscript{4},
    Mathieu Istas\textsuperscript{1},\\
    T\'omas \"Orn Rosdahl\textsuperscript{3} and
    Daniel Varjas\textsuperscript{2,3,5}}
\end{center}

\begin{center}
  \textbf{1} Univ.~Grenoble Alpes, CEA, Grenoble INP, IRIG, Pheliqs, F-38000 Grenoble, France
  \\
  \textbf{2} QuTech, Delft University of Technology, 2600 GA Delft, The Netherlands
  \\
  \textbf{3} Kavli Institute of Nanoscience, Delft University of Technology, 2600 GA Delft, The Netherlands
  \\
  \textbf{4} Department of Physics and Astronomy, University of Delaware, Newark, DE 19716, United States of America
  \\
  \textbf{5} Institute for Theoretical Solid State Physics, IFW Dresden and W\"{u}rzburg-Dresden Cluster of Excellence ct.qmat, 01069 Dresden, Germany
  \\[\baselineskip]
  $\star$ \href{mailto:christoph.groth@cea.fr}{\small christoph.groth@cea.fr}
\end{center}

\section*{\color{scipostdeepblue}{Abstract}}
\textbf{\boldmath{%
  This review is devoted to the different techniques that have been developed to compute the
  phase-coherent transport properties of quantum nanoelectronic systems connected to electrodes. Beside a review of the different algorithms proposed in the literature,
  we provide a comprehensive and pedagogical derivation of the two formalisms on which these techniques are based: the scattering approach and the (nonequilibrium) Green's function approach.
  We show that the scattering problem can be formulated as a system of linear equations and that
  different existing algorithms for solving this scattering problem amount
  to different sequences of Gaussian elimination. We explicitly prove the equivalence of the two formalisms. We discuss the stability and
  numerical complexity of the existing methods. The review ends with a selection of a few applications where numerical calculations were instrumental in shaping our understanding of the physics.
}}

\vspace{\baselineskip}

\noindent\textcolor{white!90!black}{%
\fbox{\parbox{0.96\linewidth}{%
\textcolor{white!40!black}{\begin{tabular}{lr}%
  \begin{minipage}{0.58\linewidth}%
    {\small Copyright attribution to authors. \newline
    This work is a submission to SciPost Phys. Rev. \newline
    License information to appear upon publication. \newline
    Publication information to appear upon publication.}
  \end{minipage} & \begin{minipage}{0.36\linewidth}
    {\small Received Date \newline Accepted Date \newline Published Date}%
  \end{minipage}
\end{tabular}}
}}
}

\vspace{10pt}
\noindent\rule{\textwidth}{1pt}
\tableofcontents
\noindent\rule{\textwidth}{1pt}
\vspace{10pt}

\section{Introduction}

The field of quantum nanoelectronics, known as ``mesoscopic physics'' in its early days \cite{imry2008}, was born in the early eighties with the first experiments directly illustrating the impact (beyond the atomic scale) of the wave nature of electrons on macroscopic observables such as the electrical conductance.
While it was already known that quantum mechanics plays a role in the transport properties, this role was restricted to the material band structure, i.e., to its behavior on the atomic scale, while the physics at larger scales was described by a semi-classical incoherent theory---Boltzmann equation.
Such a description was adequate until it became possible to study the transport properties of samples at ultra-low temperatures or to make very small devices.
At low temperature, the characteristic length $L_\Phi$ over which the electron wave function retains a well-defined phase becomes large and eventually exceeds the device size.
The demonstration of the quantum Hall effect \cite{klitzing1980} and the phenomena of universal conductance fluctuations or the weak localization \cite{akkermans2007} were early observations of the impact of the wave nature of electrons in electronic devices made indirectly in bulk samples.
Later, as clean room technology made it possible to pattern increasingly smaller devices, direct observation of, e.g., the Aharonov--Bohm effect in a small metal loop \cite{webb1985} or the conductance quantization in a constriction \cite{wees1988} became possible.
Nowadays, the phase coherence length has reached record, almost macroscopic, values $L_\Phi \approx \SI{250}{\um}$, making it possible to envision operating these interferometers to make flying quantum bits or other complex manipulations of quantum states \cite{bauerle2018}.

Building a theoretical understanding of the quantum transport phenomena required figuring out how one connects the macroscopic world, where one injects currents and measures voltage differences, to the mesoscopic scale described by quantum mechanics.
Landauer proposed that the problem could be formulated as a waveguide problem where the electrodes are treated as infinite waveguides with well-defined incoming waves.
This viewpoint eventually led to a number of striking predictions including the conductance quantization and the separation between the cause of finite electrical resistance (elastic scattering) and the corresponding Joule heating that takes place in the electrodes.
The theory eventually evolved into two very different looking, albeit strictly equivalent, approaches to quantum transport.
In the Landauer--Büttiker or scattering formalism, one treats the quantum-mechanical system as a scattering problem and arrives at the celebrated Landauer formula.
The other approach relies on the Keldysh perturbation theory to build the non-equilibrium Green's functions (NEGF) formalism.

Methods of computational quantum transport that aim to calculate the transport properties of phase-coherent samples numerically are almost as old as mesoscopic physics itself.
Interference effects are indeed very sensitive to microscopic details so that, as P.~W.~Anderson stated in his 1977 Nobel lecture, in the context of localization, ``one has to resort to the indignity of numerical simulations to settle even the simplest questions about it''.
Despite their indignity, numerical simulations of quantum transport have become increasingly popular and powerful.
This work aims to provide a systematic review of the techniques that were developed to perform such computations.

\subsection{Scope of this review}
\label{sec:scope}
This review focuses on computational techniques that address phase-coherent quantum transport for discrete models that consist of a finite system connected to semi-infinite quasi-one-dimensional electrodes.
We omit the calculation of local quantities (e.g., conductivities) already described in other reviews \cite{weisse2006,fan2021}.
Such local quantity calculations are based on the linear-response Kubo formula \cite{kubo1957,greenwood1958}, which may or may not capture non-local effects depending on the level of approximation \cite{baranger1989}. Here, we concentrate on techniques for computing global quantities (e.g., conductances). Also, we focus on descriptions at the single-particle level where electron-electron interactions are either not accounted for or treated at the mean field or effective level.

The goals of this article are threefold:
\begin{enumerate}
    \item provide a detailed pedagogical entry point for newcomers to the field,
    \item present a modern and comprehensive derivation of the mathematical formalism for both the scattering approach and NEGF as well as the equivalence between the two,
    \item review the relevant literature with a stress on the algorithms, as well as a few selected applications.
\end{enumerate}
The techniques introduced in this review have a broad range of applications, including devices combining multiple materials, such as semiconductors, superconductors, magnetic materials, metals, graphene, and topological insulators in a wide variety of geometries and in different dimensions.
Due to the vastness of the applications, it is not feasible for us to review all of them in detail.
Instead we will cover several illustrative examples that demonstrate the benefits provided by computational quantum transport in analyzing physical phenomena.

Although most of the material presented in this review is known, some aspects were not presented before at this level of generality and/or within a unified formalism.
We also include some new material: specifically the detailed algorithms that were developed for the Kwant package \cite{groth2014} that were not yet published.

\subsection{A short history of computational quantum transport of coherent conductors}
\label{sec:short-hist-comp-quant}

The scattering approach to quantum transport was initially defined in \cite{landauer1957} and further extended by \cite{fisher1981,buttiker1986,stone1988}.
The approach is reviewed in \cite{blanter2000} or \cite{datta1995,imry2008}.
The first formulations of the scattering problem for discrete models can be traced down to \cite{lent1990} in the nanoelectronics community and \cite{ando1991} in the physics community.
The alternative NEGF theory is based on the Keldysh formalism \cite{keldysh1964} which considerably simplifies in the context of non-interacting systems \cite{caroli1971}.
The point of reference in the field is the work \cite{meir1992}, which generalizes \cite{caroli1971} to interacting systems.

The first numerical calculation of the electrical conductance of a quantum system was made in \cite{thouless1981} and \cite{lee1981} where the recursive Green's function (RGF) algorithm was introduced for a one-dimensional system.
The RGF was soon to become the reference of the field thanks to several groups that generalized it to quasi-one-dimensional systems \cite{mackinnon1985} and developed the equivalent wave function approach \cite{ando1991}. It superseded the transfer matrix technique that was developed in the context of
Anderson localization \cite{pichard1981,pichard1981b} but suffered from numerical instabilities \cite{mackinnon1983} and has remained mostly confined to the calculation of Lyapunov exponents \cite{slevin2014}.
The initial applications of RGF focused on tight-binding systems with a square lattice and a rectangular geometry for studying Anderson localization, quantum dots, quantum billiards, or other mesoscopic systems.
This early code was considered highly useful, to the extent that its reference implementation was referred to as ``the Program'' in some research groups.

These numerical techniques were then generalized to tackle a wider class of problems including general lattices (most prominently the graphene honeycomb structure), multi-terminal systems, systems with internal degrees of freedom (\emph{e.~g.}~spin or electron/hole for superconductivity), and arbitrary geometries beyond the rectangular shape that was natural in RGF \cite{kazymyrenko2008,wimmer2009}, and extensions beyond quasi-one-dimensional systems \cite{settnes2015}.
Various strategies to accelerate the calculations were also developed, including algorithms that precalculate building blocks \cite{rotter2000,teichert2017}, parallel algorithms \cite{drouvelis2006,kuzmin2013,costa2013}, slicing strategies for recursive algorithms \cite{wimmer2009,mou2011,mason2011,thorgilsson2014,papior2017}, or nested dissection \cite{li2008,boykin2008}.
The range of applications expanded considerably to a much wider spectrum including mesoscopic superconductivity, electronic interferometers, quantum Hall effect, spintronics, graphene, molecular electronics, any combination of the above, and much more.

While most codes remain internal to the research groups where they have been developed, publicly available and open source codes have started to be developed.
These include SMEAGOL \cite{rocha2006}, nextnano \cite{birner2007}, Knit \cite{kazymyrenko2008}, Kwant \cite{groth2014}, and Quantica.jl \cite{sanjose2021b}.
There are also quantum transport extensions to ab-initio packages such as TRANSIESTA \cite{brandbyge2002,papior2017}, OpenMX \cite{ozaki2003}, NanoTCAD ViDES \cite{bruzzone2014}, GOLLUM \cite{ferrer2014}, QuantumATK \cite{smidstrup2020}, and Nemo5 \cite{huang2016}.

\subsection{What is \emph{not} in this review}
\label{sec:not_in_scope}
There have been a number of very interesting developments that are \emph{not}
covered in this review. This non-exhaustive list includes:
\begin{itemize}
\item While we limit ourselves to DC transport, the extension to AC (finite frequency) is straightforward using the well-established Floquet formalism \cite{moskalets2012}. In practice, there are simple recipes to build the AC quantum transport properties from the DC ones---instead of calculating them only at the Fermi level, one needs to compute them also at energies shifted by multiples of the drive frequency \cite{shevtsov2013,varelamanjarres2023}.
\item There are also many developments of quantum transport in the time domain, such as propagation of voltage pulses, quenches, or interaction with electromagnetic degrees of freedom.
The Landauer--Büttiker formalism (or equivalently the Non-Equilibrium Green's Function formalism) can be extended to time-dependent dynamics \cite{tuovinen2014, gaury2014} to allow such calculations. This formalism can also be used in the context of
time-dependent density functional theory \cite{varga2011,wang2015} and open source software packages start to become available \cite{kloss2021}.
\item The interplay between quantum transport and electron-electron interactions is a very active and difficult field of research. This review limits itself to non-interacting and mean-field systems and ignores all new developments around topics such as diagrammatic Monte-Carlo, tensor networks, or influence functionals. An introduction to the formalism can be found in \cite{stefanucci2013}. While in the context of this review the scattering approach and the NEGF formalism are fully equivalent, only NEGF has a natural extension to interacting problems beyond mean field, the scattering approach has not.
\item To be amenable to numerical calculations, the models must be discretized in \emph{some} way or another. Finding a proper discretization scheme is an important problem, somewhat orthogonal to the specific topic of interest here, quantum transport. While the problem will be brushed upon in the application section, we will not explicitly discuss the accuracy of e.g. different basis sets (see, e.g., \cite{driscoll2010, herrmann2010,lejaeghere2016}).
\end{itemize}

\subsection{How to read this review}
\label{sec:how-read-this}

This review first establishes the scattering matrix formalism, and uses it to derive the NEGF formalism.
Care has been taken to derive both formalisms using unified notation and to gather a comprehensive set of proofs, including several original ones.

In order to not overwhelm the readers with the technical details, we start with a pedagogical Sec.~\ref{sec:pedag-exampl-cond} that introduces the main concepts: scattering wave functions, scattering matrix, and Landauer formula using the minimal example of a square lattice in a rectangular geometry, which allows us to keep the notation and the derivation simple.

Section~\ref{sec:scatt-form-discr} introduces the scattering problem as a set of linear equations whose solution defines both the scattering wave function and the scattering matrix.
It also introduces the notation used in the rest of the review.
Solving for both the wave function \emph{and} the scattering matrix at the same time is not entirely standard in the field, and we name it the $\Psi S$-approach.
It forms the \emph{backbone} of this review from which all the other approaches are derived.
This section maintains maximal generality, only assuming discrete models and translationally invariant leads.
Section~\ref{sec:stable-form-infin} expands on Sec.~\ref{sec:scatt-form-discr} by transforming the scattering problem into a form that is more stable for numerical calculations.

We establish the relation to NEGF in Sec.~\ref{sec:green} by introducing the retarded Green's function using a linear problem approach, similar to the one that defines the scattering formalism.
While this is an uncommon approach, it has the advantage of emphasizing the close analogy between the scattering matrix and the Green's function techniques and allowing for a direct derivation of the Fisher--Lee relations that connect one with the other.

In Sec.~\ref{sec:numerics} we discuss how to numerically solve the equations introduced in Secs.~\ref{sec:scatt-form-discr}--\ref{sec:green}. To this end, we review different approaches, ranging from using sparse linear solvers to using specialized algorithms such as the RGF algorithm.

Sections~\ref{sec:scatt-form-discr}, \ref{sec:stable-form-infin} and \ref{sec:green} only deal with the single-particle quantum mechanics of infinite systems (or more precisely finite systems connected to semi-infinite electrodes) as well as the algorithms used to solve the corresponding mathematical problems.
In order to calculate observables, such as the electrical conductance, quantum mechanics must be complemented with out-of-equilibrium statistical physics.
Section~\ref{sec:phys-obs-land} introduces various observables that can be obtained from the scattering matrix ranging from the conductance (Landauer formula) to current noise and thermoelectric effects.
Section~\ref{sec:phys-obs-negf} describes the NEGF approach for calculating the observables.
Although the two approaches look very different, we establish their complete equivalence.
Sections~\ref{sec:phys-obs-land} and \ref{sec:phys-obs-negf} end the technical part of this review.

We end this review with a brief overview of different models used in computational quantum transport and a selection of prominent applications.
It is impossible for this part of the review to be exhaustive, and we restrict ourselves to a few examples that illustrate the main theoretical concepts.
In Sec.~\ref{sec:select-appl}, we thus focus on showing ``success stories'': examples where numerical quantum transport was pivotal in the development of a field.
In this way, we intend to give a broad flavor of how numerical quantum transport is employed in research.

\section{A pedagogical example: conductance of samples cut out of a two-dimensional electron gas}
\label{sec:pedag-exampl-cond}
We begin by demonstrating in a pedagogical manner how different concepts of computational quantum transport apply to a minimal example.
It allows us to introduce the main concepts of this review in a simplified way, before we treat it in all generality in Secs.~\ref{sec:scatt-form-discr}--\ref{sec:green}.
Readers who are already familiar with the scattering matrix and Landauer formula for the conductance may skip subsections~\ref{sec:land-form-cond} and \ref{sec:combination-rule-scattering}.
Because a number of sources already describe in detail the non-computational aspects of the scattering formalism of quantum transport (see, e.g., \cite{datta1995}), we restrict ourselves to a concise presentation of the basic concepts there.

We consider a fully phase-coherent quantum conductor connected to two perfect electrodes---leads---where all the dissipation takes place.
This is an idealized situation: in a real device, electrons always experience some dephasing due to, e.g., coupling to phonons or electron-electron interaction.
Nevertheless, typical dephasing lengths at dilution fridge temperature ($\sim \SI{10}{\milli\K}$) can exceed tens of microns, and as device sizes continue to shrink, this model captures the salient features of many realistic devices.

\subsection{The Schrödinger equation as a waveguide problem}
\label{sec:schr-equat-as}

\begin{figure}
  \centering
  \includegraphics[width=0.55\linewidth]{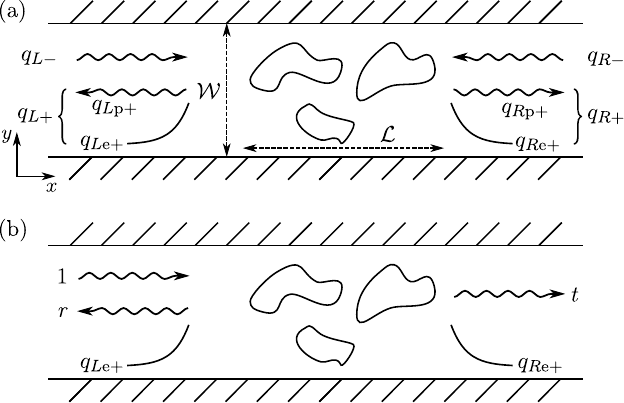}
  \caption{Sketch of a quasi one-dimensional wire of width $\mathcal{W}$. The set of closed curves in the middle symbolizes the scattering region of length $\mathcal{L}$, an arbitrary potential that scatters electrons.
    (a) A general wave function $\Psi(x,y)$ consists of waves propagating towards the scattering region as well as waves propagating away from it and waves decaying into the lead.
    (b) The wave matching problem can be solved uniquely by restricting it to a single incoming mode $\alpha_0$,  $q_{L-,\alpha} = \delta_{\alpha,\alpha_0}$, which is reflected with amplitudes $r$ and transmitted with amplitudes $t$.
  }
  \label{scattering_continuum}
\end{figure}

We consider the quantum mechanics of an electron confined in the simple geometry sketched in Fig.~\ref{scattering_continuum}:
a two-dimensional wire that is infinite in the $x$ direction, has width $\mathcal{W}$ in the $y$ direction, and contains a scattering region of size $\mathcal{L}$.
Our first task is to solve the spinless Schrödinger equation for the problem:
\begin{equation}
\label{eq:schro2D}
\frac{\hbar^2}{2m^*} \Delta \Psi(x,y) + V(x,y)\Psi(x,y)= E \Psi(x,y),
\end{equation}
where $m^*$ is the electronic effective mass, $E$ is the energy of the electron, $\Psi$ is its wave function, and $V(x,y)$ is an arbitrary potential, which is non-zero in a finite region $0<x<\mathcal{L}$ of the wire.
The boundary conditions require that $\Psi$ vanishes at the wire boundaries: $\Psi(x,y=0)=\Psi(x,y=\mathcal{W})=0$.
To construct the full solutions of this equation, we first analyze it outside of the scattering region, where $V(x,y)=0$, and the system is therefore translationally invariant.
In that case the solutions for a given energy $E$ are linear superpositions of plane waves $\exp(i k_x x + i k_y y)$ with $E = \hbar^2/(2m^*) (k_x^2 + k_y^2)$.
Imposing the boundary conditions at $y=0$ and $y=\mathcal{W}$ further restricts the wave functions to be linear combinations of $e^{ik_x x} \sin (k_y y)$, with $k_y = \alpha \pi/\mathcal{W}$, where $\alpha$ is an integer that labels lead \emph{modes}.
The dispersion relation requires that $k_x = \pm k_\alpha \equiv \pm \sqrt{2m^*E/\hbar^2 - (\pi\alpha/\mathcal{W})^2}$.
For the $N_p$ values of $\alpha$ such that $\hbar^2k_y^2/2m^* < E$, $k_\alpha$ is real and positive.
These modes are, therefore, propagating: they extend to $x\to\pm\infty$, and they define what is known as \emph{conduction channels} \cite{datta1995,imry2008}.
Infinitely many modes with $\alpha > N_p$ have imaginary $k_\alpha$, and are exponentially decaying either to the left or to the right from the scattering region.
We therefore parametrize the wave function left of the scattering region ($x<0$) using
\begin{subequations}
\label{eq:LRwave}
\begin{equation}
\label{eq:Lwave}
\Psi(x,y) = \sum_\alpha \frac{\sin \left(\pi\alpha y/\mathcal{W}\right) }{\sqrt{\mathcal{W}|k_\alpha|/2}}
\left[q_{L-,\alpha}e^{ik_\alpha x} +  q_{L+,\alpha} e^{-ik_\alpha x}\right],
\end{equation}
and accordingly on the right of the scattering region ($x>\mathcal{L}$) using
\begin{equation}
\label{eq:Rwave}
\Psi(x,y) = \sum_\alpha \frac{\sin \left(\pi\alpha y/\mathcal{W}\right) }{\sqrt{\mathcal{W}|k_\alpha|/2}}
\left[q_{R+,\alpha}e^{ik_\alpha x} +  q_{R-,\alpha} e^{-ik_\alpha x}\right].
\end{equation}
\end{subequations}
This asymptotic wave function is parametrized using coefficients $q_{L/R,\pm,\alpha}$, which are yet undetermined.
The subscripts $L/R$ label the left and right sides of the scattering region, while $-$ and $+$ label the modes propagating, respectively, towards and away from the scattering region, regardless of the side.
While the normalization constants $(W|k_\alpha|/2)^{-1/2}$ may be absorbed in the definition of $q_{L/R,\pm,\alpha}$, we introduce them for later convenience.
Furthermore, for the wave function to not diverge away from the scattering region, the solutions that diverge at $x \to \pm \infty$ must be absent: $q_{L-,\alpha} = q_{R-,\alpha} = 0$ for $\alpha > N_p$.
The coefficients collected in the vectors $q_{L/R-}$ can thus be partitioned into a vector $q_{L/R\mathrm{p}+} = (q_{L/R+,1}, \ldots, q_{L/R+,N_p})$ corresponding to propagating modes (index $\mathrm{p}$ for propagating), and a vector $q_{L/R\mathrm{e}+} = (q_{L/R-,\alpha})_{\alpha > N_p}$ corresponding to decaying modes (index $\mathrm{e}$ for evanescent).
In contrast, the coefficients $q_{L/R-}=q_{L/R\mathrm{p}-}$ only contain propagating modes, as schematically shown in Fig.~\ref{scattering_continuum}(a).

Determining the values of $q_{L/R,\pm}$ requires solving the Schrödinger equation in the difficult scattering region, and then matching the wave function in the different parts of the system by using the continuity of the wave function and its derivative.
Textbooks on the scattering formalism often write Eqs.~\eqref{eq:Lwave} and \eqref{eq:Rwave} without decaying modes, taking the value of the wave function far into the lead where the decaying modes can be neglected.
They are however \emph{crucial} for the actual wave function matching \emph{at the interface}!
Occasionally, this sort of wave function matching including evanescent modes is done in the continuum, as in Ref.~\cite{szafer1989}.
Generally, analytical solutions are difficult to obtain, and more common is a numerical approach within a discrete model---this is the main topic of this review.

Before we show how to perform the wave function matching numerically in a wire geometry, we first introduce the scattering matrix and its connection to physical observables.

\subsection{The scattering matrix}
\label{sec:scattering-matrix}

The wave matching problem is underdetermined, and therefore admits infinitely many solutions.
We therefore describe its solutions by expressing some unknown parameters $q_{L/R,\pm,\alpha}$ through the others.
Setting one of the $q_{L/R,-,\alpha_0}$ to be unity and all other $q_{L/R,-,\alpha}$ to zero, for $\alpha_0 < N_p$, allows one to determine a unique value of all the other coefficients $q_{L/R,+}$, shown schematically in Fig.~\ref{scattering_continuum}(b).
The linearity of the Schrödinger equation then implies that we can find a solution for any choice of $q_{L/R,-}$ through superposition.
Collecting these solutions into a matrix allows one to relate the amplitudes of the propagating modes using the scattering matrix $S$:
\begin{equation}
  \label{eq:scatteringS}
  \begin{pmatrix}
  q_{L\mathrm{p}+} \\
  q_{R\mathrm{p}+}
  \end{pmatrix}=
  S
  \begin{pmatrix}
  q_{L\mathrm{p}-} \\
  q_{R\mathrm{p}-}
  \end{pmatrix}\equiv
  \begin{pmatrix}
    r & t' \\
    t & r'
  \end{pmatrix}
  \begin{pmatrix}
    q_{L\mathrm{p}-} \\
    q_{R\mathrm{p}-}
  \end{pmatrix},
\end{equation}
where $q_{L/R\mathrm{p}\pm}$ are size-$N_p$ vectors containing all the coefficients of the propagating modes.
The different blocks of the $S$-matrix are the reflection ($r,r'$) and transmission ($t,t'$) matrices.

In addition to Eq.~\eqref{eq:scatteringS}, there is also a linear relationship between $q_{L/R\mathrm{e}+}$ and $q_{L/R\mathrm{p}-}$ which is typically not stated explicitly.
It will become essential in the next section where we will derive the scattering formalism for discrete models in all generality.

Let us check how current conservation applies to the $S$-matrix.
The total particle current $I_p$ flowing through the wire,
\begin{equation}
\label{eq:current_cons}
I_p = \frac{\hbar}{m^*}\int_0^\mathcal{W} dy \operatorname{Im} \Psi^*(x,y) \partial_x \Psi(x,y),
\end{equation}
is independent of $x$ for any solution of Eq.~\eqref{eq:schro2D}.
Substituting Eqs.~\eqref{eq:LRwave} into Eq.~\eqref{eq:current_cons} yields
\begin{equation}
\label{eq:current_cons2}
\sum_{\alpha =1}^{N_p} |q_{L+,\alpha}|^2 + |q_{R+,\alpha}|^2 = \sum_{\alpha =1}^{N_p} |q_{L-,\alpha}|^2 + |q_{R-,\alpha}|^2,
\end{equation}
where we have used the orthogonality of the transverse wave function,
\begin{equation*}
\int_0^\mathcal{W} dy \sin (\pi\alpha y/\mathcal{W})\sin (\pi\beta y/\mathcal{W}) = \delta_{\alpha\beta} (\mathcal{W}/2).
\end{equation*}
The decaying modes do not contribute to the current and thus do not appear in Eq.~\eqref{eq:current_cons2}.
We also see that the normalization constants in Eqs.~\eqref{eq:LRwave} are chosen so that the Eq.~\eqref{eq:current_cons2} has no prefactors.
For every solution~\eqref{eq:scatteringS} of the scattering problem to satisfy Eq.~\eqref{eq:current_cons2}, the $S$-matrix must preserve the norm of vectors, or in other words it must be unitary:
\begin{equation}
S S^\dagger = S^\dagger S = 1.
\end{equation}
Similarly to how the Hermiticity of the Hamiltonian leads to current conservation and the unitarity of $S$, other properties of the Hamiltonian give other constraints to $S$.
For example, in the presence of a magnetic field $B$ with a vector potential $\vec A(x,y)$, the Hamiltonian becomes
\begin{equation}
\label{eq:schro2D-B}
H = \frac{1}{2m^*} \left( i\hbar\vec\nabla -e\vec A\right)^2 + V(x,y).
\end{equation}
One observes that if $\Psi(x,y)$ is an eigenstate of Eq.~\eqref{eq:schro2D-B} with magnetic field $B$ and vector potential $\vec A$, then $\Psi(x,y)^*$ is an eigenstate at the same energy, field $-B$, and vector potential $-\vec A$.
While the mode wave functions in Eqs.~\eqref{eq:LRwave} are not Hamiltonian eigenstates in presence of a vector potential, a similar decomposition into lead modes still holds.
Taking the complex conjugate of Eqs.~\eqref{eq:LRwave} conjugates the amplitudes $q_{L/R\pm,\alpha}$ and exchanges the outgoing ($+$) and incoming ($-$) propagating modes.
The Eq.~\eqref{eq:scatteringS} then transforms into
\begin{equation}
\label{eq:scatteringS-B}
\begin{pmatrix}
q_{L-}^* \\
q_{R-}^*
\end{pmatrix}=
S(-B)
\begin{pmatrix}
q_{L+}^* \\
q_{R+}^*
\end{pmatrix}.
\end{equation}
Comparing this to Eq.~\eqref{eq:scatteringS} we conclude that $S^*(-B) S(B) = 1$, and additionally utilizing the unitarity of $S(B)$ we obtain
\begin{equation}
S(-B) = S^T(B).
\end{equation}
In Sec.~\ref{sec:scatt-stat-symm} we will discuss a broader class of symmetries for the scattering matrix and how to enforce them in numerical calculations.

\subsection{The Landauer formula for the conductance}
\label{sec:land-form-cond}
The scattering matrix $S(E)$ parametrizes the single particle eigenstates of the system at an energy $E$.
Because it describes the scattering of the states far away from the scattering region, it turns out to enter many macroscopic observables, such as the electrical conductance of the system.

We consider the current injected by the $L$ and $R$ electrodes described by two different thermal equilibria with chemical potentials $\mu_{L/R}$ and temperatures $\Theta_{L/R}$.
This corresponds to the different scattering solutions~\eqref{eq:scatteringS} of the Eq.~\eqref{eq:schro2D} having an occupation number $f_{L/R}(E)=[e^{(E-\mu_{L/R})/(k_B \Theta_{L/R})}+1]^{-1}$ given by the Fermi distribution of the electrode from which the incoming mode $q_{L/R,-,\alpha}$ originates.
The contribution of a state with momentum $k$ normalized to $1$ per unit length to the macroscopic observables is $(2\pi)^{-1}\int dk$ times the matrix element of the observable.
In the case of the current, this simplifies to $(e/\hbar)\sum_\alpha\int dE(k)/(2\pi)$ since the current is proportional to the group velocity $\hbar^{-1} \partial E_\alpha(k)/\partial k$.
Here
\begin{equation}
E_\alpha (k) = \hbar^2/(2m^*) \left[k^2 + (\pi\alpha/W)^2\right]
\end{equation}
is the dispersion relation of mode $\alpha$.
The electrical current flowing through the left electrode is the sum of the incoming current from the left, current that is reflected back to the left from the scattering region, and current transmitted from the right, and it equals
\begin{equation}
\label{eq:basic_landauer}
I = \frac{e}{\hbar}  \sum_{\alpha=1}^{N_p} \int \frac{dE}{2\pi}
f_L (E_\alpha) \left( 1 - \sum_{\beta=1}^{N_p} |r_{\beta\alpha}|^2 \right)
- \frac{e}{\hbar}  \sum_{\alpha=1}^{N_p} \int \frac{dE}{2\pi}
f_R (E_\alpha) \left( \sum_{\beta=1}^{N_p} |t'_{\beta\alpha}|^2 \right).
\end{equation}
This result simplifies further due to unitarity requiring that
\begin{equation}
N_p - \sum_{\alpha,\beta=1}^{N_p} |r_{\beta\alpha}|^2 = \sum_{\alpha,\beta=1}^{N_p} |t_{\beta\alpha}|^2= \sum_{\alpha,\beta=1}^{N_p} |t'_{\beta\alpha}|^2,
\end{equation}
and yields
\begin{equation}
\label{eq:landauer_simple_example}
I = \frac{e}{h} \int dE \left[ f_L(E)- f_R(E) \right] \sum_{\alpha,\beta = 1}^{N_p} |t_{\beta\alpha}(E)|^2 ,
\end{equation}
which is the general form of the celebrated Landauer formula.
In the limit of zero temperature $\Theta_{L/R}=0$ and a small bias voltage $V_b$ applied to the electrodes ($\mu_{L/R} = E_F \pm eV_b$), the Landauer formula relates the differential conductance $g=dI/dV_b$ to the scattering matrix:
\begin{equation}
\label{eq:landauer1}
g = \frac{e^2}{h} \sum_{\alpha,\beta = 1}^{N_p} |t_{\beta\alpha}(E_F)|^2= \frac{e^2}{h} \Tr t(E_F) t^\dagger (E_F).
\end{equation}
The Landauer formula relates an important measurable observable (the conductance) to the solution of a waveguide problem.
Because the matrix $t t^\dagger = 1 - rr^\dagger$ is Hermitian and positive semi-definite, it can be diagonalized to obtain the \emph{transmission eigenvalues} $0\leq T_\alpha\leq 1$.
These contain all the necessary information from the scattering matrix to obtain the conductance
\begin{equation}
\label{eq:landauer3}
g = \frac{e^2}{h} \sum_{a=1}^{N_p}\ T_\alpha (E_F).
\end{equation}
One of the first successes of the scattering theory of quantum transport is the prediction of the quantization of the conductance in units of $e^2/h$ when the conductor is fully ballistic, and $V(x,y)=0$.

Extending the above approach to other measurable quantities allows us to determine the frequency spectrum of current fluctuations, the Seebeck and Peltier coefficients as well as other thermodynamic properties, and we will discuss these in detail in Sec.~\ref{sec:phys-obs-land}.

\subsection{Combination rule of the scattering matrices}
\label{sec:combination-rule-scattering}

So far we considered the scattering matrix as the final answer of the scattering problem.
However, it is also possible to combine the scattering matrices of different parts of the system to obtain the scattering matrix of the whole system.
This becomes possible if the evanescent modes decay sufficiently strongly in the region between the two scattering regions, or in other words if the two scattering regions are separated by a sufficiently large region without scattering.

Let us consider two scattering regions in a quasi-1D conductor, with scattering matrices $S_A$ and $S_B$, with the first scattering region $A$ left of the scattering region $B$.
The scattering equations for both regions therefore read
\begin{subequations}
  \label{eq:scatteringS12}
  \begin{align}
    \begin{pmatrix}
      q_{AL+} \\
      q_{AR+}
    \end{pmatrix} & =
    S_A
    \begin{pmatrix}
      q_{AL-} \\
      q_{AR-}
    \end{pmatrix},    \\
    \begin{pmatrix}
      q_{BL+} \\
      q_{BR+}
    \end{pmatrix} & =
    S_B
    \begin{pmatrix}
      q_{BL-} \\
      q_{BR-}
    \end{pmatrix},
  \end{align}
\end{subequations}
where the new subscripts $A$ and $B$ label the scattering regions.
We observe that the modes incoming into the first scattering region from the right are the left outgoing modes of the second scattering region $q_{AR-} = q_{BL+}$, and vice versa: the modes incoming into the second scattering region from the left are the right outgoing modes of the first scattering region $q_{BL-} = q_{AR+}$.

After eliminating the variables corresponding to the region between $A$ and $B$ from the Eqs.~\eqref{eq:scatteringS12}, we obtain the expression for the scattering matrix of the two regions combined:
\begin{equation}
  \label{eq:addingS_3}
  S =
  \begin{pmatrix} r_A + t'_A r_B \frac{1}{1 - r'_Ar_B} t_A & t'_A \frac{1}{1 - r_Br'_A} t'_B \\
  t_B \frac{1}{1 - r'_Ar_B} t_A & r'_B + t_B r'_A \frac{1}{1 - r_Br'_A} t'_B \end{pmatrix}.
\end{equation}
This block matrix combination is also known as the Redheffer star product $S = S_A \star S_B$~\cite{redheffer1959}.
It also has an intuitive interpretation in terms of counting partial contributions of interfering waves. Expanding
the expression for $t$ into a power series in $r'_A r_B$ we obtain
\begin{equation}
t = t_B t_A + t_B r'_A r_B t_A + t_B r'_A r_B r'_A r_B t_A + \dots,
\end{equation}
or in other words the amplitude $t$ to be transmitted from right to left is the sum of the amplitudes
for the direct path (transmitted by $A$, then transmitted by $B$ with amplitude $t_Bt_A$), the path with two internal reflections (transmitted by $A$, then reflected by $B$ then reflected by $A$ then transmitted by $B$ with amplitude $t_Br'_Ar_Bt_A$), the path with four reflections and so on.

An alternative way to derive the Eq.~\eqref{eq:addingS_3} is to consider the \emph{transfer matrix} $M$ connecting the waves on the left of a scattering region to those on the right of it:
\begin{equation}
\label{eq:transferM}
\begin{pmatrix}
  q_{R+} \\
  q_{R-}
\end{pmatrix}=
M
\begin{pmatrix}
  q_{L-} \\
  q_{L+}
\end{pmatrix}.
\end{equation}
By using the definition of $S$, and treating $q_{R-}$ as unknown and $q_{L+}$ as known, we obtain the following expression for the transfer matrix:
\begin{equation}
  \label{eq:MfromS}
  M = \begin{pmatrix}
    t - r' t'^{-1} r & -t'^{-1}r \\
    r' t'^{-1} & t'^{-1}
  \end{pmatrix}.
\end{equation}
Because the transfer matrix relates the modes on the left to the modes on the right, the composition rule for the transfer matrices simplifies to the matrix product:
\begin{equation}
  \label{eq:addingM}
  M = M_A M_B.
\end{equation}
Inverting the Eq.~\eqref{eq:MfromS} and substituting the Eq.~\eqref{eq:addingM} once again yields the Eq.~\eqref{eq:addingS_3}.

The combination of scattering matrices~\eqref{eq:addingS_3} is a modification of Ohm's law for adding two conductors in series to the case when the conductors are coherent.
It can easily be generalized for more complicated networks of scattering matrices when the subparts have more than two leads.
Some numerical simulations use a phenomenologically defined network of scattering matrices as a starting point for the quantum transport calculations, see, e.g., \cite{chalker1988}.

\subsection{From continuum models to discrete Hamiltonians}
\label{sec:from-cont-models}
\begin{figure*}
  \centering
  \includegraphics[width=0.9\linewidth]{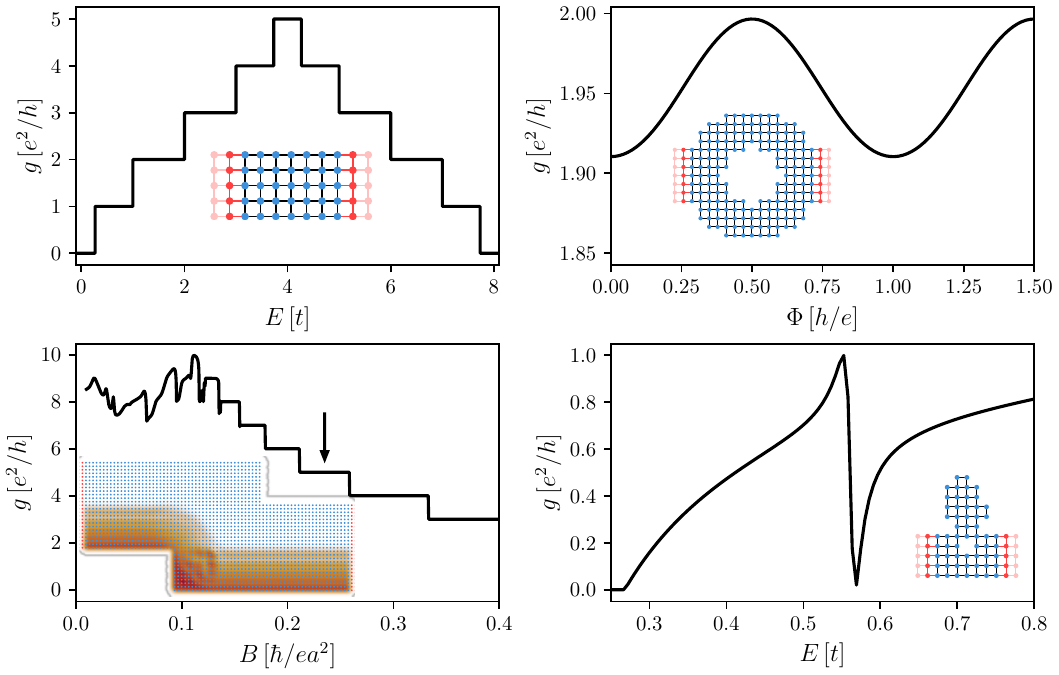}
  \caption{Examples of conductance calculations in a two dimensional square lattice. For each example the inset shows the tight-binding system that has been used with the scattering region in black and the first two layers of semi-infinite electrodes in red. Upper left: conductance quantization in a perfect ballistic quantum wire.  Conductance $g$ is shown as a function of the Fermi energy $E$ expressed in units of the nearest-neighbor hopping strength $t$. Upper right: Aharonov--Bohm effect in a ring device threaded by magnetic flux $\Phi$. Lower left: quantum Hall effect in an example shape. The conductance is shown as a function of magnetic field in units of flux quanta per surface of a lattice unit cell ($a^2$).  The inset displays the current density for one particular value of magnetic field which is indicated by the arrow. Lower right: Fano resonance in a quantum dot device.}
  \label{fig:practical_examples}
\end{figure*}

We now turn to the main focus of this review, the design of numerical techniques to compute the $S$ matrix and other related quantities.
The vast majority of these techniques are built around discretized versions of the Hamiltonian of the system.
These discrete models are obtained by various means that include the atomistic tight-binding approach where the Schr\"odinger equation is projected onto a basis of atomic orbitals, or a discretization of an effective $\mathbf{k}\cdot\mathbf{p}$ or other continuum Hamiltonians.
In this introductory section, we use a minimal three-point finite difference discretization scheme on a square lattice with lattice constant $a$:
$\partial f(x)/\partial x \approx [f(x+a) -f(x-a)]/(2a)$ and $\partial^2 f(x)/\partial x^2\approx [f(x+a) + f(x-a) - 2f(x)]/a^2$ \cite{datta1995}.
Introducing $\Psi_{n_x,n_y} =\Psi(n_x a, n_y a )$ and $V_{n_x,n_y} = V(n_x a,n_y a)$, we obtain the discretized version of Eq.~\eqref{eq:schro2D}:
\begin{equation}
\label{eq:schro2D-discrete}
\begin{multlined}
E \Psi_{n_x,n_y}= \frac{-\hbar^2}{2m^*a^2} \left[ \Psi_{n_x+1,n_y} + \Psi_{n_x-1,n_y} + \Psi_{n_x,n_y+1} \right. \\
\left. + \Psi_{n_x,n_y-1} - 4 \Psi_{n_x,n_y} \right]  + V_{n_x,n_y} \Psi_{n_x,n_y}.
\end{multlined}
\end{equation}
In general, the discretization procedure yields systems of equations of the form
\begin{eqnarray}
\sum_{m} H_{nm} \Psi_m = E \Psi_n,
\end{eqnarray}
where the indices $n$ and $m$ label the \emph{sites} and \emph{orbitals} of the system.
In our example each index consists of a tuple $(n_x,n_y)$, but in general they include other degrees of freedom such as spin or orbital index, as well as more dimensions or lattices.
The off-diagonal matrix elements $H_{nm}$ with $n\neq m$ are called \emph{hoppings}, while the diagonal ones $H_{nn}$ are the \emph{onsite energies}.
The hoppings define the underlying graph structure of the scattering problems, such as the ones shown in the different insets of Fig.~\ref{fig:practical_examples}.

To keep the algebra simple, we further restrict the construction of the solution to just one dimension and drop the subscript in the index $n_x \rightarrow n$.
Setting the unit of energy to $\hbar^2/(2m^*a^2)$ and absorbing the constant $-4$ shift in the definition of $V_n$, we seek to solve the following set of equations:
\begin{equation}
\label{eq:schro1D-discrete}
-\Psi_{n+1}  -\Psi_{n-1} + V_n \Psi_{n} = E \Psi_{n}.
\end{equation}
where $V_n \ne 0$ in the scattering region $1 \leq n \leq L$.
Similar to the continuum problem, we solve the Schrödinger equation in the electrodes first by requiring that the wave function in the left electrode region $n\le 0$ is a superposition of plane waves:
\begin{equation}
\label{eq:discrete-modes-1D-left}
\Psi_n = e^{ik n} +  r e^{-ik n},
\end{equation}
while in the right electrode
\begin{equation}
\label{eq:discrete-modes-1D-right}
\Psi_n = t e^{ik n}.
\end{equation}
Note that there are no decaying modes when there is a propagating mode in this one-dimensional problem.
This simplifies the problem significantly for this pedagogical example, and we will discuss the general case with evanescent modes in detail in Sec.~\ref{sec:scatt-form-discr}.

The dispersion relation in the electrodes for the unique mode $\alpha=1$ now takes the form
\begin{equation}
E_1(k) = -2\cos (k),
\end{equation}
and the current is $I = \Psi_n^* \Psi_{n+1} - \Psi_{n+1}^* \Psi_{n}= |t|^2 v(k)$ where the velocity now becomes $v(k) \equiv dE/dk = 2\sin k$.
The above set of equations matches its continuous version in the large wavelength limit $k\rightarrow 0$, or in other words whenever the wave function changes slowly on the scale of the lattice constant.
In the minimal example, obtaining the propagating modes \eqref{eq:discrete-modes-1D-left}, \eqref{eq:discrete-modes-1D-right} in the lead is very simple, however once the Hamiltonian becomes more complex, this calculation requires advanced algorithms that we describe in Sec.~\ref{sec:defin-infin-lead}.

To calculate the transmission amplitude $t$, the reflection amplitude $r$ and the scattering wave function $\Psi_n$ inside the scattering region, we need to ``match'' the wave function at the scattering region/electrode interface.
Using Eq.~\eqref{eq:schro1D-discrete} for $n = 0, 1$, $2\le p \le  L-1$, $L$, and $L+1$ we get a set of $L+2$ linear equations with as many unknowns,
\begin{subequations}
\label{eq:simple-PsiS}
\begin{eqnarray}
r e^{-ik}  -\Psi_1 &=& -e^{ik }, \label{eq:simple-PsiSa}\\
r + (E-V_1) \Psi_1 + \Psi_2 &=& -1, \label{eq:simple-PsiSb}\\
\Psi_{p-1} + (E-V_p) \Psi_{p} + \Psi_{p+1}  &=& 0, \label{eq:simple-PsiSc}\\
\Psi_{L-1} + (E-V_L)\Psi_{L} + t e^{ik(L+1)} &=& 0, \label{eq:simple-PsiSd}\\
\Psi_L - t e^{ikL} &=& 0. \label{eq:simple-PsiSe}
\end{eqnarray}
\end{subequations}
Solving this set of equations for the vector $(\Psi_1,\Psi_2,\dots\Psi_L, r, t)^T$ is usually done numerically using standard linear algebra routines for sparse matrices.
A large fraction of this review is devoted to setting up this linear system of equations in general situations where the geometry, lattice, number of electrodes, etc.\ is arbitrary.

Solving for the vector $(\Psi_1,\Psi_2,\dots\Psi_L, r, t)^T$ provides both the scattering matrix and the scattering wave function in one large vector, and we therefore call it the \emph{$\Psi S$-approach}.
While not historically first, we find this approach to be more mathematically direct, and we use it to derive equivalent alternative formulations.
This is also the approach used in, e.g., the Kwant package developed by some of us~\cite{groth2014}.

\subsection{Self energies and Green's functions}

The $\Psi S$-approach is mathematically equivalent to the older \emph{wave function}, \emph{mode-matching}, or \emph{open boundaries} approach~\cite{ando1991,khomyakov2005} that introduces the concept of self-energy $\Sigma(E)$.
To obtain the wave function approach, we eliminate $r$ and $t$ from the set of linear equations Eqs.~\eqref{eq:simple-PsiS}, a procedure also known as ``integrating out the leads''.
The remaining equations contain only $L$ unknowns $(\Psi_1,\Psi_2,\dots\Psi_L)^T$, and read
\begin{subequations}
\label{eq:wavefunction_approach}
\begin{eqnarray}
(E-V_1- \Sigma) \Psi_1 + \Psi_2 &=& -i\Sigma v, \\
\Psi_{p-1} + (E-V_p) \Psi_{p} + \Psi_{p+1}  &=& 0, \\
\Psi_{L-1} + (E-V_L-\Sigma)\Psi_{L}  &=& 0,
\end{eqnarray}
\end{subequations}
where we introduced the lead self-energy
\begin{equation}
\Sigma (k) = -e^{ik}.
\end{equation}
This set of equations is the simplest example of what is known in the literature as the ``wave function approach'' to the quantum transport problem.

An alternative way to arrive at the same equations is provided by the Green's function approach, which introduces the retarded Green's function~\cite{fisher1981} as a solution of
\begin{equation}
[E - H +i\eta ] G = 1.
\end{equation}
Here $\eta$ is an infinitesimal energy that sets the boundary conditions to be equivalent to those in the $\Psi S$-approach, and makes the retarded Green's function the Fourier transform of the time evolution operator between different sites of the scattering region.
The Green's function formalism is the original approach used in the early days of computation quantum transport, and it is extremely common to this day.
Writing down the Green's function equations in terms of individual elements of the Green's function yields
\begin{subequations}
  \begin{eqnarray}
  (E-V_1- \Sigma) G_{1q} + G_{2q} &=& \delta_{1q}, \\
  G_{p-1,q} + (E-V_p) G_{pq} + G_{p+1,q}  &=& \delta_{pq}, \\
  G_{L-1,q} + (E-V_L-\Sigma)G_{Lq}  &=& \delta_{Lq},
  \end{eqnarray}
\end{subequations}
where the left-hand side matrix is the same as in Eqs.~\eqref{eq:wavefunction_approach}, and the right-hand side is an identity matrix.

\subsection{Practical examples of numerical calculations}
\label{sec:practical-examples}
We end this pedagogical section with a few examples of practical systems that illustrate the kinds of calculations that can be made.
We restrict these examples to toy models of simple devices made from a two-dimensional square lattice with a single orbital per site.
More advanced examples involving more complex lattices (e.g., graphene), realistic modeling, multi-terminal devices, three dimensions, and treatment of internal degrees of freedom such as spin or particles and holes in superconductivity will be presented in the Sec.~\ref{sec:select-appl}.

Figure~\ref{fig:practical_examples} shows the conductance as a function of Fermi energy or magnetic field strength for four simple devices.
The first device is a five-site-wide perfect wire.
Due to the translation invariance of the wire all the available conductance channels have perfect transmission $T_a = 1$.
Consequently, $g(E)$, shown in the upper left panel of Fig.~\ref{fig:practical_examples}, has a staircase shape with the maximum number of propagating channels occurring in the middle of the band, where the corresponding infinite 2D system has a Van Hove singularity.
Analyzing the number of open modes in a lead is a useful part of the workflow for setting up more complex calculations.
The Hamiltonian matrix of this example has a simple graph structure that is displayed in the inset.
The Hamiltonian itself is given by Eq.~\eqref{eq:schro2D-discrete} with index $n_y$ running from $1$ to $5$ due to the finite system width.

A ring connected to two electrodes and threaded by magnetic flux, shown in the upper right panel of Fig.~\ref{fig:practical_examples}, allows us to study the Aharonov--Bohm effect.
To account for the presence of magnetic flux $\phi \hbar/e$ through the hole, we modify the Hamiltonian hopping matrix elements $H_{nm}$ along a vertical line in the lower arm of the ring $H_{nm} \rightarrow e^{i\phi} H_{nm}$.
Because of gauge invariance, the conductance does not depend on the position of the cut along which the hoppings are modified.
Other than that and the shape of the scattering region, the Hamiltonian is again given by Eq.~\eqref{eq:schro2D-discrete}.
The oscillations of $g(\phi)$ demonstrate the Aharonov--Bohm effect.

A step-shaped sample in the lower left panel of Fig.~\ref{fig:practical_examples} is subject to a uniform magnetic field, which brings it to the quantum Hall regime once the field is strong enough.
This magnetic field prevents current from flowing through the middle of the sample, and creates edge states that carry current without backscattering [shown in the inset of the lower left panel of Fig.~\ref{fig:practical_examples}].
To include a vector potential in a tight-binding system we use the Peierls substitution~\cite{peierls1933,hofstadter1976,panati2003,li2020}, which modifies the hopping according to
\begin{equation}
  \label{eq:peierls}
  H_{nm} \to H_{nm} \exp\left[\ri e/\hbar\int_{\bm{r}_n}^{\bm{r}_m} \bm{A}d\bm{r}\right].
\end{equation}
We choose the vector potential in the Landau gauge with $\bm{B} = B \hat{z}$, so that $\bm{A} = -By \hat{x}$, and the hoppings become
\begin{equation}
  H_{nm}(\Phi) = H_{nm}(0) \times \re^{-\ri \Phi (n_x - m_x) (n_y + m_y)/2},
\end{equation}
with $\Phi = B e a^2 / \hbar$ the magnetic flux per lattice unit cell in units of flux quanta, and $a$ the lattice constant.
Note that we have chosen $\mathbf{A}$ such that $H_{nm}$ does not depend on $x$.
This allows us to use the same gauge in $x$-directed electrodes while keeping the electrode Hamiltonian translation invariant.\footnote{
  If the leads are not all parallel, choosing a vector potential compatible with the lead translation symmetry becomes more complex~\cite{baranger1989}. However, the Kwant package implements a general solution to this problem.
}

A scattering setup with an irregular shape shown in the lower right panel of Fig.~\ref{fig:practical_examples} emulates a quantum dot attached to a quantum wire.
The dot traps a resonant level with energy $E_0$ and a lifetime $\Gamma$, so that interference between the direct transmission and resonant tunneling gives rise to the conductance $g(E) \sim |t_d + \Gamma/(E-E_0 + i\Gamma)|^2$, with $t_d$ the amplitude of direct transmission.
The conductance trace then has the characteristic asymmetric shape known as the Fano resonance~\cite{gores2000,clerk2001}.

\section{Scattering formalism for discrete models}          %
\label{sec:scatt-form-discr}

There are several equivalent descriptions of quantum transport that use the Green's functions, wave functions, or the scattering matrix as the main tool.
This review chooses the $\Psi S$-formalism as the starting point and defines everything in terms of the scattering wave functions and the associated scattering matrix $S$.
This section provides a comprehensive description of the $\Psi S$-approach and includes several proofs that are scattered in the literature as well as some new material.
The other mathematically equivalent approaches are derived as corollaries of the scattering approach in the later sections.

\begin{figure}
  \centering
  \includegraphics[width=0.45\linewidth]{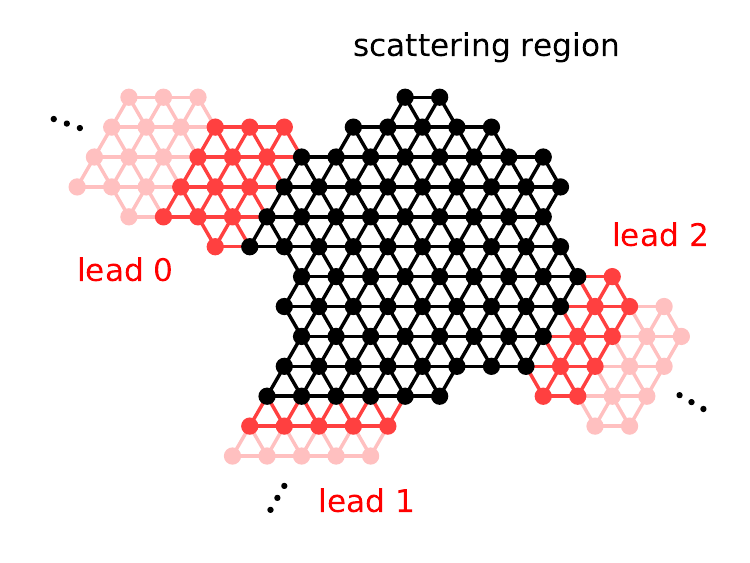}
  \caption{Schematic of a typical system for numerical simulations.
    The finite scattering region (black) is connected to semi-infinite, translationally invariant quasi-one-dimensional electrodes (red).
    The dark-light-red colored cells correspond to the first and second unit cells of the semi-infinite leads, respectively.
    Each lead can be at a different chemical potential and/or temperature.
    Reproduced from \cite{groth2014} under a CC BY 3.0 licence.}
  \label{fig:schematic_kwant}
\end{figure}

A generic system considered here is shown in Fig.~\ref{fig:schematic_kwant}.
It consists of a central finite \emph{scattering} region that has an arbitrary Hamiltonian.
The scattering region is connected to a number of semi-infinite electrodes, each consisting of an infinite number of identical repeated unit cells.

\subsection{Definition of the infinite lead problem}
\label{sec:defin-infin-lead}

We first analyze the wave function in the translationally invariant region of the leads without considering the tight-binding equations of the scattering region.
Each unit cell contains $N_\mathrm{t}$ sites and it has an onsite Hamiltonian matrix $H$ and the hopping matrix $V$ coupling one cell
 to the next.
Grouping the degrees of freedom of the unit cells of multiple disconnected leads into a single vector defines a single larger lead with the Hamiltonian and the hopping being direct sums of the Hamiltonians and hoppings of the individual leads.
Therefore without loss of generality we restrict our discussion to a single lead with the infinite Hamiltonian,
\begin{equation}
  \hat H_\text{lead} = \begin{pmatrix}
    \ddots & \ddots & \ddots  \\
    & V & H& V^\dagger  \\
    & & V & H & V^\dagger\\
    & & & V & H & V^\dagger\\
    & & & & \ddots &\ddots & \ddots\\
  \end{pmatrix},
\end{equation}
and a wave function $\hat{\phi} = [\ldots, \hat{\phi}(j-1), \hat{\phi}(j), \hat{\phi}(j+1), \ldots]^T$.
Because the tight-binding equations are translationally invariant, the wave function can be written as a superposition of the eigenvectors of translation operator, each of the form
\begin{equation}
\hat\phi(j) = \lambda^j \phi,
\end{equation}
where the solutions with $\lambda \equiv e^{ik}$ and hence $|\lambda| = 1$ are plane waves, $|\lambda| \neq 1$ are not normalizable and correspond to evanescent modes.
Applied to a translation eigenvalue, the Schrödinger equation $\hat H_\text{lead} \hat\phi = E \hat\phi$ reads
\begin{equation}
\label{eq:lead_def}
V\phi+ \lambda(H-E)\phi + \lambda^2 V^\dagger\phi=0.
\end{equation}
This is a quadratic eigenvalue problem for $\lambda$ and $\phi$ which has in general $2N_\mathrm{t}$ solutions for any energy $E$ \cite{tisseur2001}.
Introducing $\xi \equiv \lambda \phi$ recasts Eq.~\eqref{eq:lead_def} into a generalized eigenvalue problem
\begin{equation}
\label{eq:lead_def_matrix1}
\begin{pmatrix}
H-E & V^\dagger \\
1  & 0
\end{pmatrix}
\begin{pmatrix}
\phi \\ \xi
\end{pmatrix}
=
\lambda^{-1}
\begin{pmatrix}
  -V & 0 \\
  0  & 1
  \end{pmatrix}
\begin{pmatrix}
\phi \\
\xi
\end{pmatrix}.
\end{equation}
The inverse problem of Eq.~\eqref{eq:lead_def_matrix1} is finding the eigenvectors and eigenenergies of propagating waves with a given wave vector $k$:
\begin{align}
  H(k) \phi &= E(k) \phi,\\
  H(k) &= H + e^{ik} V^\dagger + e^{-ik} V.
\end{align}
This defines the band structure of the lead with the bands $E_\alpha(k)$, found by stable numerical diagonalization of a Hermitian matrix $H(k)$.
The generalized eigenvalue problem Eq.~\eqref{eq:lead_def_matrix1}, on the other hand, is not Hermitian, and therefore solving it in a numerically stable way is a nontrivial problem that we address in Sec.~\ref{sec:stable-form-infin}.

\begin{figure}
  \centering
  \includegraphics[width=0.45\linewidth]{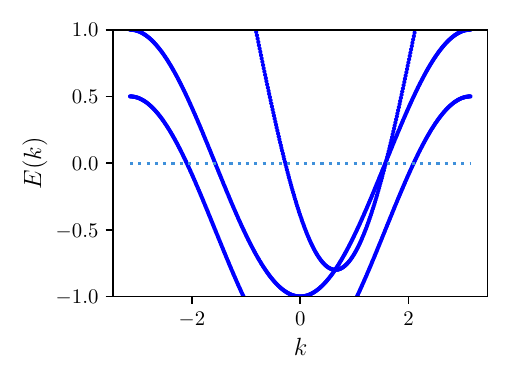}
  \caption{Example of a lead dispersion relation $E(k)$ in a simple situation where $H(k)$ is
    a $3\times 3$ matrix. For a given value of $k$, $H(k)$ always has $3$ eigenvalues. However for
    a given energy $E$, the number of different values of $k$ such that $E(k)=E$ can be strictly smaller than $6=3\times 2$. In this example for, e.g., $E=+0.75$ there are only $2\times 2$ values of $k$ ($2$ with $\partial E/\partial k >0$ and
    $2$ with $\partial E/\partial k <0$) which signifies the existence of evanescent states.
  }
  \label{fig:band}
\end{figure}

An example of dispersion relation of a $3$ sites lead is shown in Fig.~\ref{fig:band}. Note that for a given value of $k$ there always are $N_\mathrm{t}$ eigenenergies (here $3$). In contrast, for a given $E$ there are at most $2 N_\mathrm{t}$ different propagating states. In the following, we suppose that we have found the different solutions of Eq.~\eqref{eq:lead_def} at a given energy $E$ using the algorithms discussed later in this review.

Solving the general eigenvalue problem Eq.~\eqref{eq:lead_def_matrix1}
can be rather subtle if one wants to do it in a robust way, in particular
when the matrix $V$ is not invertible. This problem has been discussed in a number of publications \cite{rungger2008,sanvito1999,fujimoto2003,krstic2002,khomyakov2005,tsukamoto2014,
ono2016,tsukamoto2017,iwase2018} and will be investigated in detail below and in the next section.
A close connection exists between the lead problem (of interest here) and the problem of calculating the properties of surfaces of metallic systems that use a closely connected construction  \cite{allen1979,allen1979b,lee1981b,lee1981c,umerski1997}

\subsection{Separation of propagating and evanescent modes}
\label{sec:separ-prop-evan}
After one has solved the general eigenvalue problem Eq.~\eqref{eq:lead_def_matrix1},  the next step in the construction is to classify the different solutions $\phi_\alpha$ into propagating (p) and evanescent (e) modes, according to the value of $|\lambda|$:
propagating for $|\lambda|=1$ and evanescent for $|\lambda|\ne 1$.
Then, the propagating modes are further subclassified into outgoing (+)
modes with positive velocity $v=(1/\hbar) dE/dk>0$ and incoming (-) modes
with $v<0$. For normalized modes such that $\phi^\dagger \phi = 1$, the mode velocity can be computed through the Feynman--Hellmann equation (see Appendix~\ref{sec:deriv-feynm-hellm} for its derivation)
\begin{equation}
v = i\phi^\dagger \left[ \lambda V^\dagger - \lambda^{-1}V \right]\phi.
\end{equation}
Note that there is always the same number $N_\mathrm{p}$ of incoming and outgoing propagating channels.
This comes from the $2 \pi$-periodicity and the continuity of the bands $E(k)$.
Indeed, a horizontal line at a given energy $E$ must cross each band an even number of times, with half of the crossings at a positive slope and the other half with a negative slope, hence the equal number of modes with positive and negative velocities.
For future convenience, we normalize the propagating modes so that they carry unit velocity $\pm 1$.
The same subclassification is applied to the $N_\mathrm{e}$ evanescent modes: outgoing (+) for those decaying at $+\infty$  ($|\lambda|< 1$) and incoming (-) for the others ($|\lambda|> 1$). The evanescent modes have zero velocities, and therefore their normalization may be chosen arbitrarily.
As for propagating modes, the evanescent modes also appear in pairs: if $\lambda$ is a solution, then so is
$1/\lambda^*$. Since the total number of incoming (or outgoing) modes is equal to the number of sites in a unit cell, we have
$N_\mathrm{t} = N_\mathrm{p} + N_\mathrm{e}$.

The last step of this section is to gather the eigenstates $\phi_\alpha$ into a $N_\mathrm{t} \times 2N_\mathrm{t}$ matrix $\Phi$:
the column $\alpha$ of $\Phi$ is a vector $\phi_\alpha$, or
\begin{equation}
\Phi_{n\alpha} = (\phi_\alpha)_n.
\end{equation}
We further define the submatrices $\Phi_\mathrm{p+}$, $\Phi_\mathrm{p-}$,$\Phi_\mathrm{e+}$,$\Phi_\mathrm{t+}$\dots as the subblocks of $\Phi$ containing the corresponding modes. Similarly, we introduce the $2N_\mathrm{t} \times 2N_\mathrm{t}$ diagonal matrix $\Lambda$ that contains the eigenvalues $\lambda_\alpha$ on its diagonal. The submatrices  $\Lambda_\mathrm{p+}$, $\Lambda_\mathrm{p-}$,\dots are their restriction to the corresponding modes. In some cases, we will exclude the modes that belong to the kernel of the matrix $V$ (i.e.,
such that $V\phi_\alpha = 0$) and we note $\Phi_\mathrm{\bar t+}$, $\Lambda_\mathrm{\bar t+}$ the corresponding matrices. We denote $\Phi_\mathrm{o+}$ as the matrix that only contains those modes belonging to the kernel of $V$ ($V\Phi_\mathrm{o+}=0$). Keeping track of which modes are included in these matrices is central to all the various proofs and formulas that are derived later in this article, so that we gather this notation in Table~\ref{tab:whichmodes} for future reference.
\begin{table*}[t]
  \centering
  \small
  \begin{tabular}{c|l|c|c}
    Symbol & Included modes & Matrix size & Direction\\
    \hline
    $\Phi_\mathrm{p+}$      & propagating & $N_\mathrm{t} \times N_\mathrm{p}$ & outgoing \\
    $\Lambda_\mathrm{p+}$   & propagating & $N_\mathrm{p} \times N_\mathrm{p}$ & outgoing \\
    $\Phi_\mathrm{p-}$      & propagating & $N_\mathrm{t} \times N_\mathrm{p}$ & incoming \\
    $\Phi_\mathrm{e+}$      & evanescent  & $N_\mathrm{t} \times N_\mathrm{e}$ & decaying away \\
    $\Phi_\mathrm{\bar e+}$ & evanescent, except the $\lambda_\alpha=0$ modes & $N_\mathrm{t} \times N_\mathrm{\bar e}$ & decaying away \\
    $\Phi_\mathrm{t+}$      & $\left(\Phi_\mathrm{p+} | \Phi_\mathrm{e+}\right)$: propagating \& evanescent & $N_\mathrm{t} \times N_\mathrm{t}$ & outgoing \& decaying away\\
    $\Phi_\mathrm{\bar t+}$ & $\left(\Phi_\mathrm{p+} | \Phi_\mathrm{\bar e +}\right)$: prop.\ \& evan., except $\lambda_\alpha=0$ & $N_\mathrm{t} \times N_\mathrm{\bar t}$ & outgoing \& decaying away\\
    $\Phi_\mathrm{t-}$      & $\left(\Phi_\mathrm{p-} | \Phi_\mathrm{e+}\right)$: propagating \& evanescent & $N_\mathrm{t} \times N_\mathrm{t}$ & incoming \& decaying away\\
    $\Phi_\mathrm{\bar t-}$ & $\left(\Phi_\mathrm{p-} | \Phi_\mathrm{\bar e +}\right)$: prop.\ \& evan., except $\lambda_\alpha=0$ & $N_\mathrm{t} \times N_\mathrm{\bar t}$ & incoming \& decaying away\\
    $\Phi_\mathrm{o+}$      & only the $\lambda_\alpha=0$ modes & $N_\mathrm{t} \times N_\mathrm{o}$ & ---  \\
  \end{tabular}
  \caption{
    \label{tab:whichmodes}
    Notation for the different modes in the electrodes. The letters stand respectively for evanescent (e), propagating (p) total (t), $\lambda_\alpha = 0$ (o) and total without o modes ($\mathrm{\bar t}$). $+$ ($-$) indicates propagation or decay away from (towards) the scattering region. $N_\mathrm{t} = N_\mathrm{p} + N_\mathrm{e}= N_\mathrm{\bar t} + N_\mathrm{o}$ and $N_\mathrm{e}= N_\mathrm{\bar e}+N_\mathrm{o}$. Note that the evanescent modes $\Phi_\mathrm{e-}$ that are growing exponentially in the lead region are never needed in the calculations. The notation $\left(A | B \right)$ means stacking the columns of the $B$ matrix on the right of the columns of the $A$ matrix.}
\end{table*}

These matrices will be important ingredients of the construction of the scattering problem and allow one to express many things in a compact form. For instance, a general eigenstate of the lead Hamiltonian at a given energy can be written as a superposition of all modes which reads,
\begin{equation}
\hat\phi(j) = \Phi\Lambda^j q.
\end{equation}
Here $q$ is a vector of size $2N_\mathrm{t}$. Using the submatrices, we can form less general states. For instance, a purely
propagating state takes the form,
\begin{eqnarray}
\hat\phi(j) &=& \Phi_\mathrm{p}(\Lambda_\mathrm{p})^j q_\mathrm{p} \nonumber \\
 &=& \Phi_\mathrm{p+}(\Lambda_\mathrm{p+})^j q_\mathrm{p+} + \Phi_\mathrm{p-}(\Lambda_\mathrm{p-})^j q_\mathrm{p-}.
\end{eqnarray}
Using these matrices, Eq.~\eqref{eq:lead_def} takes the form,
 \begin{equation}
 \label{eq:lead_def2}
 V \Phi + (H-E) \Phi \Lambda + V^\dagger \Phi \Lambda^2 = 0.
 \end{equation}

\subsection{Formulation of the Scattering problem as a set of linear equations}
\label{sec:form-scatt-probl}

We now turn to the scattering problem {\it per se}:
we connect the lead to a finite scattering region sr.
The total system is now semi-infinite and is described by the Hamiltonian,
\begin{equation}\label{eq:h_sys}
  \hat H_\text{sys} = \begin{pmatrix}
    H_\text{sr} & P_\text{sr}^TV^\dagger\\
    V P_\text{sr} & H& V^\dagger  \\
    & V & H & V^\dagger\\
    & & V & H & V^\dagger\\
    & & & \ddots & \ddots &\ddots\\
  \end{pmatrix},
\end{equation}
where  $H_\text{sr}$ is the (big) $N_\text{sr}\times N_\text{sr}$ Hamiltonian of the scattering region.
Without loss of generality, we have included the first layer of the lead inside the scattering region. Also, the multi-lead problem can be treated by using a larger effective lead.
The coupling of the lead to the scattering region takes the specific form $V P_\text{sr}$ where the projector $P_\text{sr}$ is a $N_\mathrm{t} \times N_\text{sr}$ rectangular matrix with ones in the diagonal and zeros everywhere else.

We wish to calculate the scattering states $\hat\psi$ of $\hat H_\text{sys}$.
We denote $\hat{\psi}$ in the scattering region $\psi_\text{sr}$.
In the lead, the scattering states are formed by a combination of the incoming and outgoing propagating modes as well as the evanescent outgoing modes.
The general form of $\hat\psi$ in the lead reads,
\begin{equation}
\label{eq:out}
\hat\psi(j) =  \Phi_\mathrm{t+}(\Lambda_{\mathrm{t}+})^j q_\mathrm{t+} + \Phi_\mathrm{p-}(\Lambda_\mathrm{p-})^j q_\mathrm{p-},
\end{equation}
where the index $j>0$ labels the different lead unit cells.
In other words, we seek wave functions of the form
\begin{equation}
  \label{eq:psihat}
  \hat\psi =
  \begin{pmatrix}
    \psi_\textrm{sr}\\
    \Phi_\mathrm{t+}\Lambda_\mathrm{t+} q_\mathrm{t+} + \Phi_\mathrm{p-}\Lambda_\mathrm{p-} q_\mathrm{p-}\\
    \Phi_\mathrm{t+}(\Lambda_\mathrm{t+})^2 q_\mathrm{t+} + \Phi_\mathrm{p-}(\Lambda_\mathrm{p-})^2 q_\mathrm{p-}\\
    \vdots
  \end{pmatrix}.
\end{equation}

The Schrödinger equation $\hat H_\text{sys} \hat\psi = E \hat\psi$ imposes a linear relation between the incoming and outgoing modes---equivalent to the wave matching condition in the continuum---so that,
\begin{equation}
\label{eq:Stp}
q_\mathrm{t+} = S_\mathrm{tp} q_\mathrm{p-},
\end{equation}
where $S_\mathrm{tp}$ is the generalized scattering matrix. $S_\mathrm{tp}$ is a $(N_\mathrm{p}+N_\mathrm{e})\times N_\mathrm{p}$ matrix
which extends the usual definition of the scattering matrix:
it contains the outgoing propagating states {\it as well as} the outgoing evanescent ones.
The usual scattering matrix $S \equiv S_\mathrm{pp}$ is recovered by taking the submatrix consisting of the $N_\mathrm{p}$ rows
corresponding to outgoing propagating states.

Writing the Schrödinger equation $\hat H_\text{sys} \hat\psi = E \hat\psi$ in terms of the decomposition Eq.~\eqref{eq:out}, one arrives at a set of linear equations for $\psi_\text{sr}$ and $S_\mathrm{tp}$. Introducing the $N_\text{sr} \times N_\mathrm{p}$ matrix $\Psi_\text{sr}$
whose different columns contain the different solutions $\psi_\text{sr}$ corresponding to the different incoming propagating modes allows one to write the Schrödinger equation in a compact form.
Making additionally use of Eq.~\eqref{eq:lead_def2}, we arrive at the following general formulation of the scattering problem:
\begin{equation}
\label{eq:scatt_form}
\begin{pmatrix}
H_\text{sr} - E & P_\text{sr}^T V^\dagger \Phi_\mathrm{t+} \Lambda_\mathrm{t+} \\
V P_\text{sr} & -V \Phi_\mathrm{t+}
\end{pmatrix}
\begin{pmatrix}
\Psi_\text{sr}\\
S_\mathrm{tp}
\end{pmatrix}
=
\begin{pmatrix}
-P_\text{sr}^T V^\dagger \Phi_\mathrm{p-} \Lambda_\mathrm{p-} \\
V \Phi_\mathrm{p-}
\end{pmatrix}.
\end{equation}
The problem is now formally reduced to solving a set of (usually very sparse) linear equations, which can
be done very efficiently by various numerical packages.

\subsection{Dealing with non-invertible hopping matrices}

The linear system \eqref{eq:scatt_form} becomes ill-conditioned when the hopping matrix $V$ is not invertible \cite{rungger2008}, i.e., when there are modes such that $V \Phi_\mathrm{o,+} = 0$. In that case, the columns $\left(P_\text{sr}^T V^\dagger \Phi_\mathrm{o+} \Lambda_\mathrm{o+}, -V \Phi_\mathrm{o+}\right)^\dagger$ in the left-hand side of Eq.~\eqref{eq:scatt_form} become identical to zero. This is due to the fact that the $\Phi_\mathrm{o+}$-modes do not contribute explicitly to the scattering wave function. Instead, we should formulate the problem in terms of the modes $\Phi_\mathrm{\bar{t}+}$.

If this is done naively in Eq.~\eqref{eq:scatt_form}, the matrix on the left-hand side becomes rectangular, and the linear system is thus overdetermined. We can however return to a square matrix by introducing the singular value decomposition of $V=U_A D U_B^\dagger$ where $U_A$ and $U_B$ are unitary matrices and the diagonal matrix $D$ contains $N_\mathrm{\bar{t}}$ non-zero singular values. Defining  $A = U_A\sqrt{D}$, $B = U_B\sqrt{D}$
and reshaping these matrices to keep only their non-zero part, we arrive at
\begin{equation}
\label{eq:hopping_decomposition}
V = A B^\dagger,
\end{equation}
where $A$ and $B$ are $N_\mathrm{t} \times N_\mathrm{\bar{t}}$ matrices of orthogonal vectors.

With this decomposition we find
\begin{equation}
\begin{pmatrix}
H_\text{sr} - E & P_\text{sr}^T V^\dagger\Phi_\mathrm{\bar{t}+} \Lambda_\mathrm{\bar{t}+} \\
A B^\dagger P_\text{sr} & -A B^\dagger \Phi_\mathrm{\bar{t}+}
\end{pmatrix}
\begin{pmatrix}
\Psi_\text{sr}\\
S_\mathrm{tp}
\end{pmatrix}
=
\begin{pmatrix}
-P_\text{sr}^T V^\dagger \Phi_\mathrm{p-} \Lambda_\mathrm{p-} \\
A B^\dagger \Phi_\mathrm{p-}
\end{pmatrix}.
\end{equation}
Noticing that the last rows are multiplied by $A$, we find that any solution of
\begin{equation}
\label{eq:scatt_form_general}
\begin{pmatrix}
H_\text{sr} - E & P_\text{sr}^T V^\dagger\Phi_\mathrm{\bar{t}+} \Lambda_\mathrm{\bar{t}+} \\
B^\dagger P_\text{sr} & -B^\dagger \Phi_\mathrm{\bar{t}+}
\end{pmatrix}
\begin{pmatrix}
\Psi_\text{sr}\\
S_\mathrm{tp}
\end{pmatrix}
=
\begin{pmatrix}
-P_\text{sr}^T V^\dagger \Phi_\mathrm{p-} \Lambda_\mathrm{p-} \\
B^\dagger \Phi_\mathrm{p-}
\end{pmatrix}.
\end{equation}
is thus a valid solution of the scattering problem. Equation~\eqref{eq:scatt_form_general} is the most generic formulation of the scattering problem, and we will use it in the majority of this review. The name of the $\Psi S$-approach derives from the fact that one solves simultaneously for $\Psi_\text{sr}$ \emph{and} $S_\mathrm{tp}$.

\subsection{Formulation of the bound state problem}
\label{sec:form-bound-state}

Besides the states that hybridize with the continuum spectrum of the leads, there can also be bound states that decay in the leads. Those are usually outside of the leads bands, but not necessarily (a trivial example being a non connected system or different symmetries between the
scattering region and the leads). An important example of bound states are the Andreev states that form in a normal region sandwiched by superconducting leads. Another example are the edge states at the boundary of topological insulators.

The general form of $\hat\psi(j)$ for a bound state reads,
\begin{equation}
\hat\psi(j) = \Phi_{e+}(\Lambda_{e+})^j q_{e+},
\end{equation}
for $j>0$ (in the lead) and $\psi_\text{sr}$ in the scattering region. With this notation and a bit of algebra described in \cite{istas2018}, the Schrödinger equation translates into
\begin{equation}
\label{eq:bs_problem_hermitian}
\begin{pmatrix}
H_{\text{sr}} - E & P_{\text{sr}}^T V^{\dagger} \Phi_\mathrm{e+} \Lambda_{ e+} \\
\Lambda_{e+}^* \Phi_{e+}^{\dagger} V P_{\text{sr}} & - \Lambda_{ e+}^* \Phi_{e+}^{\dagger} V \Phi_{e+}
\end{pmatrix}
\begin{pmatrix}
\psi_{\text{sr}} \\
q_{e+}
\end{pmatrix}
= 0,
\end{equation}
The matrix $\Lambda_{ e+}$ can contain zero diagonal elements, which gives Eq.~\eqref{eq:bs_problem_hermitian} spurious solutions at every energy $E$. We then introduce the matrices $ \Lambda_\mathrm{\bar e+}$, where we keep only the non-zero eigenvalues, as well as $\Phi_\mathrm{\bar e+}$ and $q_\mathrm{\bar e+}$, whose columns corresponding to the zero eigenvalue have been discarded, to obtain the correct formulation of the problem
\begin{equation}
\label{eq:bs_problem_discarded}
\begin{pmatrix}
H_{\text{sr}} - E & P_{\text{sr}}^T V^{\dagger} \Phi_\mathrm{\bar e+} \Lambda_\mathrm{\bar e+} \\
\Lambda_\mathrm{\bar e+}^* \Phi_\mathrm{\bar e+}^{\dagger} V P_{\text{sr}} & - \Lambda_\mathrm{\bar e+}^* \Phi_\mathrm{\bar e+}^{\dagger}V \Phi_\mathrm{\bar e+}
\end{pmatrix}
\begin{pmatrix}
\psi_{\text{sr}} \\
 q_\mathrm{\bar e+}
\end{pmatrix}
= 0.
\end{equation}

Although Eq.~\eqref{eq:bs_problem_discarded} looks similar to Eq.~\eqref{eq:scatt_form} they are structurally different: there is no source term on the right-hand side of Eq.~\eqref{eq:bs_problem_discarded} so that we cannot simply solve a linear system. In the scattering problem, the energy $E$ is known and one seeks the solutions at that given energy, here we must first find the values of $E = E_\mathrm{BS}$ for which the matrix on the left-hand side of Eq.~\eqref{eq:bs_problem_discarded} becomes singular. For such an energy value, one looks for the corresponding kernel to find $\psi_\text{sr}$ and $q_\mathrm{\bar e+}$. Since $\Phi_\mathrm{\bar e+}$ and $\Lambda_\mathrm{\bar e+}$ depend on energy $E$,
computing the bound state energies amounts to finding the roots of a non-linear function
while computing the associated eigenstate is pure linear algebra \cite{istas2018}.

In the particular case of a semi-infinite wire with no scattering region, $H_{\text{sr}}$ is simply replaced by $H$, $P_\text{sr}$ by the identity and
$\psi_\text{sr} = \Phi_\mathrm{\bar e+} q_\mathrm{\bar e+}$.

In that case, Eq.~\eqref{eq:bs_problem_discarded} can be recast in the form,
\begin{equation}
\label{eq:bs_semiinfinite}
V \Phi_\mathrm{\bar e+} q_\mathrm{\bar e+} = 0,
\end{equation}
which implies that the matrix $V \Phi_\mathrm{\bar e+}$ does not have full rank.

\subsection{Discrete symmetries and conservation laws}
\label{sec:scatt-stat-symm}

The scattering matrix $S$ together with the definition of the scattering modes $\Phi$ describe single energy properties of an infinite system \eqref{eq:h_sys}, and therefore it is constrained by the  symmetries of the Hamiltonian.
Taking explicitly these symmetries into account when solving for the leads is necessary to obtain certain physical observables. For instance, defining a spin-resolved conductance between two normal leads requires keeping track of the spin degree of freedom in the lead. Similarly, calculating the Andreev conductance of a normal metal-superconductor junction requires keeping track of the conserved electron/hole degree of freedom in the normal lead. In the same spirit, to compute topological invariants, one must take into account the discrete symmetry class of the system.
This subsection discusses possible strategies to make use of the symmetries in numerical calculations.

Because $S$ is a map between the two vector spaces spanned by the columns of $\Phi_{p-}$ and $\Phi_{p+}$, it transforms differently from linear operators.
Specifically, applying unitary transformations $\Phi'_{p-} =  \Phi_{p-} U_{p-} $, $\Phi'_{p+} =  \Phi_{p+} U_{p+}$ transforms $S$ into $S' = U_{p+}^\dagger S U_{p-}$.
One strategy would be to describe how the symmetry constraints apply to the scattering matrix for an arbitrary choice of the basis $\Phi$.
Instead, here we utilize the arbitrariness of this choice to choose the symmetry-adapted basis where the action of symmetries on $S$ assumes a simple form, as implemented in the Kwant package \cite{groth2014}.

\subsubsection{Conservation laws}
Unitary symmetries are associated with conservation laws such as the conservation of spin.
As these conservation laws are associated with directly observable quantities, they are relatively straightforward to take into account.
We consider a conserved quantity in a lead. It is given by a Hermitian operator $\hat{A}$ that commutes with the lead Hamiltonian. More specifically, we suppose that $\hat{A}$ takes the form of a block diagonal translationally invariant operator described by a matrix $A$ inside one unit cell:
\begin{equation}
  \hat A = \begin{pmatrix}
      & \ddots &   \\
     & 0 & A& 0  \\
     & & 0 & A & 0\\
     & & & 0 & A & 0\\
     & & & & &\ddots  & \\
  \end{pmatrix}.
\end{equation}
It satisfies $[H,A] = [V,A] = 0$. For instance, the conservation of the spin along the z axis is implemented by the choice of $A = \sigma_z$ where the Pauli matrix $\sigma_z$ acts in the spin sector. The lead Hamiltonian leaves the eigensubspaces of $\hat{A}$ decoupled, and therefore any lead mode belongs to a single eigensubspace of $\hat{A}$.
Let $U^{i}$ be an orthonormal set of eigenvectors for the eigenvalue $a_i$ of $A$:
\begin{equation}
A = \sum_i U^{i} a_i U^{i\dagger},
\end{equation}
we can define the lead problem projected onto the $i$-th eigensubspace of $\hat{A}$ in terms
of the two matrices
\begin{align}
H_i &= U^{i\dagger} H U^{i} \\
V_i &= U^{i\dagger} V U^{i},
\end{align}
with the number of degrees of freedom per unit cell equal to the degeneracy of $a_i$.
Each sub-lead problem for the symmetry sector $i$ is then solved separately to obtain the
associated lead modes $\Phi_i$. Then, we obtain the full solution by just combining the $\Phi_i$ with the $U^i$ using,
\begin{equation}
    \Phi = \sum_i  U^{i} \Phi_i.
\end{equation}

\subsubsection{Discrete symmetries}
In addition to unitary symmetries that are associated with conservation laws, Hamiltonians may possess some of the discrete symmetries: time-reversal $\mathcal{T}$, particle-hole $\mathcal{P}$, and sublattice or chiral symmetry $\mathcal{C}$.
These symmetries are antiunitary ($\mathcal{T}$ and $\mathcal{P}$) and/or antisymmetries ($\mathcal{P}$ and $\mathcal{C}$). Antiunitary symmetries are defined as the product of a unitary matrix with the complex conjugate operator; they flip the sign of the momentum of propagating modes.
Antisymmetries anticommute with the Hamiltonian instead of commuting with it; they change the sign of the mode energy. These operators do not allow a simultaneous eigendecomposition with the Hamiltonian and require a separate treatment.

In the following, we focus on the case where the antiunitary symmetry operators take a form where $\mathcal{T}^2 = \pm 1$ and $\mathcal{P}^2 = \pm 1$, and the chiral symmetry takes a form such that $\mathcal{C}^2 = 1$. It can indeed be shown \cite{varjas2018} that it is always possible to construct such symmetries
(or more precisely a form where all the eigenvalues of $\mathcal{T}^2$  and $\mathcal{P}^2$
are $\pm 1$ together with $\mathcal{C}^2 = 1$.)

To see why such a construction is possible and illustrate how one can take advantage of symmetries practically, let's consider the example where
only the time reversal symmetry is present: $\mathcal{T}= U K$ where $U$ is a unitary matrix acting on the unit cell of the lead and $K$ the complex conjugation operator. One immediately obtains that
$\mathcal{T}^2 = Q \equiv U^* U$, i.e., $\mathcal{T}^2$ is a regular unitary symmetry $Q$ that commutes with the Hamiltonian as well as with $\mathcal{T}$. Let us diagonalize $Q$ first and consider an eigenstate $\mathbf{q}$ with eigenvalue $e^{iq}$, i.e., $Q\mathbf{q}=e^{iq}\mathbf{q}$. We find that $\mathcal{T}\mathbf{q}$ is also an eigenstate of $Q$ with eigenvalue  $e^{-iq}$. If $q=0$ or $\pi$ then we have
$\mathcal{T}^2\mathbf{q} = \pm\mathbf{q} $ and the construction of the new time-reversal operator $\mathcal{T'}$ is trivial in  this sector: $\mathcal{T'}=\mathcal{T}$. In all other situations,
$e^{-iq}\ne e^{iq}$ and the two states $\mathbf{q}$ and $\mathcal{T}\mathbf{q}$ are orthogonal. Now,
we define a new unitary matrix $R$ inside the space spanned by $\mathbf{q}$ and $\mathcal{T}\mathbf{q}$ by $R \mathbf{q} = e^{-iq} \mathbf{q}$ and $R \mathcal{T}\mathbf{q} = \mathcal{T}\mathbf{q}$. From $R$, we can define
a new time-reversal operator $\mathcal{T'}= R \mathcal{T}$. This new operator is obviously the product of a unitary matrix $R U$ times $K$ and commutes with the Hamiltonian, hence it is a proper time-reversal operator.
It is easy to verify that it satisfies $\mathcal{T'}^2 \mathbf{q} = \mathbf{q}$ and
$\mathcal{T'}^2 \mathcal{T}\mathbf{q}= \mathcal{T}\mathbf{q}$, i.e., that $\mathcal{T'}^2=1$ which proves our statement.

These discrete symmetries act on the Bloch Hamiltonian as follows:
  \begin{subequations}
    \begin{align}
      \mathcal{T}H(\boldsymbol{k})\mathcal{T}^{-1} &= H(-\boldsymbol{k}),\\
      \mathcal{P}H(\boldsymbol{k})\mathcal{P}^{-1} &= -H(-\boldsymbol{k}),\\
      \mathcal{C}H(\boldsymbol{k})\mathcal{C}^{-1} &= -H(\boldsymbol{k}).
    \end{align} \label{eq:disc_trans}
  \end{subequations}
Since $\hbar v = \partial E/ \partial k$, $\mathcal{T}$ and $\mathcal{C}$ flip the sign of $v$, while $\mathcal{P}$ leaves it unchanged. It follows that $\mathcal{T}$ and $\mathcal{C}$ can be used to define the outgoing modes from the incoming ones.

\subsubsection{Construction of the lead modes using discrete symmetries}

The strategy to take advantage of the discrete symmetries is to use the symmetries in the construction of the modes. Whenever a discrete symmetry maps one propagating mode onto another, we define one of the two modes in the pair by applying the symmetry operator to the other one.
More precisely, we use the symmetries (if present!) in the following order: if a conservation law exists, it is used first, as described in the previous section. Second we use $\mathcal{T}$ that allows one to define the outgoing modes from the incoming ones. Third, only at $E=0$, we can use $\mathcal{P}$ which allows to define the modes with negative $-\pi < k < 0$ from the mode with positive $0 < k < \pi$.
Last, only at $E=0$, the $\mathcal{C}$ symmetry provides the outgoing mode from the incoming one.

If a discrete symmetry coexists with a conservation law $\hat{A}$, it must map an eigensubspace of $\hat{A}$ onto another eigensubspace of $\hat{A}$, possibly the same one \cite{varjas2018}.
If different eigensubspaces are related by one or more discrete symmetries, we define the modes $\Phi$ in one of these subspaces by applying the discrete symmetry $\mathcal{T}$, $\mathcal{P}$, and/or $\mathcal{C}$ (in the order defined above)  to the modes in the other subspace.

When $\mathcal{P}^2 = -1$, its presence guarantees a Kramers-like degeneracy of propagating modes with the same velocity at $E=0$ for the high-symmetry momenta $k=0$ and $k=\pi$.
We determine a pair of modes related by particle-hole symmetry at these momenta by selecting an arbitrary mode $\phi$, computing its particle-hole partner $\mathcal{P} \phi$, and projecting the remaining modes at this momentum onto the subspace orthogonal to this pair.

When $\mathcal{P}^2=1$ at a high-symmetry momentum $k=0$ or $k=\pi$ and $E=0$, we cannot use the strategy of choosing the wave function of one mode in a pair by acting with the symmetry on another mode.
To implement a symmetry-adapted basis in this case, we choose the mode wave functions such that they are mapped by $\mathcal{P}$ onto themselves.
To do so, we compute the action of $\mathcal{P}$ within the space of modes with the same velocity at a high-symmetry momentum: $\Omega = \Phi^\dagger \mathcal{P} \Phi$.
Transforming $\Phi \rightarrow \Phi \Omega^{1/2}$ through a diagonalization of the $\Omega$ matrix achieves the desired result $\phi = \mathcal{P} \phi$ at a high-symmetry momentum.

The combination of the presence/absence of different symmetries gives rise to many possible scenarios for the properties of the obtained scattering matrix using the above construction \cite{fulga2012}. As an example, in the case where only time reversal symmetry is present, without an additional conservation law, we find
\begin{equation}
S_\mathrm{pp} = \mathcal{T}^2 S_\mathrm{pp}^T.
\end{equation}
Similarly the presence of the chiral symmetry alone leads to
\begin{equation}
S_\mathrm{pp}(E=0) = S_\mathrm{pp}^\dagger\,.
\end{equation}
The presence of a global particle-hole symmetry
$\mathcal{P}$ in the Bogoliubov--De Gennes equation for superconductors combined with the conservation law $A=\tau_z$  (The Pauli matrix $\sigma_z$ acts in the Nambu electron-hole space) leads to
\begin{equation}
S_\mathrm{pp}(E=0) = \begin{pmatrix} 0 & 1\\ \mathcal{P}^2&0\end{pmatrix}S_\mathrm{pp}^*\begin{pmatrix} 0 & \mathcal{P}^2\\ 1&0\end{pmatrix}\, ,
\end{equation}
where $S_\mathrm{pp}$ has a $2\times 2$ block structure in electron-hole space. The conservation law guarantees charge conservation in a non-superconducting lead.
More details and derivations can be found in Appendix~\ref{sec:deriv-symmetry}.
When combining symmetries in different ways, yet more complex situations may arise.

\subsection{Other related wave function approaches}
\label{sec:mixed-wavef-appr}

There are many closely related ways one can calculate the scattering wave functions and in this review, we have focused on the one used in
the Kwant package \cite{groth2014} developed by some of us. Historically,
the wave function approach was pioneered in the work of \cite{ando1991} who integrated out the leads degrees of freedom to obtain
\begin{equation}\label{eq:scattwf_ando}
[ E - H_\text{sr} - P_\text{sr}^T\Sigma P_\text{sr}]\Psi_\text{sr}  = P_\text{sr}^T V^\dagger \Phi_{\mathrm{p}-} \Lambda_{\mathrm{p}-} - P_\text{sr}^T \Sigma \Phi_{\mathrm{p}-}.
\end{equation}
From the perspective of this review, the Ando approach lies halfway between the $\Psi S$-approach and the Green's function approach. A natural derivation of this equation will be done in Sec.~\ref{sec:green} where we will introduce the self-energy matrix $\Sigma(E)$. Variations of this technique have been implemented by several groups in different contexts such as metallic spintronics \cite{khomyakov2005} or constrictions \cite{zhang2017}.

\section{Stable formulation of the lead and scattering problems}%
\label{sec:stable-form-infin}

The purpose of this section is to design a stable algorithm to
find the solutions of the generalized quadratic equation~\eqref{eq:lead_def} and
construct the matrices $\Phi$ and $\Lambda$ that
are needed for the scattering problem.

Quadratic equations like Eq.~\eqref{eq:lead_def},
\begin{equation}
\label{eq:def_lead}
V\Phi+ \lambda(H-E)\Phi + \lambda^2 V^\dagger\Phi=0,
\end{equation}
can be recast into generalized eigenvalue problems with the simple change of
variable $\xi \equiv \lambda \Phi$:
\begin{equation}
\label{eq:lead_def_matrix}
\begin{pmatrix}
H-E & V^\dagger \\
1  & 0
\end{pmatrix}
\begin{pmatrix}
\Phi \\ \xi
\end{pmatrix}
=
\lambda^{-1}
\begin{pmatrix}
-V\Phi \\
\xi
\end{pmatrix}
\end{equation}
Note that we have formulated the eigenproblem in terms of $\lambda^{-1}$ since we are
interested in $|\lambda|\leq 1$, and eigensolvers find large eigenvalues better
than small ones. General eigensolvers are not always stable, however, and the rest of this section is devoted
to the derivation of a robust formulation of the problem.

\subsection{Simple case: \texorpdfstring{$V$}{V} is invertible}
\label{sec:simple-case:-v}

In the simplest situation, the matrix $V$ is invertible.
Multiplying the first line of Eq.~\eqref{eq:lead_def_matrix} with $V^{-1}$, we arrive at
an ordinary eigenproblem,
\begin{equation}
\label{eq:lead_def_matrix_inv}
\begin{pmatrix}
-V^{-1}(H-E) & -V^{-1} V^\dagger \\
1  & 0
\end{pmatrix}
\begin{pmatrix}
\Phi \\ \xi
\end{pmatrix}
=
\lambda^{-1}
\begin{pmatrix}
\Phi \\
\xi
\end{pmatrix},
\end{equation}
that can therefore be solved by usual linear algebra routines. Although this case might appear to be rather generic,
there are many important examples, e.g., graphene, where the matrix $V$ is not invertible.

\subsection{Eigendecomposition in the general case}
\label{sec:elim-zero-infin}
If the hopping matrix $V$ is not fully invertible, the matrix on the left-hand side of Eq.~\eqref{eq:lead_def_matrix} is singular and often ill-conditioned.
Moreover, solving the generalized eigenproblem \eqref{eq:lead_def_matrix} also yields the solutions $\Phi_{o+}$ such that $V \Phi_{o+} = 0$.
In fact, every singular value of $V$ gives rise to an eigenvalue $\lambda=0$ and $\lambda=\infty$ in the generalized eigenproblem.
However, the modes $\Phi_{o+}$ do not contribute to the scattering problem as we observed in Sec.~\ref{sec:form-scatt-probl}.
Computing them explicitly is thus unnecessary and inefficient.

We can use the decomposition introduced in Eq.~\eqref{eq:hopping_decomposition}, $V=A B^\dagger$, with $A$ and $B$ $N_\mathrm{t} \times N_{\bar{\mathrm{t}}}$ matrices of orthogonal vectors, to alleviate both problems.
With this decomposition, we can now rewrite Eq.~\eqref{eq:def_lead} into the
following form,
\begin{equation}
\label{eq:psi_tilde}
 (E-H)(\lambda\Phi) = A(B^\dagger \Phi)+ \lambda B (\lambda A^\dagger\Phi),
\end{equation}
which is naturally expressed in terms of the variables $\Phi_A$ and $\Phi_B$
that we shall use from now on:
\begin{equation}
\label{eq:def_phi_ab}
\Phi_A = \lambda A^\dagger \Phi \quad , \quad \Phi_B = B^\dagger\Phi.
\end{equation}
The next step is to express $\lambda \Phi$ in terms of $\Phi_A$ and $\Phi_B$. Since $E-H$ is not necessarily invertible, we first add a self-energy-like term
$+i (AA^\dagger +BB^\dagger)(\lambda\Phi)$ on both sides of Eq.~\eqref{eq:psi_tilde}. This term
shifts the eigenvalues of $H$ away from the real axis, guaranteeing that
the matrix $E-H+i (AA^\dagger +BB^\dagger)$ is always invertible except
when $H-E$ and $V$ have a joint zero eigenvector (or in other words that
there is a flat band at exactly the energy $E$). The problem now reads,
\begin{eqnarray}
  \label{eq:psi_tilde2}
\left[E-H+i (AA^\dagger +BB^\dagger) \right](\lambda\Phi) = \nonumber \\
(\lambda B +i A) \Phi_A + (A+i\lambda B) \Phi_B.
\end{eqnarray}
Introducing
\begin{equation}
G_{AB} = \left[E-H+i (AA^\dagger +BB^\dagger) \right]^{-1},
\end{equation}
we get,
\begin{equation}
\label{eq:lambdaPhi}
\lambda\Phi =G_{AB} (\lambda B +i A) \Phi_A + G_{AB} (A+i\lambda B) \Phi_B.
\end{equation}
Last, multiplying Eq.~\eqref{eq:lambdaPhi} by $A^\dagger/\lambda$ and $B^\dagger/\lambda$ we arrive at the following eigenvalue problem that only involves $\Phi_A$ and $\Phi_B$,
\begin{eqnarray}
\label{eq:stable_leads}
\begin{pmatrix}
A^\dagger G_{AB} B & iA^\dagger G_{AB} B \\
-B^\dagger G_{AB} B & 1-iB^\dagger G_{AB} B
\end{pmatrix}
\begin{pmatrix}
\Phi_A \\
\Phi_B
\end{pmatrix}=
\frac{1}{\lambda}
\begin{pmatrix}
1-iA^\dagger G_{AB} A & -A^\dagger G_{AB} A \\
iB^\dagger G_{AB} A & B^\dagger G_{AB} A
\end{pmatrix}
\begin{pmatrix}
\Phi_A \\
\Phi_B
\end{pmatrix}.
\end{eqnarray}
Equation~\eqref{eq:stable_leads} provides a stable generalized eigenproblem that can be solved with standard linear algebra routines.
Additionally, this generalized eigenproblem \eqref{eq:stable_leads} is smaller than the one in Eq.~\eqref{eq:lead_def_matrix}: $2N_{\bar{\mathrm{t}}} \times 2 N_{\bar{\mathrm{t}}}$ instead of $2N_\mathrm{t} \times 2N_\mathrm{t}$.
Equation~\eqref{eq:stable_leads} is used in particular in the Kwant package\cite{groth2014}.
Alternative techniques to reduce the eigenproblem in the case of singular hopping matrices have been developed in Refs.~\cite{rungger2008, sanjose2021}.

\subsection{Link with the scattering problem}
\label{sec:schur-decomposition}
We now go back to the scattering problem Eq.~\eqref{eq:scatt_form_general}.
This scattering problem can be rewritten in terms of the $\Phi_A$ and $\Phi_B$ matrices only,
\begin{equation}
\label{eq:scatt_form_stable2}
\begin{pmatrix}
H_\text{sr} - E & P_\text{sr}^T B \Phi_{A,\mathrm{ \bar t+}}  \\
B^\dagger P_\text{sr} & -\Phi_{B,\mathrm{\bar t+}}
\end{pmatrix}
\begin{pmatrix}
\Psi_\text{sr}\\
S_\mathrm{\bar tp}
\end{pmatrix}
=
\begin{pmatrix}
-P_\text{sr}^T B \Phi_{A,\mathrm{p-}}  \\
 \Phi_{B,\mathrm{p-}}
\end{pmatrix}.
\end{equation}

In a last step, we slightly modify Eq.~\eqref{eq:scatt_form_stable2} in order to make the left-hand side better conditioned for numerical purposes.
Indeed, one observes that in some situations the columns of the matrix
\begin{equation}
  \begin{pmatrix}
    \Phi_{A,\mathrm{\bar e+}}\\
    \Phi_{B,\mathrm{\bar e+}}
  \end{pmatrix},
\end{equation}
can be nearly linearly dependent due to the eigenvalue problems \eqref{eq:lead_def_matrix_inv} and \eqref{eq:stable_leads} being non-Hermitian.
In that case the matrix of the left-hand side of Eq.~\eqref{eq:scatt_form_stable2} is ill-conditioned \cite{wimmer2010}.
To avoid performing an unstable eigenvalue decomposition, we use the generalized Schur decomposition (QZ decomposition) \cite{golub1996} of Eq.~\eqref{eq:stable_leads} instead.
The transformation performs a joint decomposition of a pair of matrices
(matrix pencil) as $(X, Y) = (QSZ^\dagger, QTZ^\dagger)$, with matrices $Q$ and $Z$ unitary and $S$ and $T$ upper triangular.
The diagonal entries of $S$ and $T$ are related to the eigenvalues of the generalized eigenproblem $X\psi = \lambda Y \psi$ by $\lambda_i = S_{ii}/T_{ii}$.
In other words, the first vectors of the $Z$ matrix form an orthogonal basis in the eigensubspace corresponding to the first eigenvalues appearing on the diagonals of $S$ and $T$.
We then use standard LAPACK functions to extract the eigenvectors corresponding to propagating modes from the
Schur decomposition, as well as an orthogonal basis corresponding to the evanescent modes.
The evanescent modes are related to this basis $Z_{A/B}\mathrm{ \bar e+}$ by the relation
\begin{equation}
\label{eq:scatt_form_stable3}
\begin{pmatrix}
 \Phi_{A,\mathrm{ \bar e+}}  \\
\Phi_{B,\mathrm{\bar e+}}
\end{pmatrix}
=
\begin{pmatrix}
Z_{A,\mathrm{ \bar e+}} \\
Z_{B,\mathrm{\bar e+}}
\end{pmatrix}
R,
\end{equation}
where $R$ is an upper triangular matrix.
Effectively, the QZ decomposition computes the unitary matrix of the QR-decomposition of the evanescent modes, however it avoids an unstable step of obtaining the eigenvectors $\Phi_\mathrm{\bar e+}$ from the eigenvalue problem Eq.~\eqref{eq:stable_leads}.
Furthermore, obtaining a subset of the eigenvectors from Schur does not introduce a computational overhead because the Schur decomposition is performed by LAPACK as the first step of solving the generalized eigenvalue problem anyway.
Using the orthogonal basis for the evanescent states, we arrive at the following $2\times 3$ block structure
which forms our final form of the scattering problem.
\begin{eqnarray}
\begin{pmatrix}
H_\text{sr} - E & P_\text{sr}^T B \Phi_{A,\mathrm{ \bar p+}} & P_\text{sr}^T B Z_{A,\mathrm{ \bar e+}}  \\
B^\dagger P_\text{sr} & -\Phi_{B,\mathrm{\bar p+}} & -Z_{B,\mathrm{\bar e+}}\\
\end{pmatrix}
\begin{pmatrix}
\Psi_\text{sr}\\
S_\mathrm{pp} \\
R S_\mathrm{\bar ep}
\end{pmatrix}
=
\begin{pmatrix}
-P_\text{sr}^T B \Phi_{A,\mathrm{p-}}  \\
 \Phi_{B,\mathrm{p-}}
 \label{eq:scatt_form_stable5}
\end{pmatrix}.
\end{eqnarray}
Since we are typically not interested in $S_\mathrm{\bar ep}$, we do not need to compute the matrix $R$ explicitly.
Practical calculations, for instance in the Kwant package, are performed by numerically solving the linear system Eq.~\eqref{eq:scatt_form_stable5}.

\subsection{Diagonalization of current operator and proper modes}
\label{sec:diag-curr-oper}

\begin{figure}
  \centering
  \includegraphics[width=0.7\linewidth]{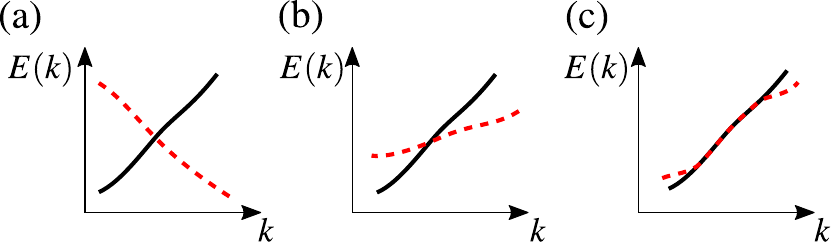}
  \caption{Different scenarios for the band structure $E(k)$ that give rise to degenerate eigenvalues $\lambda_{i_n}$ for the mode eigenproblem in Eqs.~\eqref{eq:lead_def_matrix_inv} and \eqref{eq:stable_leads}.}
  \label{fig:bs_crossings}
\end{figure}

The mode eigenproblems in Eqs.~\eqref{eq:lead_def_matrix_inv} and \eqref{eq:stable_leads} may have degenerate eigenvalues $\lambda_{i_n}$ (with $n=1, \dots, N_\text{deg}$, where $N_\text{deg}$ is the degeneracy).
In this case, any linear superposition of eigenvectors is also an eigenvector, and in general numerical algorithms will indeed return an arbitrary superposition.

For propagating modes, $\lambda_{i_n} = e^{i k}$ with real $k$, this corresponds to a crossing or degeneracy of bands $E(k)$ at a given value of $k$.
A couple of scenarios that give rise to this situation are shown in Fig.~\ref{fig:bs_crossings}.
This case needs special treatment:

\begin{itemize}
\item In the derivation of the Landauer--Büttiker formula in Sec.~\ref{sec:land-form-cond} we need to assume that the scattering states are orthogonal.
  This also requires that the lead eigenstates are orthogonal, and for lead modes with the same $k$, this implies that $\phi_{i_n}$ need to form an orthogonal set.
  Additionally, the derivation of the Landauer--Büttiker formula requires that the lead modes diagonalize the current operator.
  This follows automatically for modes corresponding to different $\lambda_i \neq \lambda_j$ (see Appendix~\ref{app:hermit}), but not for degenerate $\lambda_{i_n}$.
\item For both Landauer--Büttiker and the definition of retarded Green's function in Sec.~\ref{sec:def-of-green} we need to reliably separate in- and outgoing propagating modes.
  To this end, the velocity of a numerically computed mode needs to be continuous.
  This is also guaranteed by imposing the condition of modes $\phi_{i_n}$ being orthogonal \emph{and} diagonalizing the current operator, as both conditions are true away from the degeneracy point and result in modes that are unique up to a phase, if the velocities $v_{i_n}$ are all different.\footnote{Note that this does not resolve the uniqueness of the modes for the scenario sketched in Fig.~\ref{fig:bs_crossings}(c) where there are modes with equal velocities.
    However, any linear superposition of modes with the same velocity is compatible with both the Landauer--Büttiker and the Green's function approach.
    Hence this does not pose a problem.}
\end{itemize}

The algorithm is thus as follows: Let $\Phi_{\mathrm{p}, \text{deg}}$ be the $N_\mathrm{t} \times N_\text{deg}$ matrix consisting of eigenvectors of Eqs.~\eqref{eq:lead_def_matrix_inv} and \eqref{eq:stable_leads} corresponding to the degenerate eigenvalue $\lambda_{i_n}$ with
$|\lambda_{i_n}| = 1$.
We first find an orthogonal basis for the space spanned by the eigenvectors using a QR decomposition:
\begin{equation}
  \Phi_{\mathrm{p},\text{deg}} = Q_{\mathrm{p},\text{deg}} R\,.
\end{equation}
We then use this orthogonal basis to compute the $N_\text{deg} \times N_\text{deg}$ velocity matrix
\begin{equation}
  \mathcal{V} = i Q_{\mathrm{p},\text{deg}}^\dagger (\lambda_{i_n} V^\dagger - \lambda_{i_n}^{-1} V) Q_{\mathrm{p},\text{deg}}.
\end{equation}
Being Hermitian, $\mathcal{V}$ can be diagonalized with real eigenvalues as
\begin{equation}
  U^\dagger \mathcal{V} U =\begin{pmatrix}
  v_{i_1}&&\\
  &\ddots&\\
  &&v_{i_{N_\text{deg}}}
  \end{pmatrix}\,.
\end{equation}
$\bar{\Phi}_{\mathrm{p},\text{deg}} = Q_{\mathrm{p},\text{deg}} U$ then forms an orthogonal set of propagating modes that diagonalizes the current operator, as demanded by the requirements above.
This procedure has to be repeated for every cluster of degenerate eigenvalues corresponding to a propagating mode.

\section{Green's function formalism for the quantum problem}
\label{sec:green} %

In the preceding sections, we have introduced quantum transport from the point of view of the scattering matrix formalism.
The scattering matrix approach expresses the quantum transport problem as a waveguide problem which is very appealing conceptually.
In particular, it provides a natural explanation for the quantization of conductance.
It is also a very effective formulation for numerical purposes, since, as we have seen, it allows one to map the problem to the solution of a (sparse) linear problem.
In this section, we introduce another, yet fully equivalent, formalism in terms of Green's functions.
In fact, the Green's functions approach was introduced first and was for a long time the preferred approach for numerics through the celebrated RGF algorithm.

This section contains an introduction to the Green's function approach and to its connection with the scattering formalism. We focus here on the retarded Green's function which can be put in direct correspondence with the scattering matrix: the retarded Green's function provides the amplitude for the propagation between two sites while the scattering matrix gives the amplitude for the propagation between two lead modes. We defer to Sec.~\ref{sec:phys-obs-negf} the derivation of the non-equilibrium Green's function (NEGF) formalism that allows one to calculate the actual physical observables from the knowledge of the retarded Green's function.

In Sec.~\ref{sec:def-of-green} we first provide the general definitions of Green's functions.
Since electron-electron interactions are only considered at the mean-field level in this review,
we restrict ourselves to quadratic Hamiltonians for which Green's functions take a much simpler form.
Section~\ref{sec:gener-form-as} shows how the retarded Green's function can be obtained from the solution of a linear problem similar to the one defined in the preceding sections for the scattering matrix.
We proceed in Secs.~\ref{sec:self-energy-leads} and \ref{sec:gamma} by defining the self-energy of a lead, an important concept of the Green's function approach.
Interestingly, we shall find that the self-energy satisfies a self-consistent equation that could allow one to calculate it
without the construction of the lead modes, allowing alternative approaches.
The relation between the retarded Green's function and the scattering matrix, known as the Fisher--Lee relation, is worked out in Sec.~\ref{sec:fisher-lee}.
We end with Sec.~\ref{sec:similarity}  where we discuss
general relations that unveil common mathematical structures that occur in different parts of the formalism.

\subsection{Definitions of Green's functions}
\label{sec:def-of-green}

The Fourier transform $\hat G(E)$ of the retarded Green's function takes the general formal form
\begin{equation}
\hat G(E) = \int dt \hat G(\tau) e^{iE\tau} = \lim_{\eta\rightarrow 0}\frac{\hat 1}{E + i\eta - \hat H_\text{sys}},
\end{equation}
where the real-time retarded Green's function is defined as
\begin{equation}
\hat G(\tau)= i \Theta (t)e^{-i\hat H_\text{sys}\tau}.
\end{equation}
Equivalently, one can define the retarded Green's function in energy directly as the solution of the following equation:
\begin{equation}
\label{eq:def_G}
 [(E +i \eta) \hat 1 - \hat H_\text{sys}] \hat G(E) = \hat 1.
\end{equation}
Note that Eq.~\eqref{eq:def_G} is only well defined (i.e. has a unique solution)
in the presence of the infinitely small positive imaginary energy $\eta>0$ that defines the retarded Green's function. Computing $\hat G(E)$ is the central problem of Green's-function-based methods.

In the following, we follow a route very similar to the one taken to define the scattering matrix. This allows us to obtain the Green's function as the solution of a set of linear equations and also to obtain a direct connection between the Green's function and the scattering matrix (referred to as the Fisher--Lee relation).
A different, more traditional route makes no reference to the scattering problem and makes use of the Dyson equation to calculate the retarded Green's function directly. This route will be followed in the next section. It is completely equivalent to the first.

\subsection{General formulation as a linear problem}
\label{sec:gener-form-as}

Having defined the retarded Green's function, our first task is to show that it can be obtained through the solution of a linear system very similar to the one that defines the scattering matrix.
Let us define $G(j,k)$ as the various subblocks of $\hat G(E)$ on the different unit cells (we keep the energy $E$ as implicit).
Remember that $j>0$ corresponds to the semi-infinite leads and $j=0$ to the scattering region.
Let us further suppose that we are only interested in $G(j,0)$ and mostly in $G_\text{sr} \equiv G(0,0)$.
Writing explicitly the block structure of Eq.~\eqref{eq:def_G} gives
\begin{gather}
 [E + i\eta - H_\text{sr}] G_\text{sr} - P^T_\text{sr} V^\dagger G(1,0) = 1, \\
 -V P_\text{sr} G_\text{sr} +[E + i \eta- H] G(1,0) - V^\dagger G(2,0) = 0,\\
 \label{eq:def_G2}
 -V G(j-1,0) +[E + i \eta - H] G(j,0) - V^\dagger G(j+1,0) = 0
\end{gather}
where the last equation holds for $j>1$.
Equation~\eqref{eq:def_G2} is in fact identical to the equation found for $\Psi(j)$ so that a decomposition similar to Eq.~\eqref{eq:out} applies.
To keep the problem normalizable, only decaying and propagating modes can contribute.
However, the presence of the small imaginary $\eta$ means that also propagating modes now become either decaying or growing exponentially away from the scattering region.
This can be seen by considering $\lambda_\alpha = \lambda_\alpha(E) = e^{i k_\alpha(E)}$ as a function of complex energy and then performing a Taylor expansion,
\begin{equation}
  \begin{split}
    k_\alpha(E + i\eta) = & k_\alpha(E) + i \frac{d k_\alpha(E)}{dE} \eta + \mathcal{O}(\eta^2) \\
    = & k_\alpha(E) + \frac{i\eta}{\hbar v_\alpha(E)} + \mathcal{O}(\eta^2),
  \end{split}
\end{equation}
where $v_\alpha = \frac{1}{\hbar} \frac{d E}{d k}$ is the velocity of the mode.
Hence, all outgoing propagating modes with $v_\alpha>0$ become decaying ($\eta>0$ implies $|e^{i k_\alpha(E+i\eta)}|<1$), and we can write
\begin{equation}
  \label{eq:retarded_bc}
	G(j,0) = \Phi_{\mathrm{t}+} (\Lambda_{\mathrm{t}+})^j G_{\mathrm{t}+},
\end{equation}
where the matrix $G_{\mathrm{t}+}$ sets the weight associated to the corresponding modes.
Note that the advanced Green's function $G^A = G^\dagger$ is defined by adding a small \emph{negative} contribution $-i \eta$ to the energy $E$.
The advanced Green's function would thus involve the \emph{incoming} modes $\Phi_\mathrm{p -}$ and the evanescent modes $\Phi_\mathrm{e +}$
(which together form $\Phi_\mathrm{t -}$ as defined in Table~\ref{tab:whichmodes}).

Following the same calculation as for the scattering problem, we
arrive at a linear problem similar to Eq.~\eqref{eq:scatt_form}:
\begin{equation}
\label{eq:calc_G}
\begin{pmatrix}
E - H_\text{sr}  & -P^T_\text{sr} V^\dagger \Phi_{\mathrm{t}+} \Lambda_{\mathrm{t}+} \\
-V P_\text{sr} & V \Phi_{\mathrm{t}+}
\end{pmatrix}
\begin{pmatrix}
G_\text{sr} \\
G_{\mathrm{t}+}
\end{pmatrix}
=
\begin{pmatrix}
1 \\
0
\end{pmatrix},
\end{equation}
where the role of the ``source'' is now taken by the identity matrix
(in site space) instead of the different incoming modes.
Note that we can already now take the limit of $\eta\rightarrow 0$ in this finite matrix problem, as we already used $\eta$ to choose the contributing modes in the leads.
If $V$ is non-invertible, we arrive in analogy to the wave function in
Eq.~\eqref{eq:scatt_form_general} at
\begin{equation}
\label{eq:calc_G_general}
\begin{pmatrix}
E - H_\text{sr}  & -P^T_\text{sr} V^\dagger \Phi_{\mathrm{\bar t}+} \Lambda_{\mathrm{\bar t}+} \\
-B^\dagger P_\text{sr} & B^\dagger \Phi_{\mathrm{\bar t}+}
\end{pmatrix}
\begin{pmatrix}
G_\text{sr} \\
G_{\mathrm{\bar t}+}
\end{pmatrix}
=
\begin{pmatrix}
1 \\
0
\end{pmatrix}.
\end{equation}
Both linear systems can be solved directly by standard numerical methods.

\subsection{Self-energy of the leads}
\label{sec:self-energy-leads}

Our next step to make contact with the standard formulation of the Green's function approach
is to introduce the self-energy of the lead,
i.e., to eliminate the $G_{\mathrm{\bar t}+}$ matrix in Eq.~\eqref{eq:calc_G_general}.
From the second row of this equation, we find
\begin{equation}
  \label{eq:G_tbar_plus}
  G_{\mathrm{\bar t}+} = \frac{1}{B^\dagger \Phi_{\mathrm{\bar t}+}} B^\dagger P_{\textrm{sr}} G_\textrm{sr}.
\end{equation}
The matrix $B^\dagger \Phi_{\mathrm{\bar t}+}$ is invertible unless the energy corresponds to a bound state of the semi-infinite lead. This can only happen for discrete values of the energy $E$. The proof of this statement can be found in Appendix~\ref{sec:inverse_exist}.

Inserting \eqref{eq:G_tbar_plus} into the first row of Eq.~\eqref{eq:calc_G_general}, we arrive at
\begin{equation}
\label{eq:green-from-self}
\left[ E - H_\text{sr}  -P^T_\text{sr} \Sigma(E) P_\text{sr} \right] G_\text{sr}  = 1,
\end{equation}
i.e., the retarded Green's function is simply given by the inverse of the scattering region Hamiltonian to which one has added a self-energy term $\Sigma=\Sigma(E)$. Here we have introduced the self-energy as
\begin{equation}
\Sigma= V^\dagger \Phi_{\mathrm{\bar t}+} \Lambda_{\mathrm{\bar t}+}
\frac{1}{B^\dagger \Phi_{\mathrm{\bar t}+}}B^\dagger.
 \label{eq:self_energy_out_modes}
\end{equation}
This equation allows for an efficient computation of the self-energy through the stable formulation of the lead eigenproblem, as shown in Appendix~\ref{sec:stable_selfenergy}.

The above equation can also be reformulated by noticing that
$\frac{1}{B^\dagger \Phi_{\mathrm{\bar t}+}}B^\dagger \Phi_{\mathrm{\bar t}+} = 1$ and
$\frac{1}{B^\dagger \Phi_{\mathrm{\bar t}+}}B^\dagger \Phi_{\mathrm{o}+} = 0$. Hence,
\begin{equation}
  \label{eq:first_rows_of_inverse_mode_matrix}
\frac{1}{B^\dagger \Phi_{\mathrm{\bar t}+}}B^\dagger = 1_\mathrm{\bar{t}\times t} \Phi_{\mathrm{t}+}^{-1},
\end{equation}
with
\begin{equation}
\mathrm{1}_\mathrm{\bar t \times t} =
 \left(
  \begin{array}{cccc|cc}
  \bovermat{N_\mathrm{\bar t}}{1 & 0 & 0 & \cdots}& \bovermat{N_\mathrm{o}}{0 & \cdots} \\
  0 & 1 & 0 &  & \vphantom{\int^0}\smash[t]{\vdots} \\
  0 & 0 & 1 &  &  \\
  \vphantom{\int^0}\smash[t]{\vdots} & & & \ddots &  \\
  \end{array}.
 \right)
 \label{eq:Id_tbar_t}
\end{equation}
In other words, $\frac{1}{B^\dagger \Phi_{\mathrm{\bar t}+}}B^\dagger$ is equal to the first $N_\mathrm{\bar t}$
rows of $\Phi_{\mathrm{t}+}^{-1}$. Since $\Phi_{\mathrm{\bar t}+} \Lambda_{\mathrm{\bar t}+} 1_\mathrm{\bar t \times t} =
\Phi_{\mathrm{t}+} \Lambda_{\mathrm{t}+}$, we thus find
\begin{equation}
\label{eq:def-self}
\Sigma  = V^\dagger \Phi_{\mathrm{t}+} \Lambda_{\mathrm{t}+} (\Phi_{\mathrm{t}+})^{-1}.
\end{equation}
Note that Eq.~\eqref{eq:def-self} is valid for both invertible and non-invertible hopping matrices.

The retarded Green's function $G(i,j)$ has the asymptotic behavior \eqref{eq:retarded_bc} of only outgoing modes in the lead. The advanced Green's function $G^\dagger(i,j)$ can be obtained in the same fashion, but now with an asymptotic behavior of only \emph{incoming} modes in the lead. From this we find
\begin{equation}
\label{eq:selfenergy_advanced}
\Sigma^\dagger = V^\dagger \Phi_{\mathrm{t}-} \Lambda_{\mathrm{t}-} (\Phi_{\mathrm{t}-})^{-1}.
\end{equation}

Let us now consider the surface Green's function $G_{\text{lead}}$ of a standalone semi-infinite lead (i.e., before it is connected to the scattering region). By surface, we mean the diagonal part of the
Green's function on the last unit cell. No additional calculation is required to obtain $G_{\text{lead}}$, it is a particular case of Eq.~\eqref{eq:green-from-self} where $H_\text{sr}$ is replaced with $H$:
\begin{equation}
\label{eq:surface-from-self}
\left[ E - H  - \Sigma  \right] G_\text{lead}  = 1,
\end{equation}
or alternatively can be defined as a linear system similar to Eq.~\eqref{eq:calc_G}.
Our last goal for this subsection is to show that $G_\text{lead}$ and the self-energy
$\Sigma$ are very simply related.

Using Eq.~\eqref{eq:lead_def2}, one can rewrite Eq.~\eqref{eq:def-self} above as
\begin{equation}
\Sigma  \Phi_{\mathrm{t}+} \Lambda_{\mathrm{t}+} = (E-H) \Phi_{\mathrm{t}+} \Lambda_{\mathrm{t}+},
- V \Phi_{\mathrm{t}+},
\end{equation}
from which one gets
\begin{equation}
\label{eq:alter-mode-def}
\Phi_{\mathrm{t}+} \Lambda_{\mathrm{t}+} = G_\text{lead} V \Phi_{\mathrm{t}+},
\end{equation}
and eventually obtain the sought-after connection,
\begin{equation}
\label{eq:fermi-golden-rule}
\Sigma  = V^\dagger G_\text{lead} V.
\end{equation}
Equation~\eqref{eq:fermi-golden-rule} can be seen as a generalization of the Fermi golden rule.
Its importance stems from the fact that we now have a closed set of equations~\eqref{eq:green-from-self}, \eqref{eq:surface-from-self} and \eqref{eq:fermi-golden-rule} that does not make use of the mode decomposition in terms of $\Phi_{\mathrm{t}+}$ and $\Lambda_{\mathrm{t}+}$, and hence can be amenable to calculations through a different class of algorithms. From this perspective,
one can see Eq.~\eqref{eq:alter-mode-def} as an eigenvector problem for the matrix $G_\text{lead} V$ which provides an implicit definition for $\Phi_{\mathrm{t}+}$ and $\Lambda_{\mathrm{t}+}$. In this review we have chosen to emphasize the ``constructive approach'' that starts from the scattering problem and constructs the Green's function approach as a consequence. The alternative approach where one first defines the Green's function is fully equivalent.

\subsection{Properties of the linewidth matrix \texorpdfstring{$\Gamma$}{Gamma}}
\label{sec:gamma}
We now introduce the linewidth matrix $\Gamma$ that plays an important role in the
Green's function formalism. It is a Hermitian matrix defined as
\begin{equation}
\label{eq:def-Gamma}
\Gamma = i \left[ \Sigma - \Sigma^\dagger\right].
\end{equation}
Multiplying the above equation by $\Phi_{\mathrm{t}+}^\dagger$ ($\Phi_{\mathrm{t}+}$) on the left (right), we find using Eq.~\eqref{eq:def-self} that
\begin{equation}
\Phi_{\mathrm{t}+}^\dagger \Gamma \Phi_{\mathrm{t}+} = i\left[
\Phi_{\mathrm{t}+}^\dagger V^\dagger \Phi_{\mathrm{t}+} \Lambda_{\mathrm{t}+}
- \Lambda_{\mathrm{t}+}^* \Phi_{\mathrm{t}+}^\dagger V \Phi_{\mathrm{t}+} \right].
\end{equation}
We recognize the expressions for the velocity calculated in Eqs.~\eqref{eq:veloc_ee}, \eqref{eq:veloc_pe}, \eqref{eq:veloc_ep} and \eqref{eq:veloc_pp}
from which it follows that $\Phi_{\mathrm{p}+}^\dagger \Gamma \Phi_{\mathrm{p}+} = 1$,
while all the other blocks vanish:
\begin{equation*}
\Phi_{\mathrm{p}+}^\dagger \Gamma \Phi_{\mathrm{e}+} =
\Phi_{\mathrm{e}+}^\dagger \Gamma \Phi_{\mathrm{p}+} =
\Phi_{\mathrm{e}+}^\dagger \Gamma  \Phi_{\mathrm{e}+} = 0.
\end{equation*}
Introducing the diagonal matrix $\mathrm{1}_\mathrm{p}$ whose diagonal entries are unity for the propagating $\mathrm{p}$ sector and zero for the evanescent $\mathrm{e}$ sector, we therefore arrive at the very compact expression,
\begin{equation}
\label{eq:Gamma}
\Gamma =   (\Phi_{\mathrm{t}+}^\dagger)^{-1}    \mathrm{1}_\mathrm{p} (\Phi_{\mathrm{t}+})^{-1},
\end{equation}
from which one can directly get
\begin{equation}
\label{eq:Gamma2}
\Gamma = \Gamma \Phi_{\mathrm{t}+}\Phi_{\mathrm{t}+}^\dagger \Gamma = \Gamma \Phi_{\mathrm{p}+} \Phi_{\mathrm{p}+}^\dagger \Gamma\,,
\end{equation}
where the last equality follows from $\Gamma \Phi_{\mathrm{e}+} = 0$, as evident from Eq.~\eqref{eq:Gamma}.
Similar expressions can be obtained for incoming modes. Indeed, they satisfy
$\Sigma^\dagger (E) \Phi_{\mathrm{t}-} = V^\dagger \Phi_{\mathrm{ t}-} \Lambda_{\mathrm{t}-}$. Following the same route as above, and using Eq.~\eqref{eq:veloc_pp_incoming} one arrives at
\begin{equation}
\label{eq:Gamma_inc}
\Gamma =   (\Phi_{\mathrm{t}-}^\dagger)^{-1}    \mathrm{1}_\mathrm{p} (\Phi_{\mathrm{t}-})^{-1},
\end{equation}
from which one can directly get
\begin{equation}
\label{eq:Gamma2_inc}
\Gamma = \Gamma \Phi_{\mathrm{t}-}\Phi_{\mathrm{t}-}^\dagger \Gamma = \Gamma \Phi_{\mathrm{p}-}\Phi_{\mathrm{p}-}^\dagger \Gamma.
\end{equation}
Equations~\eqref{eq:Gamma2} and \eqref{eq:Gamma2_inc} are central for showing the equivalence of the Green's function and scattering wave function approach.

\subsection{Fisher--Lee relation}
\label{sec:fisher-lee}

The Landauer--Büttiker formula derived in Sec.~\ref{sec:land-form-cond} relates the conductance to the intuitive scattering formalism.
In the following section we derive an expression that demonstrates the equivalence of the scattering approach to the Green's function formalisms, originally demonstrated in~\cite{fisher1981}.

From the second row of Eq.~\eqref{eq:scatt_form_general}, we find
\begin{equation}
  \label{eq:smat_from_wf}
  \begin{split}
  S_\mathrm{\bar t p} & = \frac{1}{B^\dagger \Phi_{\mathrm{\bar t}+}}B^\dagger \left(P_\text{sr}\Psi_\text{sr}
  - \Phi_\mathrm{p -}\right)\\
  & = 1_\mathrm{\bar t \times t} \Phi_{\mathrm{t}+}^{-1} \left(P_\text{sr}\Psi_\text{sr}
  - \Phi_\mathrm{p -}\right).
 \end{split}
\end{equation}
Inserting this into the first row of Eq.~\eqref{eq:scatt_form_general} we then arrive at an
equation for the wave function $\Psi_\text{sr}$,
\begin{equation}
  \label{eq:psi_sr_self_energy}
(E - H_\text{sr} - P_\text{sr}^T \Sigma P_\text{sr}) \Psi_\text{sr} = P_\text{sr}^T \left(V^\dagger \Phi_\mathrm{p -} \Lambda_\mathrm{p -} - \Sigma \Phi_\mathrm{p -} \right),
\end{equation}
where we used Eq.~\eqref{eq:self_energy_out_modes}.
In Eq.~\eqref{eq:psi_sr_self_energy}, we recognize the Ando mixed wavefunction approach introduced in Sec.~\ref{sec:mixed-wavef-appr}.

From Eq.~\eqref{eq:green-from-self} we know that $G_\text{sr} = \left(E - H_\text{sr} - P_\text{sr}^T \Sigma P_\text{sr}\right)^{-1}$. Additionally, from Eq.~\eqref{eq:selfenergy_advanced} we have $\Sigma^\dagger(E) \Phi_\mathrm{p -} = V^\dagger \Phi_\mathrm{p -} \Lambda_\mathrm{p -}$. Hence,
\begin{equation}
\Psi_\text{sr} = i G_\text{sr}P_\text{sr}^T\Gamma\Phi_\mathrm{p-}.
\label{eq:wf_simplified}
\end{equation}
Inserting this identity back into Eq.~\eqref{eq:smat_from_wf}, we find
\begin{equation}
\label{eq:fisher-lee_with_evanescent}
S_\mathrm{\bar{t}p} = \mathrm{1}_\mathrm{\bar t \times t} \Phi_\mathrm{t+}^{-1}
\left[ i P_\text{sr} G_\text{sr} P^T_\text{sr} \Gamma - 1\right] \Phi_\mathrm{p-}.
\end{equation}
Equation~\eqref{eq:fisher-lee_with_evanescent} is a generalization of the original Fisher--Lee relations that connect the retarded Green's function to the scattering matrix. If we restrict the scattering matrix to only propagating modes, we can use the fact that $\Phi_\mathrm{p +}^\dagger \Gamma = 1_\mathrm{p \times t} \Phi_\mathrm{t +}^{-1}$ to simplify this expression further to obtain
\begin{equation}
\label{eq:fisher-lee}
S_\mathrm{p p} = \Phi_\mathrm{p+}^{\dagger}
\left[ i \Gamma P_\text{sr} G_\text{sr} P^T_\text{sr} \Gamma - \Gamma\right] \Phi_\mathrm{p-}.
\end{equation}

\subsection{Common underlying structure of the site elimination problem}
\label{sec:similarity}

The Fisher--Lee relation is a particular example of a family of relations. In this subsection, we identify common mathematical patterns in the different algebraic routes that have been followed so far. This subsection does not contain new material but rather offers a global point of view on various operations and algorithms that might look disconnected at first sight.

Our starting point is a general tight-binding equation that connects two subsystems
$1$ and $2$. It corresponds to one line of the Schrödinger equation written in block form and is therefore under-determined,
\begin{equation}
\label{eq:underdet}
    H_{11} \Psi_1 + H_{12} \Psi_2 = 0,
\end{equation}
where $H_{11}$ is a square matrix (that incorporates the energy $E$ for concision) and $H_{12}$ is rectangular. Since we cannot fully solve it, we will rather express the value of some variables in terms of others.

The first step is to write a low-rank representation of $H_{12}$ using, e.g., a singular value decomposition. $H_{12}$ factorizes as $H_{12} = A B^\dagger$ and we further introduce $\Psi_A = A^\dagger \Psi_1$, $\Psi_B = B^\dagger \Psi_2$. It is straightforward to find that these two vectors are related through,
\begin{equation}
\Psi_A = -A^\dagger \frac{1}{H_{11}} A \Psi_B.
\end{equation}
To continue, we want to choose a ``double'' basis on which we will decompose $\Psi_A$ and $\Psi_B$. We refer to the first basis as ``known'' and the second one as ``unknown''; we will see some concrete examples below. The different vectors of these bases are stacked together so that the block matrix
\begin{equation}
\Phi = \begin{pmatrix}
\Phi^\textrm{kn}_{A} & \Phi^\textrm{un}_{A} \\
\Phi^\textrm{kn}_{B} & \Phi^\textrm{un}_{B} \\
\end{pmatrix},
\end{equation}
forms a basis for the vector $(\Psi_A,\Psi_B)^T$.
Now, searching for the solution of Eq.~\eqref{eq:underdet} of the form
$\Psi_{A/B} = \Phi^\textrm{un}_{A/B} S -\Phi^\textrm{kn}_{A/B}$ where the $S$ matrix is some sort of generalized scattering matrix, we arrive at
\begin{equation}
\label{eq:split}
\begin{pmatrix}
H_{11} & A \Phi_{B}^\textrm{un} \\ -A^\dagger & \Phi_{A}^\textrm{un}
\end{pmatrix}
\begin{pmatrix}
\Psi_1 \\ S
\end{pmatrix} =
\begin{pmatrix}
A\Phi_{B}^\textrm{kn} \\ \Phi_{A}^\textrm{kn}
\end{pmatrix}.
\end{equation}
Equation~\eqref{eq:split} is very general and accounts for many familiar situations depending on how one splits the system and which basis is used for the decomposition between known and unknown states.

The first example is for the actual scattering matrix of the system. In that case $\Psi_2$ refers to the leads, the ``unknown'' basis corresponds to the output states and the ``known'' to the input. However
Eq.~\eqref{eq:split} is not limited to this case. For instance, choosing $\Phi^\textrm{un}_{A} = \Phi^\textrm{kn}_{B}  = I_d$ and $\Phi^\textrm{un}_{B} = \Phi^\textrm{kn}_{A}  = 0$, we arrive
at $\Sigma = B S B^\dagger$ where $\Sigma$ is the self-energy due to subsystem $1$ on the sites of
subsystem $2$. A third example is the transfer matrix $M$ of the system.
Suppose that we equip the space spanned by $\Psi_A$ and $\Psi_B$ with a left/right block structure (this structure might actually correspond to parts of subsystem $2$ situated on the left and on the right of subsystem $1$,  but this definition is more general). Then we choose $\Phi^\textrm{kn}_{A} = \Phi^\textrm{un}_{B}  = 1_L \oplus 0_R$ and $\Phi^\textrm{un}_{A} = \Phi^\textrm{kn}_{B}  = 0_L \oplus 1_R$, where $1_{L/R}$ ($0_{L/R}$) is the identity (null) matrix acting on the left/right. The obtained $S$ from this partition is actually the transfer matrix $M$ of the system. More examples could be constructed in the same way. For instance, one may define virtual modes that are eigenstates of the current operator. The associated virtual leads would allow one to split the study of the system into two (or more) regions that are recombined at the end. This could be advantageous when one part of the system needs to be updated more often than the other (e.g., the disordered part for statistics) or in the presence of a bottleneck in the system (e.g., a quantum point contact). A possible choice for this is $\Phi^\textrm{un}_{A} = \Phi^\textrm{kn}_{A}  = I_d$ and $\Phi^\textrm{un}_{B} = -\Phi^\textrm{kn}_{B}  = i I_d$.

The above construction has the advantage of automatically providing the Fisher--Lee relations between different objects. Indeed, the knowledge of $S$ allows one to compute
\begin{equation}
\label{eq:F1}
A^\dagger H_{11}^{-1} A = -\left(\Phi^\textrm{un}_{A} S -\Phi^\textrm{kn}_{A}\right)
\left(\Phi^\textrm{un}_{B} S -\Phi^\textrm{kn}_{B}\right)^{-1},
\end{equation}
and in turn the knowledge of $A^\dagger H_{11}^{-1} A$ allows one to compute $S$ in a potentially different basis,
\begin{equation}
\label{eq:F2}
S = \left(\Phi^\textrm{un}_{A} + A^\dagger H_{11}^{-1} A \Phi^\textrm{un}_{B}\right)^{-1}
\left(\Phi^\textrm{kn}_{A}  + A^\dagger H_{11}^{-1} A \Phi^\textrm{kn}_{B}\right).
\end{equation}
Inserting Eq.~\eqref{eq:F1} with $S$ and $\Phi$ into Eq.~\eqref{eq:F2} for a different splitting
$S'$ and $\Phi'$ provides a cumbersome yet fully explicit ``Fisher--Lee'' relation between $S$ and $S'$.

\section{Numerical algorithms}
\label{sec:numerics}

We have now introduced two equivalent formalisms, the $\Psi S$ scattering wavefunction
and the Green's function formalism. The initially infinite eigenvector problem $\hat{H}_\text{sys}\hat{\psi}=E\hat{\Psi}$ has been cast onto a finite linear problem of the form $Ax=b$ where $A$ is a large sparse matrix. Solving this problem numerically amounts to doing some sort of Gaussian elimination for which there are various strategies.

In this section, we go through the main algorithms that have been developed to solve the quantum transport problem.
In Sec.~\ref{sec:lead-problem} we comment on how to solve the lead problem that enters into the linear system.
Section~\ref{sec:solving_linear_systems} discusses using direct sparse solvers to solve a general quantum transport problem, and why these solvers are often the method of choice.
Section~\ref{sec:dyson-equat-gluing} introduces the Dyson equation, an effective tool for practical calculations of the retarded Green's function, and the basis of many Green's function-based algorithms. Section~\ref{sec:orig-recurs-green} uses it for explaining the original RGF algorithm and its extensions.
We end this section with a broader discussion in Sec.~\ref{sec:discussion-algorithms} of a wide range of algorithms used in quantum transport including those that do not fully fit within the Green's function or the scattering matrix approach.

\subsection{The lead problem}
\label{sec:lead-problem}

All the central formulas of this review, such as Eq.~\eqref{eq:scatt_form_general} for computing the scattering wave function or Eqs.~\eqref{eq:green-from-self} and \eqref{eq:def-self} to compute the Green's function, rely on first computing the lead modes.

If a lead has a simple structure it may be possible to determine the corresponding $\phi$ and $\lambda$ analytically.
One such example is a square lattice with only a single orbital \cite{datta1995}.
In the general case, however, it is necessary to follow the procedure outlined in Sec.~\ref{sec:stable-form-infin}.
In particular, this involves solving the (generalized) eigenproblems in Eqs.~\eqref{eq:lead_def_matrix_inv} and \eqref{eq:stable_leads}, as well as the singular value decomposition of the matrix $V$.
For each of these problems there are standard routines in LAPACK \cite{anderson1999}.
Several implementations of LAPACK and BLAS also allow for parallelization through OpenMP.

Solving the (generalized) eigenproblem of a dense matrix with LAPACK yields all eigenvalues and eigenvectors.
Recent works have employed contour-integral methods to only compute eigenvalues close in a finite region around the unit circle, an approach that also is amenable to parallelization \cite{laux2012, iwase2017, iwase2018}.

Before eigenproblems were used to compute lead modes, the self-energy or surface Green's function of the lead was often computed iteratively \cite{sancho1985}.
In this approach, repeated doubling of the lead unit cell gives a fast convergence towards a truly semi-infinite lead (this scheme is related to the RGF method described in Sec.~\ref{sec:orig-recurs-green}).
However, it requires the introduction of a small but finite imaginary part
$\eta$ to the energy $E\rightarrow E+i\eta$.
Hence, it requires some fine tuning to perform the calculations with high accuracy \cite{velev2004}. It can be an interesting option in cases (not considered in this review) where the electrode is \emph{not} invariant by translation.

\subsection{Solving the linear system of \texorpdfstring{Eq.~\eqref{eq:scatt_form_general}}{the general scattering problem}}
\label{sec:solving_linear_systems}

Equation~\eqref{eq:scatt_form_general} (and \eqref{eq:calc_G_general} for the Green's function) are typically sparse linear systems of equations.
The direct, numerical solution of sparse linear systems has been studied extensively in the past decades (for a comprehensive review see \cite{davis2006, davis2016}).
For the solution of the quantum transport problem, we can heavily lean on these developments.
In fact, there are several publicly available software packages of direct solvers, such as MUMPS \cite{amestoy2001, amestoy2019}, UMFPACK \cite{davis2004}, or SuperLU \cite{li2005}.

The crucial step in solving a sparse system of linear equations is a sparse $LU$-decomposition of the coefficient matrix.
The time and memory required for this task depends directly on the number of non-zeros in the factors $L$ and $U$.
It is known that fill-in, i.e., additional non-zeros in $L$ and $U$ compared to the original coefficient matrix, depends crucially on the order of decomposition, and thus on the ordering of the coefficient matrix \cite{davis2016}.
Many different heuristic algorithms for finding fill-in-reducing orderings have been developed, such as approximate minimum degree ordering \cite{amestoy1996} or nested dissection \cite{george1973, lipton1979}.
Several of these heuristic algorithms are distributed with the direct sparse solver packages mentioned above, or are available as separate software packages, such as Metis \cite{karypis1998} or SCOTCH \cite{pellegrini1996}.

Out of these, nested dissection is particularly interesting for sparse linear systems arising from a real-space discretization.
For example, it has been shown \cite{george1973} for a two-dimensional $M\times M$ square grid that solving the linear system of equations takes $\mathcal{O}(M^3)$ operations and $\mathcal{O}(M^2 \log(M))$ storage, a scaling that has been shown to be optimal \cite{hoffman1973}.
For comparison, the traditional RGF algorithm needs $\mathcal{O}(M^4)$ operations and $\mathcal{O}(M^3)$ storage for the same problem (see Sec.~\ref{sec:orig-recurs-green}).
In fact, for practical examples, using direct sparse solvers was shown to be significantly faster than the RGF algorithm \cite{luisier2008, boykin2008}, and scaling more favorably \cite{groth2014}.

We derived Eqs.~\eqref{eq:scattwf_ando} and \eqref{eq:green-from-self} by eliminating parts of the unknowns in Eqs.~\eqref{eq:scatt_form_general} and \eqref{eq:calc_G_general}.
It may seem advantageous to solve these smaller linear systems instead.
However, the self-energy $\Sigma$ entering these equations can diverge (see Sec.~\ref{sec:self-energy-leads}), although the original systems are well-behaved.
It is thus advantageous to rather solve Eqs.~\eqref{eq:scatt_form_general} and \eqref{eq:calc_G_general}, as direct sparse solvers use pivoting to enhance numerical stability.
If only a part of the solution is needed, it can be advantageous to use sparse methods to compute a Schur complement directly \cite{lin2022}. In Appendix~\ref{sec:stab-schur}, we explain how the Schur complement may be stabilized in the case where the matrix one wants to pivot upon is singular.

Historically, dedicated solvers have been developed to solve the quantum transport problem.
However, for a general problem---with arbitrary structure---direct sparse solvers are superior to dedicated solvers in terms of numerical stability, flexibility and efficiency, given the amount of research in that field.
Hence, a direct sparse solver often is the method of choice, e.g., in Kwant \cite{groth2014}.
Still, taking into account the explicit structure of a specific problem can be advantageous and reduce, e.g., memory requirements or overhead, and in the following we review such approaches.

\subsection{Dyson equation and the gluing sequence}
\label{sec:dyson-equat-gluing}

The set of equations~\eqref{eq:green-from-self}, \eqref{eq:surface-from-self} and \eqref{eq:fermi-golden-rule}
is very appealing because it reduces the task of finding the inverse of an infinite matrix to the inverse of a finite one \eqref{eq:green-from-self} together with the self-consistent equations~\eqref{eq:surface-from-self} and \eqref{eq:fermi-golden-rule}.

We now introduce the Dyson equation, which provides a practical route for such a calculation.
We do not consider here the original Dyson equation of quantum field theory, but a much weaker
form that rather belongs to linear algebra. In this section we switch to notation that uses Gothic letters such as $\mathds{H}$ or
$\mathds{G}$. A matrix $\mathds{H}$ may refer to the actual Hamiltonian matrix $H$ but more likely will refer to a sub-block of certain matrix elements of it. For instance, it may be the Hamiltonian matrix of certain sites in the absence of certain hopping matrix elements. Many such matrices need to be considered in an actual algorithm.

Suppose that we split a Hamiltonian $\mathds{H}$ into two parts $\mathds{H} = \mathds{H}_0+\mathds{H}_1$ (usually referred to as the unperturbed Hamiltonian and the perturbation) and want to calculate the Green's function $\mathds{G}=1/(E - \mathds{H})$. We further suppose that $\mathds{G}_0=1/(E - \mathds{H}_0)$ is already known.
Elementary algebra shows that
\begin{equation}
\label{eq:general-dyson}
\mathds{G}= \mathds{G}_0+\mathds{G}_0\mathds{H}_1 \mathds{G} = \mathds{G}_0 + \mathds{G}\mathds{H}_1\mathds{G}_0
\end{equation}
(as proved by multiplying Eq.~\eqref{eq:general-dyson} by $E - \mathds{H}$). Equation~\eqref{eq:general-dyson} is known as the Dyson equation.
\begin{figure}
  \centering
  \includegraphics[scale=0.5]{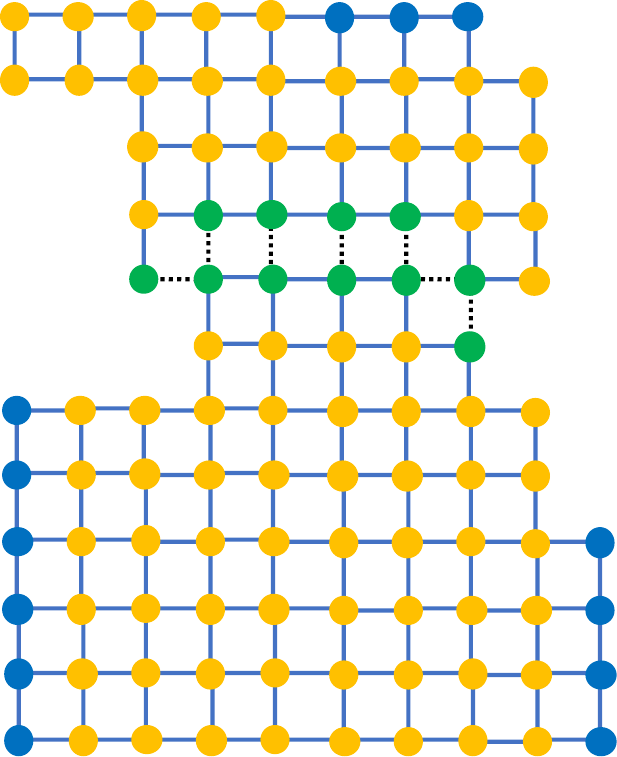}
  \caption{Schematic of the ``gluing sequence'' that uses the Dyson equation to incorporate new matrix elements (symbolized by the dotted black lines) into the calculation. The first step of the calculation involves only the (green) ``connected'' sites that are directly affected by the new matrix elements. In the second and last steps, one updates the Green's function for the blue sites, a subset of all the possible sites. The interest of Dyson equation stems from the fact that only the blue and green sites are involved in the calculation.}
  \label{fig:glueing}
\end{figure}
Its power originates from the fact that one can often choose the matrix $\mathds{H}_1$ such that $(\mathds{H}_1)_{mn}\neq 0$ only for a small number of ``connected'' sites $m,n$.
The Dyson equation projected onto these connected sites gives a closed set of equations. Hence, it is not necessary to invert the (potentially infinite) matrix $E - \mathds{H}$ but a much smaller version.

In particular, the Dyson equation can be used in a three-step process to ``glue'' together two parts of a system that are initially disconnected, see a cartoon in Fig.~\ref{fig:glueing}.
Introducing the superscript $\text{C}$ (green sites) and $\text{D}$ (blue and orange sites) for the corresponding connected and disconnected blocks (i.e., sites belonging or not to the group of connected sites), one finds that Eq.~\eqref{eq:general-dyson} takes a closed form for the $\text{CC}$ block,
\begin{equation}
\label{eq:general-dyson-connected}
\mathds{G}^\text{CC}= \mathds{G}_0^\text{CC}+\mathds{G}_0^\text{CC}\mathds{H}_1\mathds{G}^\text{CC}.
\end{equation}
Solving Eq.~\eqref{eq:general-dyson-connected} amounts to inverting a matrix whose size is given only by the number of connected sites,
\begin{equation}
\label{eq:general-dyson-connected-solution}
\mathds{G}^\text{CC}= 1/[(\mathds{G}_0^\text{CC})^{-1} - \mathds{H}_1].
\end{equation}
Once $\mathds{G}^\text{CC}$ has been computed for the connected sites, one can solve
the Dyson equation for the $\text{CD}$ block,
\begin{equation}
\label{eq:general-dyson-connected2}
\mathds{G}^\text{CD}= \mathds{G}_0^\text{CD} + \mathds{G}^\text{CC} \mathds{H}_1 \mathds{G}_0^\text{CD}.
\end{equation}
through a simple matrix product. Last, one solves the Dyson equation for the $\text{DD}$ block
\begin{equation}
\label{eq:general-dyson-connected3}
\mathds{G}^\text{DD}= \mathds{G}_0^\text{DD} + \mathds{G}_0^\text{DC} \mathds{H}_1 \mathds{G}^\text{CD}.
\end{equation}
Together, the three equations~\eqref{eq:general-dyson-connected}
\eqref{eq:general-dyson-connected2} and \eqref{eq:general-dyson-connected3}
form the three steps of a gluing sequence \cite{kazymyrenko2008} that allows one to
progressively introduce new non-zero matrix elements and update the elements
of $\mathds{G}$ accordingly. Note that Eqs.~\eqref{eq:general-dyson-connected2} and \eqref{eq:general-dyson-connected3} can be restricted to only compute the Green's function in parts of the $\text{D}$ sector:
For example, to compute the conductance, only the Green's function at the leads is sufficient as seen in Eq.~\eqref{eq:fisher-lee}. In the example of Fig.~\ref{fig:glueing} this could mean that only the blue sites will be considered in the $\text{D}$ sector so that a large number of (orange) sites need not be considered in the calculation.

\begin{figure}
  \centering
  \includegraphics[scale=0.5]{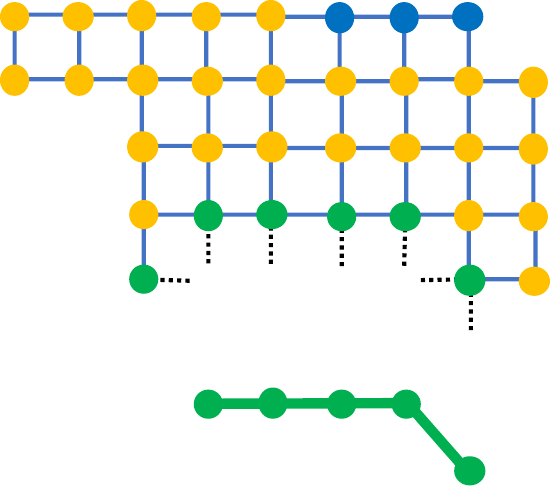}
  \caption{Illustration of the concept of self-energy for the system shown in Fig.~\ref{fig:glueing}. After integrating out the lower part of Fig.~\ref{fig:glueing}, there remains the upper part of the system (upper panel)
    connected to a self-energy term (lower panel) to be added to the Hamiltonian.}
  \label{fig:self}
\end{figure}

An alternative way for using the Dyson equation is to ``integrate out'' some of the sites present in the system, which we now illustrate.
To this end, we will use an additional structure, namely the fact that the two different parts of Fig.~\ref{fig:glueing} (as separated by the dotted lines) are unconnected before
$\mathds{H}_1$ connect them. In other words $\mathds{G}^\text{CC}_0$ is block diagonal in this upper/lower (denoted as up/lo, respectively) block substructure while $\mathds{H}_1$ is purely block off-diagonal:
\begin{equation}
\mathds{G}^{\text{CC}}_0 = \begin{pmatrix}
\mathds{G}_0^\text{up} & 0 \\
0 & \mathds{G}_0^\text{lo}
\end{pmatrix},
\end{equation}
\begin{equation}
\mathds{H}_1 = \begin{pmatrix}
0 & \mathds{V}^\dagger \\
\mathds{V} & 0
\end{pmatrix}.
\end{equation}
With this additional structure Eq.~\eqref{eq:general-dyson-connected-solution} takes the form
\begin{equation}
  \label{eq:general-dyson-uplo-blockstructure}
  G_0^\text{CC} = \begin{pmatrix}
    \left(\mathds{G}_0^\text{up}\right)^{-1}&-\mathds{V}^\dagger\\
    -\mathds{V}&\left(\mathds{G}_0^\text{lo}\right)^{-1}
    \end{pmatrix}^{-1}.
\end{equation}
Using $ \left(\mathds{G}_0^\text{up}\right)^{-1} = E - \mathds{H}_\text{up}$ and using the block inverse formula we find
\begin{equation}
\mathds{G}^{\text{up}} = \frac{1}{E - \mathds{H}_{\text{up}} - \Sigma_{\text{lo}}(E)}
\end{equation}
where the self-energy $\Sigma_{\text{lo}}(E)$ is given by,
\begin{equation}
\Sigma_{\text{lo}}(E) = \mathds{V}^\dagger \mathds{G}_0^{\text{lo}} \mathds{V}.
\end{equation}
In other words, the lower part has been condensed into a self-energy term within the Hamiltonian of the upper part. In contrast to the Hamiltonian, however, the self-energy is a dense matrix (that connects all the green sites of the lower part together), is non-Hermitian and depends on the energy (schematically shown in Fig.~\ref{fig:self}). The self-energy approach can be used to ``decimate'' the sites one after the other until one is left with only the sites of interests \cite{pastawski2001}.
However, one should keep in mind an important point in the definition of the retarded Green's functions: for infinite systems, the matrix elements of the Green's functions are smooth functions of the energy.
For finite systems they are, however, essentially the sum of Dirac functions positioned at the eigenenergies of the system, and therefore ill-defined numerically.
Hence, the decimation is usually performed starting from a semi-infinite lead. Alternatively, one can calculate the Green's function slightly away from the real axis but that leads to algorithms that require extrapolation and are not very robust.

Finally, we note that the Dyson equation can be used to recover some of the important equations derived earlier. We consider the case where $\mathds{H}_1$
is the part of the Hamiltonian matrix that connects the lead to the scattering region.
$\mathds{G}_0$ is the Green's function of the disconnected scattering region + lead system. Then
$\mathds{G}^\text{up}_0 = (E - H_\text{sr})^{-1}$, $\mathds{G}_0^\text{lo} = G_\text{lead}$, and
$\mathds{V} = V P_\text{sr}$.
With this, Eq.~\eqref{eq:general-dyson-uplo-blockstructure} can be recast as
\begin{equation}
\label{eq:dyson-lead-sr}
\begin{pmatrix}
E- H_\text{sr} & -P_\text{sr}^T V^\dagger  \\
-V P_\text{sr} & G_\text{lead}^{-1}
\end{pmatrix}
\mathds{G}^\text{CC} = 1
\end{equation}
which we recognize as Eq.~\eqref{eq:green-from-self} before block-matrix inversion.
Hence, we see that we can recover the definition of the self-energy without any prior knowledge of the lead modes. In fact, this derivation is even more general since it does not explicitly assume that the lead has translational symmetry.

\subsection{The recursive Green's function (RGF) algorithm and its extensions}
\label{sec:orig-recurs-green}

The RGF algorithm plays an important role in quantum transport as one of the first available numerical techniques that were first developed in one dimension \cite{lee1981,thouless1981} and then extended to quasi-one-dimensional systems \cite{mackinnon1985} for addressing disordered systems and later to ballistic multi-terminal systems
\cite{baranger1991}. Here we step away from the historical formulations and explain the RGF algorithm in a slightly more general way in terms of the gluing sequence of the preceding section.

\begin{figure}
  \centering
  \includegraphics[scale=0.5]{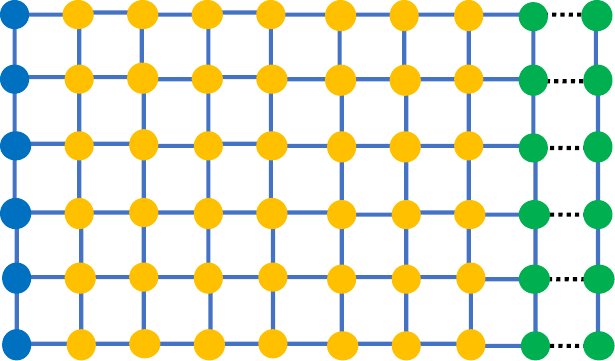}
  \caption{Schematic of one step of the original recursive Green's Function (RGF) algorithm. The gluing sequence is used iteratively to add new sites slice by slice. The site coloring is the same as in Fig.~\ref{fig:glueing}.}
  \label{fig:rgf}
\end{figure}

Figure~\ref{fig:rgf} shows a schematic of the RGF algorithm for a rectangular sample of width $W$ and length $L$. One first needs to calculate the Green's function of the translationally invariant semi-infinite electrode. These matrix elements need only be computed for the last slice of sites (the so-called surface Green's function), and can be computed with one of the methods explained in Sec.~\ref{sec:lead-problem}.

Once one has the electrode surface Green's function, RGF simply uses the gluing sequence to iteratively add one slice after the other of the scattering region from left to right. When all slices have been added, the right electrode is finally glued to the system. The overall complexity of RGF scales as $W^3 L$ as can be seen from the equations of the gluing sequence (one $W\times W$ matrix inversion per slice added). It is therefore very efficient for systems that are close to being one-dimensional $L\gg W$.

The original RGF algorithm has been generalized to different strategies for a parallel implementation \cite{drouvelis2006} and adding slices of different shapes in an optimized way. This can be done using algorithms for analyzing the connectivity graph of the Hamiltonian matrix \cite{wimmer2009, mason2011,mou2011,lima2018}, using minimum slices with just a single site \cite{kazymyrenko2008} or circular slicing \cite{thorgilsson2014}. These more recent algorithms have the advantage of being easy to deploy on multi-terminal systems, arbitrary geometries and lattices but have a scaling similar to RGF\@. Other approaches take advantage of a structure of the problem, such as some parts of the device being defect free, to speed up the calculation \cite{teichert2017}. Reference~\cite{metalidis2005} extends RGF to calculate other quantities than the transmission probability such as the local density of states.

\begin{figure}
  \centering
  \includegraphics[scale=0.5]{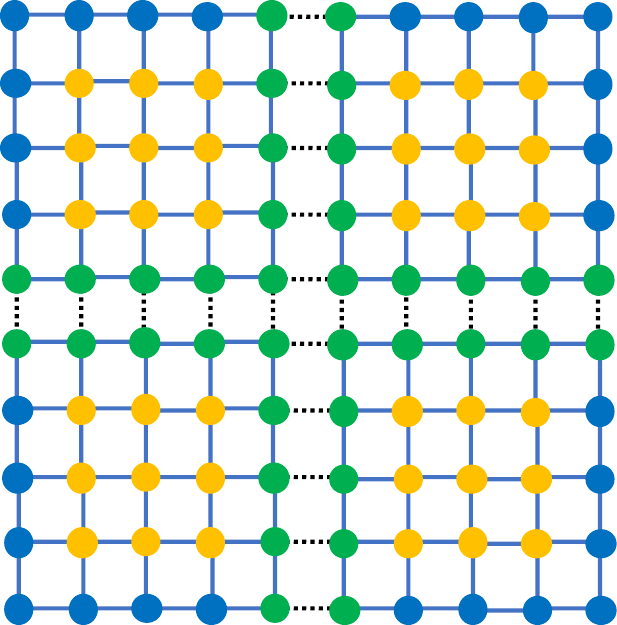}
  \caption{Schematic of one step of the nested dissection algorithm. The gluing sequence is used to recursively double the size of a $L\times L$ square system to a $2L\times 2L$ system. The site coloring is the same as in Fig.~\ref{fig:glueing}}
  \label{fig:nested}
\end{figure}

For systems whose different dimensions are of similar sizes ($L\approx W$), an efficient way to build the system is to use the nested dissection algorithm \cite{kuzmin2013,li2008}. In nested dissection one recursively double the system size by using the gluing sequence according to the schematic of Fig.~\ref{fig:nested}. For a
$L\times L$ square system, the computing time reduces to $L^3$ which is parametrically faster than the $L^4$ of RGF\@. However, implementing the nested dissection algorithm in a stable way is not easy as it implies gluing blocks that
are not connected to the electrodes (hence finite with badly conditioned Green's functions, see the discussion above).
It can thus be beneficial to use one of the existing direct sparse solvers with a nested dissection ordering, as discussed in Sec.~\ref{sec:solving_linear_systems}.

There have been many algorithms related to RGF that all rely on some form of the gluing sequence. Those include the modular Green's function algorithm \cite{rotter2000, rotter2003}, the (parallel) patchwork algorithm \cite{costa2013} or matching methods \cite{bergeron2005} that glue together various precalculated subparts of the system. Other applications include multiscale modeling \cite{calogero2019} where a central part of the system is described accurately within DFT while the periphery is modeled by a parametrized tight-binding model.

A very closely related class of algorithm uses ``decimation'' to integrate out sites that are not of direct interest. The sites integrated out appear through self-energy terms. Despite being apparently different, these algorithms are very close to RGF kind of algorithms. Early algorithms include \cite{grosso1989,pastawski2001}.

\subsection{Discussion of alternative approaches and extensions.}
\label{sec:discussion-algorithms}

In its early days RGF faced competition from an alternative algorithm expressed in terms of transfer matrices Eq.~\eqref{eq:transferM} \cite{pichard1981,pichard1981b}.
Compared to RGF the transfer matrix approach is conceptually simpler:
since the transfer matrix relates the wave-function in one layer to that in the next \cite{usuki1995},
putting two systems in series is simply equivalent to multiplying their transfer matrices (see Eq.~\eqref{eq:addingM}).
Such chaining of transfer matrices requires stabilization against accumulation of errors \cite{mackinnon1983},
since unitarity is not preserved otherwise \cite{usuki1994}.
Although this approach can be used to calculate
transmission probabilities \cite{pendry1992},
its principal domain of application has been the computation of Lyapunov exponents (inverse localization lengths) in the context of Anderson localization.
It remains competitive there \cite{slevin2014}, but otherwise has been mostly abandoned in favor of other methods such as RGF.

The RGF class of algorithms has a computational effort that scales
as $L^{3d-2}$. These algorithms are very efficient in one dimension or for
very elongated systems. For systems where all dimensions have similar sizes,
approaches based on nested dissection are parametrically faster and in
practice outperform RGF for $L \ge 100$ \cite{groth2014}. In two dimensions
a calculation of a square system of $10^6$ sites takes less than 1 hour
on a single core, making 2D calculations easily tractable.

In three dimensions, the situation is more complex since it is difficult to
calculate systems large enough for the calculations not to be dominated by
finite size effects. An alternative idea to RGF-type calculations is to
use a different starting point: instead of starting from vacuum and adding
sites layer by layer, one may start from a pristine material and use the Dyson
equation to modify a finite number of Hamiltonian matrix elements \cite{ostrovsky2010,settnes2015}.
This approach allows one to start directly in the thermodynamic limit and hence address
difficult 2D or 3D systems. In some simple cases, the pristine Green's function may be
obtained analytically through complex contour integration \cite{ostrovsky2010,schelter2011,settnes2015}.
The method has been generalized by \cite{istas2019} to arbitrary pristine systems by a numerical calculation
of the poles and residues of the problem.

When the current is forced to go through a small constriction with very few propagating channels, one
may take advantage of this fact to build an effective quasi-1D problem that is much faster to solve
\cite{darancet2009,darancet2010} even if the original electrodes are fully 3D.

For quantities that are mostly local such as the electronic density, local density of states or the conductivity (as opposed to the full conductance of a device that is essentially global), linear scaling
methods such as the Kernel Polynomial Method (KPM) can be very efficient \cite{weisse2006}. Linear scaling methods for transport have been recently reviewed in \cite{fan2021,joao2020}.

Another aspect of simulation that becomes increasingly important is the ability to take advantage of
high performance computing (HPC) infrastructure.
This requires being able to express the algorithms in a parallel (typically through the
Message Passing Interface MPI) and/or a vectorized way (for GPUs or TPUs).
Efforts in this direction include \cite{drouvelis2006,kuzmin2013,feldman2014} for parallelization, \cite{jeong2021} for GPUs,
and \cite{steiger2011,sawant2025} for full HPC architectures.
Most of the software listed in Sec.~\ref{sec:short-hist-comp-quant} utilize parallelism to some degree.

This review has focused on discrete models. In some specific cases, it is also possible to perform wave-matching directly in the continuum \cite{szafer1989,usuki1994,crawford2002} or by treating part of the system (e.g., the leads) directly in the continuum \cite{lent1990}.

Last, let us mention two alternative techniques that are slightly different from the mainstream
approach. The Complex Absorbing Potential (CAP) technique can be used to bypass solving the lead problem by including a finite part of the lead in the calculation in the presence of a smoothly varying complex potential \cite{zhang2007,driscoll2008, bubin2010,zhang2013,xie2014}. This imaginary potential absorbs the incoming waves before they can be reflected by the finite boundary. The accuracy of this method is controlled by the smoothness of the imaginary potential (hence the size of the part of the lead which is added) since any abrupt variation generates reflections \cite{weston2016}.
In the Contact Block Reduction (CBR) approach, one approximates the Green's function of the isolated scattering region part by selecting only a finite fraction of the eigenstates of the closed system \cite{mamaluy2003}.

\section{Physical observables in the scattering (Landauer--Büttiker) approach}%
\label{sec:phys-obs-land}

So far, we have been concerned solely with one-body quantum mechanics. We now turn to expressing the out-of-equilibrium observables, such as the conductance, in terms of the solution of the quantum mechanical problem, i.e., we now incorporate statistical physics.

Introducing the destruction $c_n$ (creation $c^\dagger_n$) operators on site $n$ for an electron, we are now considering Hamiltonians in second-quantized form,
\begin{equation}
{\cal H} = \sum_{nm} \hat H_{nm} c^\dagger_n c_m,
\end{equation}
where the matrices $\hat H$ correspond to the various infinite matrices of the one-body problem.

\subsection{Many-body steady states}

The solutions of the scattering problem $\hat\psi(E)$ computed in Sec.~\ref{sec:form-scatt-probl} together with the bound states from Sec.~\ref{sec:form-bound-state} form a complete basis of single-particle wave functions of the Hamiltonian~\eqref{eq:h_sys}.

Because we focus on time-independent quantum transport, we limit our consideration to steady many-body states, where each of the single particle states has a specific occupation number $f_{\alpha}(E)$.
Requiring additionally that the fermion occupation numbers are uncorrelated, we obtain the general form of the density matrix of such states
\begin{align}
  \label{eq:steady_state_dos}
  \rho &= \prod_{\alpha E}\left([1-f_{\alpha}(E)](1-\Psi_{\alpha E}^\dagger\Psi_{\alpha E})
  + f_{\alpha}(E)\Psi_{\alpha E}^\dagger\Psi_{\alpha E}\right),\\
  \Psi_{\alpha E} &= \sum_n \hat\psi_{\alpha E,n} c_n.
\end{align}

The single particle density matrix for such a state is
\begin{equation}
  \label{eq:expectation_values_inf}
  \langle c^\dagger_n c_m \rangle = \sum_{\alpha}\int \frac{dE}{2\pi}  \hat\psi_{\alpha E,n}^*
   \hat\psi_{\alpha E,m} f_{\alpha}(E),
\end{equation}
where we remember that the density of states of the one-dimensional channels has already been incorporated (up to the factor $2\pi$) in the normalization of the wave function. Indeed, we normalize the states to carry unit current and in one dimension the velocity $(1/\hbar) dE/dk$ is the inverse of the density of states. When the sites $n,m$ belong to the scattering region, we arrive at
\begin{equation}
  \label{eq:expectation_values}
  \langle c^\dagger_n c_m \rangle = \sum_{\alpha}\int \frac{dE}{2\pi}  (\Psi_\text{sr}^*)_{n\alpha}
   (\Psi_\text{sr})_{m\alpha} f_{\alpha}(E),
\end{equation}
which is expressed in terms of quantities that we learned to calculate.
We narrow our scope further and consider one particularly important class of steady states where each of the leads has fixed chemical potential $\mu_a$ and temperature $\Theta_a$.
Then the modes $\hat\psi_{\alpha E}$ incoming from the lead $a$ have an occupation number given by the Fermi distribution of that lead
\begin{equation}
  \label{eq:fermi}
  f_{a}(E) = \frac{1}{e^{\beta_a(E - \mu_a)} + 1},
\end{equation}
with $\beta=1/k_B\Theta$. It is worth noting that, although such a steady state is
not a thermal state (the different leads can have different temperatures and/or chemical potentials), Eq.~\eqref{eq:steady_state_dos} is the thermal state of the effective nonlocal Hamiltonian $H_\mathrm{eff}$,
\begin{equation}
  \label{eq:effective_hamiltonian}
  \rho = \frac{1}{Z} e^{-H_\mathrm{eff}},\quad
  H_\mathrm{eff}=\sum_{\alpha E}\beta_a(E - \mu_a)\Psi^\dagger_{\alpha E}\Psi_{\alpha E}.
\end{equation}
This form makes it explicit that the Wick theorem is applicable to this density matrix. ($Z$ is the partition function that ensures normalization).

\subsection{Landauer formula and its generalizations}
\label{sec:calc-observ}

All the observable quantities follow from the solution of the scattering problem, namely from the quantities $\Phi$, $\Lambda$, $S_\mathrm{tp}$, $\Psi_\text{sr}$.
The most important object is the scattering matrix $S = S_\mathrm{pp}$ between propagating modes.
We use the abbreviated notation $S_{\alpha,\beta} = [S_\mathrm{pp}]_{\alpha\beta}$ for the corresponding matrix elements between channel $\alpha$ and $\beta$. Below we review the formulae used for the most common observables calculated in quantum transport.

\subsubsection{Electric current and conductance}
\label{sec:conductance}

In analogy to the derivation of Eq.~\eqref{eq:landauer_simple_example}, the out-of-equilibrium current that flows through lead $a$ reads,
\begin{equation}
\label{eq:general_landauer}
I_a = \frac{e}{h} \sum_{b}\int dE
\sum_{\alpha\in a, \beta\in b} \left| S_{\alpha, \beta} \right|^2  \left[ f_b(E)-f_a(E) \right]\,.
\end{equation}
Here, we use the convention that current $I_a>0$ means current flowing towards the scattering region and the electron charge is $-e$ with $e>0$.
Using the unitarity of the scattering matrix $S$, this expression can be rewritten as
\begin{equation}
\label{eq:current_general_T}
I_a = \frac{e}{h} \sum_{b}\int dE\, g_{ab}(E)\,  f_b(E)\,,
\end{equation}
where $g_{ab}(E)$ is defined as
\begin{equation}
g_{ab}(E) = \sum_{\alpha\in a,\beta\in b}
( |S_{\alpha,\beta}|^2 - \delta_{\alpha,\beta} \delta_{a,b} )\,.
\end{equation}
For $a\neq b$, $g_{ab}(E)$ is the total transmission from lead $b$ to lead $a$. $g_{aa}(E)$ is negative, with $|g_{aa}(E)|$ equal to the total transmission from lead $a$ to all other leads $b$.

While Eq.~\eqref{eq:current_general_T} is now very well established and has been validated in particular by the observation of the quantization of conductance in quantum point contacts, several other formulas were initially proposed \cite{buttiker1985}.

When all the leads have equal temperatures and the applied voltages $V_a$ are
small, $\mu_a = E_F - e V_a$, we get in linear response,
\begin{equation}
I_a = \sum_b \conductance_{ab} V_b\,, \label{eq:conductance matrix}
\end{equation}
with the conductance matrix defined as,
\begin{equation}
  \label{eq:landauer_finite_T}
\conductance_{ab} = \frac{e^2}{h}  \int \ dE g_{ab}(E) \ \frac{\partial f}{\partial E}(E)\,.
\end{equation}
At zero temperature, this expression further reduces to what is known as the multi-terminal Landauer formula \cite{buttiker1986}.
\begin{equation}
  \label{eq:landauer}
\conductance_{ab} = - \frac{e^2}{h} g_{ab}(E_F)\,.
\end{equation}

The $N_\mathrm{leads} \times N_\mathrm{leads}$ conductance matrix $\conductance_{ab}$ satisfies current conservation, $\sum_{a} \conductance_{ab} = 0$. Additionally, gauge invariance requires that adding a constant to all voltages does not change currents; therefore, $\sum_b \conductance_{ab} = 0$.
In other words, the vector $\mathbb{1} = (1, 1,\ldots, 1)^T$ is both a left and a right eigenvector of the $\conductance_{ab}$ matrix with eigenvalue zero. Note that both relations are due to the unitarity of the scattering matrix $S$.

A commonly occurring measurement technique of multiterminal samples is to send a current between two electrodes, while measuring all the possible voltage differences.
In particular, the four terminal resistance is
\begin{equation}
\label{eq:4_terminal_resistance}
R_{ab,cd} = \frac{V_a - V_b}{I}\,,
\end{equation}
with $V_a$ and $V_b$ corresponding to $I_c = -I_d = I$, and all the other currents equal to 0. The four terminal resistance can correspond to longitudinal or Hall resistance depending on the measurement configuration.
For $c=a$ and $d=b$ it corresponds to a two-terminal resistance.
Because the conductance matrix has a zero eigenvalue, straightforward inversion of Eq.~\eqref{eq:conductance matrix} to calculate $R_{ab,cd}$ is impossible.
A possible workaround is eliminating a single row (the corresponding current can be calculated at a later stage using current conservation) and a column (this is equivalent to fixing the voltage on the corresponding terminal to 0) of the conductance matrix and invert the truncated conductance matrix.
An alternative solution is to use our knowledge of the zero eigenvector of $\conductance$ to construct the resistance matrix as a pseudoinverse. Introducing the projector $\mathbb{P} = \mathbb{1}\mathbb{1}^T/N_\mathrm{leads}$ onto the kernel of $\conductance$, one can simply shift the matrix $\conductance$ by an arbitrary constant $a\ne 0$ times $\mathbb{P}$ to make it invertible. This allows one to define the voltage $V$ as,
\begin{equation}
  \label{eq:resistance matrix}
   V = (\conductance - a \mathbb{P})^{-1} I\,.
\end{equation}
Indeed, when the vector $I$ satisfies current conservation (i.e., $\mathbb{P} I = 0$), Eq.~\eqref{eq:resistance matrix} implies that the average of $V$ vanishes (i.e., $\mathbb{P} V = 0$ since, i.e., $\mathbb{P} \conductance = 0$) and it follows that $V$ satisfies
$I = \conductance V$.

Note that the conductance matrix from Eq.~\eqref{eq:conductance matrix} only takes into account the device described by the scattering region. In an experiment, the device will be connected to electric lines that have a resistance that can be comparable to the one of the device. In that situation, reconstructing the conductance matrix needs additional care as one needs to embed it inside a classical electric circuit. This can be further complicated if nonlinear effects start to play a role.

\subsubsection{Conductance in normal/superconductor hybrids}

If a grounded superconducting electrode is present in the system, the conductance can still be described with the scattering formalism using the Bogoliubov--De Gennes wave function (see also Sec.~\ref{sec:mesosc-superc}). In this case, charges may enter the superconductor in Cooper pairs, resulting in Andreev reflection, a process in which an incoming electron $e$ converts into an outgoing hole $h$ in the same lead via Andreev reflection, or in a different lead via crossed Andreev reflection.

The current in a normal lead $a$ is then given as \cite{lambert1991}
\begin{equation}
  \label{eq:current_BdG}
  I_a = \frac{e}{h} \sum_{b}\int dE g_{ab}(E) \left[ f_b(E)-f_\text{S}(E) \right]\,,
\end{equation}
where the summation is over all normal leads $b$, $f_\text{S}(E)$ is the Fermi--Dirac distribution function of the grounded superconductor, and energy $E$ is defined with respect to the Fermi level of the superconductor.
\begin{equation}
  \label{eq:BTK}
g_{ab}(E) =
\sum_{\alpha\in a,\beta\in b} |S_{\alpha e,\beta e}|^2 - |S_{\alpha h,\beta e}|^2 - \delta_{\alpha e, \beta e} \delta_{a,b}\,.
\end{equation}
Here the index $e$ and $h$ refer to the Nambu electron and hole modes, respectively.
This conductance matrix is only valid for the normal leads, and alone does not exhibit current conservation. The reason for this is the presence of the superconductor, that not only allows current in the quasi-particle continuum but also in the Cooper pair condensate.

In typical experiments, bias voltages are on the order of the superconducting gap. Since the conductance of a normal/superconductor hybrid changes drastically over this energy scale \cite{blonder1982}, conductance is usually not considered in linear response in this case. Instead, it is obtained by taking the derivative of Eq.~\eqref{eq:current_BdG} with respect to the voltage $V_b$. In the limit of temperatures much lower than the superconducting gap, the Fermi--Dirac distribution is well approximated by a step function and we find
\begin{equation}
\frac{d I_a}{d V_b} = -\frac{e^2}{h} g_{ab}(-e V_b)\,.
\end{equation}

Cases with more than one superconducting lead are beyond the scope of this review.
Then, currents can flow in equilibrium (DC Josephson effect) and voltage differences can lead to time-dependence (AC Josephson effect, multiple Andreev reflection \cite{averin1995}).

\subsubsection{Thermoelectric effects}
\label{sec:therm-effects}

Equation~\eqref{eq:current_general_T} can also be used to calculate Seebeck effects by introducing temperature gradients \cite{sivan1986}. In the limit of small gradients,
$k_B\Theta_a = k_B\Theta_\mathrm{av} +k_B\delta\Theta_a$, we find
\begin{equation}
I_a = \sum_b \conductance^S_{ab}\, \delta\Theta_b\,, \label{eq:seebeck_matrix}
\end{equation}
with the Seebeck conductance matrix defined as
\begin{equation}
  \label{eq:seebeck_finite_T}
\conductance^S_{ab} = \frac{e}{h \Theta_\mathrm{av}}   g_{ab}^{(1)}\,,
\end{equation}
where we have introduced
\begin{equation}
  \label{eq:seebeck2_finite_T}
g_{ab}^{(n)}= \int dE \ g_{ab}(E)\left(-\frac{\partial f}{\partial E}\right)(E)(E-E_F)^n\,.
\end{equation}
At low temperatures $\Theta_\mathrm{av}$, the expression simplifies after a Sommerfeld expansion to
\begin{equation}
\conductance^S_{ab} = \frac{e \pi^2 k_B^2 \Theta_\mathrm{av}}{3h} \frac{\partial g_{ab}}{\partial E}(E_F)\,.
\end{equation}
In analogy with the charge current, one can further introduce the heat current
as the energy flowing in the lead, counted from the chemical potential,
\begin{equation}
\label{eq:current_heat}
I_a^H = -\frac{1}{h} \sum_{b}\int dE\,
g_{ab}(E) (E-\mu_a) \left[ f_b(E)-f_a(E) \right]\,.
\end{equation}
Here we use sign conventions such that $I_a^H>0$ means heat flowing towards the scattering region.
In the limit of small bias voltages, one gets the Peltier effect
\begin{equation}
I_a^H = \sum_b \conductance^P_{ab}\,\, V_b, \label{eq:peltier_matrix}
\end{equation}
with the Peltier conductance matrix defined as
\begin{equation}
  \label{eq:landauer_Peltier}
\conductance^P_{ab} = \frac{e}{h}  g_{ab}^{(1)} = \Theta_\mathrm{av}  \conductance^S_{ab}\,.
\end{equation}
The direct correspondence with the Seebeck effect is a consequence of the thermodynamic Onsager relations.
Last, in the limit of small temperature gradients, we get
\begin{equation}
I_a^H = \sum_b \conductance^H_{ab}\, \delta \Theta_b\,,
\end{equation}
with the heat conductance matrix defined as,
\begin{equation}
  \label{eq:landauer_Heat_cond}
\conductance^H_{ab} = \frac{1}{h \Theta_\mathrm{av}}   g_{ab}^{(2)}\,.
\end{equation}
In the limit of low temperature, the heat conductance simplifies to
\begin{equation}
  \label{eq:landauer_Heat_cond_low_T}
\conductance^H_{ab} = -\frac{\pi^2 k_B^2 \Theta_\mathrm{av}}{3 h}  g_{ab}(E_F)\,.
\end{equation}
We note that in this limit, we recover
\begin{equation}
\frac{\conductance^H_{ab}}{\Theta_\mathrm{av} \conductance_{ab}} = \frac{\pi^2 k_B^2}{3 e^2}\,,
\end{equation}
the mesoscopic version of the Wiedemann--Franz law.

\subsubsection{Spin currents}
A Landauer formula can also be derived for spin currents. It can be used to
calculate, e.g., the spin torque exerted on a magnetic layer by an injected current \cite{waintal2000,borlenghi2011}. The definition of the spin current, which provides an explicit link between spin current and spin torque, is obtained from the continuity equation of the current \cite{waintal2002}. Introducing
the vector of Pauli matrices $\vec\sigma = (\sigma_x,\sigma_y,\sigma_z)^T$, the spin current $\vec I$ flowing through a surface $\Sigma$ takes the form $\propto \int_{\vec r\in\Sigma} d\vec r \ \mathrm{Im}\
\sum_{\eta\eta'} \Psi_{\eta}(\vec r)\vec \sigma_{\eta\eta'} \vec\nabla\Psi_{\eta'}(\vec r)$ where $\Psi_\eta(\vec r)$ is the wave function for the spin component $\eta$. In other words, the spin current is formally equivalent to the charge current with an addition of Pauli matrices in its definition.
In the limit of zero temperature and small bias voltages, the spin current takes the form
\begin{equation}
\label{eq:current_general}
\vec I_a = \sum_b \vec \conductance_{ab}\, V_b\,,
\end{equation}
where the spin conductance matrix $\vec \conductance_{ab}$ between non-magnetic leads $a$ and $b$ takes the form
\begin{equation}
\label{eq:mixing}
\vec \conductance_{ab} = \frac{e}{4\pi} \sum_{\eta\eta'\eta''}
\sum_{\alpha\in a,\beta\in b}
\vec\sigma_{\eta\eta'}
S_{\alpha\eta',\beta\eta''}(E_F) S_{\alpha\eta,\beta\eta''}^*(E_F)\,.
\end{equation}
Here, the scattering matrix is defined for states with a well-defined spin component $\eta$, i.e., this expression is only applicable to leads where spin is a conserved quantity.
Spin currents, being transfer of spin quanta $\hbar$ per unit time, have the dimensions of energy. Note that a variant of Eq.~\eqref{eq:mixing} first appeared in the so-called \emph{generalized} \cite{bauer2003} circuit theory of spin transport \cite{tserkovnyak2005} through a quantity known as the spin-mixing conductance.
This semi-classical theory is derived from the Keldysh formalism and is equivalent to a combination of the scattering matrix approach with random matrix theory \cite{rychkov2009}.

Note that in contrast to charge current, there is no conservation of spin current in a magnetic system.
Hence, it may be needed to calculate the spin current inside the system as well as in the lead.
This can be done using Eq.~\eqref{eq:expectation_values}.
There is also no gauge invariance, so spin current can flow in the system {\it at equilibrium}.
In the absence of spin-orbit coupling, such a spin current can be identified with magnetic exchange interaction \cite{bruno1995,waintal2002}.

\subsubsection{Injectivities and Emissivities}
Two secondary concepts of the Landauer--Büttiker formalism are the emissivity and the injectivity
\cite{buttiker1993,buttiker1995}. The injectivity $\partial\langle c_n^\dagger c_n\rangle/\partial\mu_a$ expresses how much the electronic density $\langle c_n^\dagger c_n\rangle$ varies when one raises the chemical potential $\mu_a$ of lead $a$. The emissivity
$\partial I_a/\partial (H_\text{sr})_{nn}$ is the variation of current $I_a$ upon a change of electric potential $(H_\text{sr})_{nn}$.  These two concepts arise naturally when one extends quantum transport to AC regimes or non-linear regimes to introduce a minimal treatment
of the effect of electron-electron interactions, see the discussion around Eq.~\eqref{eq:landauer_dipole}.

The injectivity arises naturally in the formalism and reads at zero temperature,
\begin{equation}
\frac{\partial\langle c_n^\dagger c_n\rangle}{\partial\mu_a} =
\frac{1}{2\pi} \sum_{\alpha\in a} |\Psi_\text{sr}(\mu_\alpha)|_{n\alpha}^2\,,
\end{equation}
where the scattering wave function $\Psi_\text{sr}$ is calculated at $E=\mu_\alpha$.
Other generalizations can be obtained from Eq.~\eqref{eq:expectation_values} in a straightforward manner. The local density of states $\ldos$ is simply the sum of the injectivities coming from the different leads:
\begin{equation}
\label{eq:ldos_Psi}
\ldos = \frac{1}{2\pi}\sum_{\alpha} |\Psi_\text{sr}(E)|_{n\alpha}^2.
\end{equation}

To obtain the emissivity, one differentiates Eq.~\eqref{eq:scatt_form} and arrives at the
following linear system,
\begin{equation}
  \label{eq:automatic_diff}
\begin{pmatrix}
H_\text{sr} - E & P_\text{sr}^T V^\dagger \Phi_\mathrm{t+} \Lambda_\mathrm{t+} \\
V P_\text{sr} & -V \Phi_\mathrm{t+}
\end{pmatrix}
\begin{pmatrix}
\frac{\partial\Psi_\text{sr}}{\partial (H_\text{sr})_{nm}}\\
\frac{\partial S_\mathrm{tp}}{\partial (H_\text{sr})_{nm}}
\end{pmatrix}
=
\begin{pmatrix}
-\Xi \Psi_\text{sr}\\
0
\end{pmatrix},
\end{equation}
that must be solved numerically. Here, the matrix $\Xi$ is defined as
$\Xi_{ij} = \delta_{in}\delta_{jm}$.
Equation~\eqref{eq:automatic_diff} can be used to implement automatic differentiation
schemes in quantum transport calculations. Equation~\eqref{eq:automatic_diff} has the same structure as
Eq.~\eqref{eq:calc_G} from which it follows that the emissivity can also be obtained from the knowledge of the Green's function,
\begin{eqnarray}
\frac{\partial\Psi_\text{sr}}{\partial (H_\text{sr})_{nm}} = G_\text{sr} \Xi \Psi_\text{sr}\\
\frac{\partial S_\mathrm{tp}}{\partial (H_\text{sr})_{nm}} = G_\mathrm{t+} \Xi \Psi_\text{sr}.
\end{eqnarray}

\subsection{Quantum noise}
The observables defined in the preceding sections are all one-body observables,
i.e., mean observables. Another class of important observables are
the quantum fluctuations around the mean and in particular the quantum fluctuation of the current \cite{blanter2000}. The general expression of the noise correlation
$\bar S_{ab}(\omega)$ between lead $a$ and $b$ (noise power when $a=b$) at finite frequency $\omega$ reads,
\begin{equation}
\begin{split}
\label{eq:general_noise}
\bar S_{ab}&(\omega) =   \frac{e^2}{h} \sum_{c,d} \sum_{\gamma\in c, \delta \in d} \int dE\\
&A^{cd}_{\gamma\delta} (a,E,E+\hbar\omega) A^{dc}_{\delta\gamma} (b,E+\hbar\omega,E) \\
&\{ f_c(E) [1-f_d(E+\hbar\omega)] + f_d(E+\hbar\omega)[1-f_c(E)]\},
\end{split}
\end{equation}
where $A^{cd}_{\gamma\delta} (a,E,E')$ is defined in terms of the scattering matrix as,
\begin{equation}
A^{cd}_{\gamma\delta} (a,E,E') = \delta_{ac} \delta_{ad} \delta_{\gamma\delta} - \sum_{\alpha\in a} S_{\alpha,\gamma}^*(E) S_{\alpha,\delta}(E').
\end{equation}
Equation~\eqref{eq:general_noise} has a number of simplified forms, notably at zero frequency, small bias, zero temperatures, etc. We refer to \cite{blanter2000} for a comprehensive review. See also the next section for a simple route for deriving such formula.

\section{Physical observables: the Non-equilibrium Green's function approach}%
\label{sec:phys-obs-negf}

In this section, we take an alternative, yet equivalent, route compared to the one in the preceding section to calculate out-of-equilibrium physical observables such as the electric conductance from the solution of the quantum mechanical problem. While, in the preceding section, the latter was expressed in terms of the scattering matrix and
scattering states, here quantum propagation is encoded in the retarded Green's function.

The connection from the retarded Green's function to the physical observable is done through the so-called Keldysh formalism \cite{keldysh1964}. As we do not consider electron-electron interactions, but simpler quadratic Hamiltonians, the Keldysh formalism considerably simplifies into what is often known as the NEGF formalism \cite{caroli1971,meir1992}. NEGF is mathematically equivalent to the Landauer--Büttiker approach, as we will explicitly demonstrate in the following.
It is perhaps less intuitive but has the advantage that once a few basic equations have been established, everything follows from straightforward algebra.

\subsection{Keldysh formalism in a nutshell}

In this subsection, we summarize the main definitions and results of the Keldysh
formalism that will be needed. We refer the reader to textbooks such as \cite{rammer1986,stefanucci2013} for the derivation of the results that we state here without proofs.

The Keldysh formalism introduces two independent Green's functions, the ``lesser'' ($<$)
and ``greater'' ($>$) Green's functions
\begin{eqnarray}
\hat { G}^<_{nm}(\tau) &=& i
\langle c^\dagger_m e^{i{\cal H}\tau} c_n e^{-i{\cal H}\tau} \rangle,\\
\hat { G}^>_{nm}(\tau) &=& -i
\langle  e^{i{\cal H}\tau} c_n e^{-i{\cal H}\tau} c^\dagger_m \rangle.
\end{eqnarray}
These Green's functions are very convenient since at $\tau =0$ they correspond directly to the observables (e.g., density or current) of interest. From these two Green's functions, one may construct the ``time-ordered'' ($T$), ``anti-time ordered'' ($\bar T$), ``retarded'' ($R$) and ``advanced'' ($A$) Green's functions as,
\begin{eqnarray}
\hat { G}^T_{nm}(\tau) &=&
\theta(\tau) \hat { G}^>_{nm}(\tau) +  \theta(-\tau) \hat { G}^<_{nm}(\tau),\\
\hat { G}^{\bar T}_{nm}(\tau) &=&
\theta(\tau) \hat { G}^<_{nm}(\tau) +  \theta(-\tau) \hat { G}^>_{nm}(\tau),\\
\label{eq:retarded_gf_many_body}
\hat { G}^{R}_{nm}(\tau) &=&
\theta(\tau) \hat { G}^>_{nm}(\tau) - \theta(\tau) \hat { G}^<_{nm}(\tau),\\
\hat { G}^{A}_{nm}(\tau) &=&
\theta(-\tau) \hat { G}^<_{nm}(\tau) - \theta(-\tau) \hat { G}^>_{nm}(\tau).
\end{eqnarray}
Note that, in the preceding sections, we encountered only one type of Green's function, the retarded Green's function, and the subscript $R$ has been omitted.
These Green's functions are linearly dependent. Below, we will focus on the Green's functions $\hat G^<$, $\hat G^R$ and $\hat G^A$ using the fact that $\hat G^< =\hat G^T -\hat G^R = \hat G^{\bar T} +\hat G^A$ as can be proven trivially from the above definitions.

In this review, we only consider many-body Hamiltonians ${\cal H}$ restricted to be quadratic in the destruction $c_n$ and creation $c^\dagger_n$ operators on site $n$,
\begin{equation}
{\cal H} = \sum_{nm} \hat H_{nm} c^\dagger_n c_m.
\end{equation}
The equivalence between the definition of the retarded Green's function
used in the preceding sections and the one in Eq.~\eqref{eq:retarded_gf_many_body} is not entirely trivial and indeed valid only for quadratic Hamiltonians.
In the following, we make use of this simplification.

In the Keldysh formalism, these Green's functions naturally appear in the form of
a $2\times 2$ matrix associated with the so-called Keldysh contour,
\begin{align}
& \mathbf{\hat{G}}_{nm}(\tau) = \begin{pmatrix}
\hat { G}^T_{nm}(\tau) & \hat { G}^<_{nm}(\tau)\\
\hat { G}^>_{nm}(\tau) & \hat { G}^{\bar T}_{nm}(\tau)
\end{pmatrix}.
\end{align}
Since this review is concerned with d.c.\ quantum transport (i.e., time-independent Hamiltonians), we work chiefly in energy representation,
\begin{equation}
\mathbf{\hat{G}}_{nm}(E) = \int d\tau \mathbf{\hat{G}}_{nm}(\tau) e^{iE\tau}.
\end{equation}

Now that we have defined the mathematical objects of interest, we can state
the very few results from non-equilibrium many-body theory that will be needed
to establish the basic equations of quantum transport. The equations of motion
read
\begin{align}
\sum_p
\begin{pmatrix}
E \delta_{mp}- \hat H_{mp} &  0 \\
0 & E \delta_{mp} - \hat H_{mp}
\end{pmatrix}
\mathbf{\hat{G}}_{pn}(E)
=
\begin{pmatrix}
\delta_{mn} &  0 \\
0 & \delta_{mn}
\end{pmatrix},
\end{align}
or in compact notation,
\begin{align}
\begin{pmatrix}
E - \hat H &  0 \\
0 & E - \hat H
\end{pmatrix}
\mathbf{\hat{G}}(E)
=
\begin{pmatrix}
1 &  0 \\
0 &  1
\end{pmatrix}.
\end{align}
Splitting the Hamiltonian matrix $\hat H$ into an ``unperturbed'' matrix $\hat H_0$ and a ``perturbation'' $\hat W$,
\begin{equation}
\hat H = \hat H_0 + \hat W,
\end{equation}
and denoting $\mathbf{\hat{G}}_0(E)$ the Green's function associated with $\hat H_0$, one can write a form of Dyson equation \cite{rammer1986,stefanucci2013} as
\begin{align}
\label{eq:keldysh_dyson}
\mathbf{\hat{G}}(E)
=
\mathbf{\hat{G}}_0(E) +
\mathbf{\hat{G}}_0(E)
\begin{pmatrix}
\hat W &  0 \\
0 &  -\hat W
\end{pmatrix}
\mathbf{\hat{G}}(E),
\end{align}
and also
\begin{align}
\label{eq:keldysh_dyson2}
\mathbf{\hat{G}}(E)
=
\mathbf{\hat{G}}_0(E) +
\mathbf{\hat{G}}(E)
\begin{pmatrix}
\hat W &  0 \\
0 &  -\hat W
\end{pmatrix}
\mathbf{\hat{G}}_0(E).
\end{align}
Note that the partition of $\hat H$ into $\hat H_0$ and $\hat W$ is arbitrary,
although in what follows the $\hat W$ matrix will contain the matrix elements that connect the leads to the scattering region and $\hat H_0$ the rest of the Hamiltonian. Identifying the different components of the $2\times 2$ Dyson equation Eq.~\eqref{eq:keldysh_dyson}, one may write
\begin{eqnarray}
\hat{G}^R &=& \hat{G}^R_0 + \hat{G}^R_0 \hat W \hat{G}^R, \\
\label{eq:dysonadv}
\hat{G}^A &=& \hat{G}^A_0 + \hat{G}^A_0 \hat W \hat{G}^A, \\
\label{eq:dysonlesser}
\hat{G}^< &=& \hat{G}^<_0 + \hat{G}^R_0 \hat W \hat{G}^< + \hat{G}^<_0 \hat W \hat{G}^A.
\end{eqnarray}
The first equation (or equivalently the second one) encodes the quantum mechanics
and forms a closed set of equations for the retarded Green's function alone. The third equation encodes the statistical physics. Similarly, Eq.~\eqref{eq:keldysh_dyson2} leads to
\begin{eqnarray}
\hat{G}^R &=& \hat{G}^R_0 + \hat{G}^R \hat W \hat{G}^R_0, \\
\label{eq:dysonadv2}
\hat{G}^A &=& \hat{G}^A_0 + \hat{G}^A \hat W \hat{G}^A_0, \\
\label{eq:dysonlesser2}
\hat{G}^< &=& \hat{G}^<_0 + \hat{G}^R \hat W \hat{G}^<_0 + \hat{G}^< \hat W \hat{G}^A_0.
\end{eqnarray}
For a system at thermal equilibrium, one
has
\begin{equation}
\label{eq:keldysh_eq}
\hat{G}^<(E) = -f(E) \left[  \hat{G}^R(E) - \hat{G}^A(E) \right],
\end{equation}
which completes the basic formalism that we shall need in this section.

\subsection{Main NEGF equation}
\label{sec:MainNEGF}
Within the NEGF formalism, computing out-of-equilibrium amounts to performing two tasks.
First, one computes the retarded Green's function (i.e., solving the quantum mechanical problem) using, e.g., any algorithm from Sec.~\ref{sec:numerics}. Second, one must compute the lesser Green's function (i.e., the observables). In this subsection, we derive the main result of NEGF that relates the lesser Green's function to the retarded one in the context of quantum transport. The derivation is a generalization of the original work by Caroli et al.\ \cite{caroli1971}. These results have later been generalized and popularized by Wingreen and Meir \cite{meir1992}.

We start by splitting and applying the Dyson equation to $\hat H_\text{sys}$. The perturbation
$\hat W$ contains the matrix elements that connect the leads to the scattering region.
$\hat H_0$ contains the Hamiltonian of the disconnected elements of the system: the semi-infinite leads and the isolated scattering region. Let us write the Dyson equation in
block form where the block index indicates the lead ($a=1,2,\dots$) or
the scattering region ($0$). By construction, the disconnected Green's functions
$(\hat{G}_0)_{kl} = \delta_{kl} (\hat{G}_0)_l$
are diagonal with respect to the block index.
Equation~\eqref{eq:dysonlesser} for the scattering region block ${G}^<_\text{sr}=\hat{G}^<_{00}$ reads
\begin{equation}
\label{eq:NEGF1}
\hat{G}^<_\text{sr} = (\hat{G}^<_0)_0 + (\hat{G}^R_0)_0 \hat W_{0a} \hat{G}^<_{a0} + (\hat{G}^<_0)_0 \hat W_{0a} \hat{G}^A_{a0},
\end{equation}
with an implicit summation over leads $a>0$. Note that in the rest of this section we have absorbed the projection matrices $P_\text{sr}$ into $\hat W$, such that, e.g., $\hat{W}_{a0} = V_a P_\text{sr}$, where $V_a$ is the hopping matrix of lead $a$.
Multiplying Eq.~\eqref{eq:NEGF1} on the left by $E+i\eta-H_\text{sr}$ and using $[E-H_\text{sr}](\hat{G}^<_0)_0 = 0$ and
$[E-H_\text{sr}](\hat{G}^R_0)_0 = 1$ we obtain
\begin{equation}
\label{eq:NEGF4}
\left[E-H_\text{sr} \right]{G}^<_\text{sr} =  \hat W_{0a} \hat{G}^<_{a0}.
\end{equation}
To close the above equation, we write Eqs.~\eqref{eq:dysonlesser} and \eqref{eq:dysonlesser2} for the $a0$ and $0a$ blocks respectively,
\begin{equation}
\label{eq:NEGF2}
\hat{G}^<_{a0} =  (\hat{G}^R_0)_a  \hat W_{a0} {G}^<_\text{sr}
+ (\hat{G}^<_0)_a \hat W_{a0} {G}^A_\text{sr},
\end{equation}
\begin{equation}
\label{eq:NEGF2bis}
\hat{G}^<_{0a} =  {G}^R_\text{sr} \hat W_{0a} (\hat{G}^<_0)_a  + {G}^<_\text{sr} \hat W_{0a} (\hat{G}^A_0)_a.
\end{equation}
Inserting Eq.~\eqref{eq:NEGF2} into \eqref{eq:NEGF4}, we arrive at
\begin{equation}
\label{eq:NEGF3}
\left[E-H_\text{sr} - \hat W_{0a}(\hat{G}^R_0)^a \hat W_{a0} \right]{G}^<_\text{sr} =  \hat W_{0a}(\hat{G}^<_0)_a \hat W_{a0} {G}^A_\text{sr}.
\end{equation}
We recognize Eq.~\eqref{eq:green-from-self} on the left hand side, so that,
\begin{equation}
\label{eq:NEGF5}
{G}^<_\text{sr} = {G}^R_\text{sr} \hat W_{0a}(\hat{G}^<_0)_a \hat W_{a0} {G}^A_\text{sr}.
\end{equation}
To conclude, we use Eq.~\eqref{eq:keldysh_eq} assuming each lead remains at equilibrium
with its Fermi function $f_a(E)$. We arrive at the main result of NEGF theory,
\begin{equation}
\label{eq:NEGF}
{G}^<_\text{sr}(E) = {G}^R_\text{sr}(E) \Sigma_\text{sr}^<(E) {G}^A_\text{sr}(E),
\end{equation}
with the lesser self-energy $\Sigma_\text{sr}^<$ defined as
\begin{equation}
\label{eq:lesserself}
\Sigma_\text{sr}^<(E) = -\sum_a f_a(E)
\left[\Sigma_{a}^R(E)-\Sigma_{a}^A(E)\right],
\end{equation}
where
\begin{equation}
\Sigma^R_{a} = \hat W_{0a}(\hat{G}^R_0)_a \hat W_{a0},
\end{equation}
is the partial contribution to the retarded self-energy $\Sigma^R(E) = \sum_a \Sigma^R_{a}$.
Since ${G}^A_\text{sr}= ({G}^R_\text{sr})^\dagger$ and ${\Sigma}^A_{a}= ({\Sigma}^R_{a})^\dagger$, we can equivalently write Eq.~\eqref{eq:lesserself} as
\begin{equation}
\label{eq:lesserself2}
\Sigma_\text{sr}^<(E) = i \sum_a f_a(E) \Gamma_a(E)\,,
\end{equation}
using the definition of the linewidth~\eqref{eq:def-Gamma} for lead $a$, $\Gamma_a = i ( \Sigma^R_a - \Sigma^A_a)$.

Equation~\eqref{eq:NEGF} provides a direct expression of the lesser Green's function in terms of the
retarded one and the occupation function in the leads. From there, one can directly calculate
physical observables as $\langle c^\dagger_m c_n \rangle = -i G^<_{nm}(\tau=0)$
which leads to the one-body density matrix,
\begin{equation}
\label{eq:1body_dst_matrix}
\langle c^\dagger_m c_n \rangle = -i \int \frac{dE}{2\pi} (G^<_\text{sr})_{nm}(E),
\end{equation}
which generalizes the concept of local density of states to off-diagonal observables and out-of-equilibrium situations.

Inserting Eq.~\eqref{eq:Gamma2_inc} into Eqs.~\eqref{eq:lesserself2} and \eqref{eq:NEGF} as well as using Eq.~\eqref{eq:wf_simplified} one can relate the lesser Green's function to the scattering wave-functions; one arrives at
\cite{wimmer2010, gaury2014}
\begin{equation}
\label{eq:NEGF_scattering}
G^<_{\rm sr}(E)_{nm} = i\sum_a f_a(E) \sum_{\alpha \in a} \Psi_\text{sr}(E)_{n\alpha} \Psi_\text{sr}^*(E)_{m\alpha}.
\end{equation}
At equilibrium, this equation simplifies into
$G^<_{\rm sr}(E) = i\Psi_\text{sr}(E) \Psi_\text{sr}^\dagger(E) f(E)$ which can be derived straightforwardly using the definition of $G^<$ \cite{stefanucci2013}.

It is worth noting that there is a subtle point in the derivation above: the limit $\eta\rightarrow 0$ has to be taken at the end of the calculation, not at the beginning, whenever the energy $E$ matches a bound state of the system, owing to the emergence of (diverging) Dirac functions at these energies.
It follows that, strictly speaking, one should add the bound states contributions to Eq.~\eqref{eq:NEGF} or \eqref{eq:NEGF_scattering} which only contain contributions from the scattering states.
For example, Eq.~\eqref{eq:NEGF_scattering} then reads
\begin{equation}
\begin{split}
G^<_{\rm sr}(E)_{nm} &= i\sum_a f_a(E)\sum_{\alpha \in a} \Psi_\text{sr}(E)_{n\alpha} \Psi_\text{sr}^*(E)_{m\alpha} \\
&+ i\sum_j f_j(E) \delta(E-E_j) \left(\Psi_\text{sr}\right)_{nj} \left(\Psi_\text{sr}^*\right)_{mj},
\end{split}
\end{equation}
where the index $j$ now labels bound states at energy $E_j$ with corresponding wave function $(\Psi_\text{sr})_{nj}$.
These contributions are usually omitted because (i) they do not contribute to transport properties and (ii) the filling factor $f_j(E)$ depends on the history of the system and must be decided based on physics not described within our framework \cite{stefanucci2013}.
These bound states may play a role nevertheless in certain situations \cite{profumo2015}.

At equilibrium, the expression for ${G}^<_\text{sr}$ can be simplified.
To do so, we use the relation for the retarded and advanced Green's functions obtained by (i) multiplying the equation $ {G}^R_\text{sr} (E-H_\text{sr} -\Sigma^R)= 1$ by ${G}^A_\text{sr}$ on the right,
(ii) multiplying its Hermitian conjugate $(E-H_\text{sr} -\Sigma^A) {G}^A_\text{sr}  = 1$ by ${G}^R_\text{sr}$ on the left and (iii) subtracting the second equation from the first.
One arrives at
\begin{equation}
\label{eq:optic}
{G}^R_\text{sr} [\Sigma^R - \Sigma^A] {G}^A_\text{sr} = {G}^R_\text{sr} - {G}^A_\text{sr}.
\end{equation}
Inserting Eq.~\eqref{eq:optic} into Eq.~\eqref{eq:NEGF}, we get
\begin{equation}
\label{eq:optic2}
{G}^<_\text{sr} = -f(E) [ {G}^R_\text{sr} - {G}^A_\text{sr} ],
\end{equation}
recovering Eq.~\eqref{eq:keldysh_eq}.
Calculating the number of electrons on one site at equilibrium using Eq.~\eqref{eq:1body_dst_matrix}, we get back the usual expression in terms of the local density of states
$\ldos$:
\begin{equation}
\langle c^\dagger_n c_n \rangle = \int dE f(E) \ldos,
\end{equation}
with
\begin{equation}
\label{eq:ldos_GR}
\ldos  = -\frac{1}{\pi}  \mathrm{Im} (G^R_\text{sr})_{nn}(E).
\end{equation}

\subsection{Landauer formula within NEGF}
\label{sec:non-equil-greens-1}
Let us now define the current flowing towards the scattering region and establish the Green's function version of the Landauer formula. Introducing the total number of electrons in the scattering region
$\mathcal{Q}_\text{sr}(t)=\langle {\cal Q}\rangle$ with ${\cal Q} = \sum_{n\in \text{sr}} c^\dagger_n c_n$, the evolution of
${\cal Q}$ (in Heisenberg representation) $\partial {\cal Q}/\partial t = i [{\cal H},{\cal Q}]$ provides the continuity equation
\begin{equation}
\frac{\partial \mathcal{Q}_\text{sr}}{\partial t} = \sum_a I_a(t)\,.
\end{equation}
In steady state, i.e., time-independent current flow $I_a(t) = I_a$, we find the current $I_a$ from lead $a$ towards the scattering region as
\begin{equation}
I_a = \frac{e}{h}\int dE\Tr [ G^<_{a0} \hat W_{0a} - \hat W_{a0} G^<_{0a}].
\end{equation}
Using Eq.~\eqref{eq:NEGF2}, Eq.~\eqref{eq:NEGF2bis}, and the definitions of the self-energies we get
\begin{equation}
\label{eq:pre-negf-landauer}
I_a = \frac{e}{h}\int dE\Tr \left[ G^<_\text{sr}( \Sigma^R_a - \Sigma^A_a) - (G^R_\text{sr} -G^A_\text{sr})\Sigma^<_a \right].
\end{equation}
Further using Eqs.~\eqref{eq:optic} and \eqref{eq:lesserself2}, we finally arrive at
 \begin{equation}
\label{eq:negf-landauer}
I_a = \frac{e}{h}\sum_b \int dE\Tr \left[ G^R_\text{sr} \Gamma_b G^A_\text{sr} \Gamma_a \right]
\left[f_b(E)-f_a(E) \right].
\end{equation}
This equation was first derived in \cite{caroli1971} and is the NEGF counterpart to the Landauer formula. Nevertheless it is often (somewhat abusively) referred to as the Landauer formula. It automatically satisfies two important requirements: the absence of incoming currents at equilibrium (note that there may be persistent currents
inside the scattering region), and the global conservation of current $\sum_a I_a = 0$.

To conclude this section, we end with the proof that Eq.~\eqref{eq:negf-landauer} is indeed equivalent
to the (original) scattering matrix version of the Landauer formula Eq.~\eqref{eq:general_landauer}.
Inserting Eqs.~\eqref{eq:Gamma2} and \eqref{eq:Gamma2_inc} in Eq.~\eqref{eq:negf-landauer} and using the invariance of the trace under cyclic permutation, we get
\begin{equation}
\label{eq:negf-landauer2}
\begin{split}
& \Tr \left[ G^R_\text{sr} \Gamma_b G^A_\text{sr} \Gamma_a \right]  \\
&=\Tr \left[ G^R_\text{sr} \Gamma_b \Phi_{\mathrm{p}-}^b\Phi_{\mathrm{p}-}^{\dagger b}\Gamma_b G^A_\text{sr}
\Gamma_a \Phi_{\mathrm{p}+}^a\Phi_{\mathrm{p}+}^{\dagger a} \Gamma_a\right] \\
&=\Tr \left[
\left(i \Phi_{\mathrm{p}+}^{\dagger a} \Gamma_a  G^R_\text{sr} \Gamma_b \Phi_{\mathrm{p}-}^b \right)
\left(-i \Phi_{\mathrm{p}-}^{\dagger b}\Gamma_b G^A_\text{sr} \Gamma_a \Phi_{\mathrm{p}+}^a \right)
\right] \\
&=\Tr \left[ S^{ab} (S^{ab})^\dagger\right],
\end{split}
\end{equation}
where the $\Phi_{\mathrm{p}+}^{a(b)}$ matrix is restricted to the modes of lead $a$ ($b$). We
have introduced a new object $S^{ab}$,
\begin{equation}
S^{ab} = i \Phi_{\mathrm{p}+}^{\dagger a} \Gamma_a  G^R_\text{sr} \Gamma_b \Phi_{\mathrm{p}-}^b\,.
\end{equation}
From the Fisher--Lee relation~\eqref{eq:fisher-lee}, we can indeed identify
\begin{equation}
S^{ab} = S^{ab}_{\text{p}\text{p}},
\end{equation}
i.e., the scattering matrix between lead $a$ and lead $b$, restricted to the propagating channels, concluding our proof of equivalence between the scattering approach and the NEGF approach.
Note that in the sum of Eq.~\eqref{eq:negf-landauer} only terms with $b\neq a$ contribute, and hence Eq.~\eqref{eq:negf-landauer2} is valid only for $a\neq b$, i.e., describes transmission from lead $b$ to $a$.

\subsection{Higher-order observables}
The observables defined in the preceding sections are all one-body observables
(except for the quantum noise), i.e., quadratic in the creation/destruction operators.
To calculate higher-order observables, one can make use of the fact that, even though the system is out of equilibrium, the density matrix is the exponential of a (highly non-local) quadratic Hamiltonian, see Eq.~\eqref{eq:effective_hamiltonian}.
It follows that the Wick theorem applies, allowing expression of, e.g., quartic terms in terms of quadratic ones,
\begin{eqnarray}
&\langle c^\dagger_n(t)c_m(t)& c^\dagger_p(t')c_q(t')\rangle = \\
&\langle c^\dagger_n(t)c_m(t)\rangle&\langle c^\dagger_p(t')c_q(t')\rangle
+\langle c^\dagger_n(t)c_q(t')\rangle\langle c_m(t)c^\dagger_p(t')\rangle, \nonumber
\end{eqnarray}
i.e., in terms of a sum of products of lesser and greater Green's functions. This approach can be used directly to calculate quantities such as shot noise \cite{gaury2016}.

The same approach can be used to derive expressions relating higher-order observables to the scattering wavefunction. For this, one uses the following dictionary \cite{gaury2014} :
\begin{equation}
  \label{eq:G<_Psi}
  [G^<(\tau)]_{nm} = i\sum_{\alpha}\int \frac{dE}{2\pi} (\Psi_\text{sr})_{n\alpha} (\Psi_\text{sr}^*)_{m\alpha} e^{-iE\tau} f_{\alpha}(E),
\end{equation}
and
\begin{equation}
  \label{eq:G>_Psi}
  [G^>(\tau)]_{nm} = -i\sum_{\alpha}\int \frac{dE}{2\pi} (\Psi_\text{sr})_{n\alpha} (\Psi_\text{sr}^*)_{m\alpha} e^{-iE\tau} [1-f_{\alpha}(E)].
\end{equation}
Note that likewise, the retarded Green's function can be expressed as
\begin{equation}
  \label{eq:GR_Psi}
  [G^R(\tau)]_{nm} = -i\theta(\tau)\sum_{\alpha}\int \frac{dE}{2\pi} (\Psi_\text{sr})_{n\alpha} (\Psi_\text{sr}^*)_{m\alpha} e^{-iE\tau}.
\end{equation}

\section{Selected applications}%
\label{sec:select-appl}

We have now completed the technical part of this review where we presented the mathematical formalisms and the associated algorithms used to make actual numerical calculations.
We are left to discuss how these techniques are used for real-world applications.
Numerical quantum transport calculations are by now very common in a wide variety of fields ranging from molecular electronics through spintronics to nanoelectronics.
Like any numerical tool, numerical quantum transport can play different roles:
It can be used to support an analytical theory, can test the validity of a hypothesis, can explore the behavior of a system to gain understanding through numerical experiments, or can predict effects that are later observed in experiment.

In this section, we will present selected success stories of numerical quantum transport calculations, i.e., applications where it has played a pivotal role.
In doing so, we hope to give the reader a sense of the breadth of applications numerical quantum transport is used for in practice.
We want to showcase the kind of questions it can help answer, but also highlight some potential pitfalls and limitations.
The choice of examples is necessarily subjective, and we will frequently refer to other reviews.

\subsection{A hierarchy of discrete models}
\label{sec:hierarchy}

\begin{table}
\begin{center}
    \begin{NiceTabular}{c|c}[cell-space-limits=8pt]
        \hline
        \Block{3-1}{\rotate Optionally \\ self-consistent} & \Block[fill=[HTML]{98db65}]{1-1}{Discretized $\mathbf{k}\cdot\mathbf{p}$ Hamiltonians}\\
        \cline{2-2}
        & \Block[fill=[HTML]{c7db65}]{1-1}{Tight-binding models (TBM)}\\
        \cline{2-2}
        & \Block[fill=[HTML]{dbc365}]{1-1}{Semi-empirical tight-binding models \\ (SE-TBM)}\\
        \hline\hline
        \Block{2-1}{\rotate Self-consistent} & \Block[fill=[HTML]{dba065}]{1-1}{First principles tight-binding models\\ (FP-TBM)}\\
        \cline{2-2}
        & \Block[fill=[HTML]{db8165}]{1-1}{Density functional theory (DFT)}\\
        \hline
    \end{NiceTabular}
\caption{Schematic of the different levels of modeling used in numerical quantum transport, see text.}
\label{fig:hierarchy}
\end{center}
\end{table}

The first step for a numerical quantum transport calculation is to select the appropriate level of modeling for the
Hamiltonian of the scattering region as well as of the (semi-infinite) electrodes and of the system-electrode interface.
There exists a hierarchy of discrete Hamiltonians that can be used to describe a system, and physicists must select an appropriate level of modeling, requiring a trade-off between accuracy and numerical tractability.
Table~\ref{fig:hierarchy} shows a schematic of this hierarchy. The models can be roughly divided into two large classes: those that aim at calculating the physics at the atomic
scale (hence at predicting the underlying band structure of the conductors) and those that focus only on the mesoscopic scale, hence need only to \emph{account} for the atomic level.
The former correspond to the lower two layers of Table~\ref{fig:hierarchy} and the latter to the upper three.

The bottom layer corresponds to calculations performed entirely within the framework of density functional theory.
In the absence of strong electronic correlations, DFT provides an accurate description of the system from first principles and has \emph{de facto} become the reference approach to the calculation of realistic materials \cite{kurth2017}.
We note, however, that for quantum transport care must be taken: the ground state Kohn--Sham Hamiltonian may not describe the non-equilibrium situation of quantum transport well, and we refer the reader to Refs.~\cite{thoss2018, evers2020} for a detailed discussion.

Next come the first-principles tight-binding models (FP-TBM) where a DFT calculation is followed by the projection of the corresponding (self-consistent) Hamiltonian onto a set of localized states such as maximally localized Wannier functions.
Semi-Empirical Tight-Binding Models (SE-TBM) are also models that are derived from a DFT calculation.
However, in contrast to FP-TBM where the model is constructed in a systematic way, the SE-TBM are parametrized and then fitted to properly reproduce the DFT calculations (typically the band structure) and/or experimentally known properties of the material (such as the band gap).
The orbitals and terms in the Hamiltonian included in the SE-TBM are typically dictated by symmetry and physics considerations.
When the Tight-Binding Model (TBM) is not tightly bound to a first-principles calculation, we simply call it a general TBM.
These TBMs can be studied from a purely theoretical perspective or in situations where a simple model is sufficient to account for the physics.
These model systems can also come from the discretization of a continuum model such as a $\mathbf{k}\cdot\mathbf{p}$ model or a simple effective mass model.

The $\mathbf{k}\cdot\mathbf{p}$ Hamiltonians are the highest level of our hierarchy. They are low-energy effective theories valid in the vicinity of some high-symmetry $k$-point. They describe the physics at large scale, not at the atomic scale, which can be advantageous for the simulation of large systems.
There exists a large literature on the $\mathbf{k}\cdot \mathbf{p}$ method \cite{voon2009}.
For example, Ref.~\cite{kormanyos2015} describes the particular case of two-dimensional transition metal dichalcogenide semiconductors while Ref.~\cite{marconcini2011} focuses on graphene.
$\mathbf{k}\cdot\mathbf{p}$ Hamiltonians can account for magnetic field as well as spin-orbit effects \cite{winkler2003}.

Another important ingredient is the electromagnetic and in particular electrostatic environment in which these electrons propagate.
In fact, the electrostatic energy is usually the largest energy of the problem. In first-principles calculations, the electrostatic problem is solved self-consistently with the quantum mechanical one. It is required to properly account for the electronic distribution at the atomic scale. In some systems, for instance semiconducting devices, charge redistribution can also take place at the mesoscopic scale, hence also necessitating a self-consistent calculation. However, in model systems (general TBM and $\mathbf{k\cdot p}$ models) it is common to assume that the electric potential is known and has a simple form.

In the remainder of this section, we will give a few selected examples to illustrate cases where computational quantum transport had an important impact on a field. We mainly focus on ``model'' systems (the three upper types of models in Table~\ref{fig:hierarchy}) but also show examples on more ``realistic'' models (the two lower levels in Table~\ref{fig:hierarchy}).

The distinction between ``model'' and ``realistic'' modeling is not only one of accuracy but more deeply corresponds to two different and complementary approaches. In the ``realistic'' modeling, one tries to describe the system microscopically as well as possible to perform what could be called ``numerical experiments''. The more precise the simulation, the better, and one tries to improve the simulations by adding more physical effects (e.g. spin-orbit coupling, proper modeling of the interfaces, better density functionals...). In contrast, in the ``model'' system, one often simplifies the model until only the minimum necessary physical ingredients to obtain an effect are taken into account.
These simulations are often performed in conjunction with analytical calculations and would be better described as ``computer-assisted theory'' than ``numerical experiments''.

When comparing both kinds of simulations to actual experiments, it is tempting to expect the ``realistic'' models to be more accurate. It can be the case, but there are also numerous situations where the role of the microscopic structure is to provide a few effective parameters that can be considered as fitting parameters or measured directly. In that case, the model systems can be actually very accurate. Examples include electronic interferometers
where the role of different quantum point contact (``beam splitters'') is merely to provide tunable reflection/transmission amplitudes and need not necessarily be modeled precisely.
Another important class of systems are the topological insulators and quantum Hall physics where the topological protection makes many predictions insensitive to the microscopic details. Our review is clearly biased towards the ``model'' systems due to the background of most of the authors. The role of our discussion of ``realistic'' systems is mostly to illustrate the breadth of applicability of numerical quantum transport, highlight a few success stories and give pointers to the further literature.

\subsection{Applications to the two-dimensional electron gas}
\label{sec:early-applications}

\subsubsection{The early days: Disordered systems and Anderson localization}
\label{sec:anderson-localization}

An important application of numerics to quantum transport is the study of
the effect of static disorder on conductance as well as of the associated Anderson localization phenomena.
This is perhaps the historically first application of the techniques described in this review and also one of the most successful ones.

The model is the discretized effective mass Schrödinger equation Eq.~\eqref{eq:schro2D-discrete} in the presence of static disorder. In two dimensions the model reads
\begin{eqnarray}
\label{eq:schro2D-discrete-bis}
 \Psi_{n_x+1,n_y} &+& \Psi_{n_x-1,n_y} + \Psi_{n_x,n_y+1} + \Psi_{n_x,n_y-1}  \nonumber \\
  &+& V_{n_x,n_y} \Psi_{n_x,n_y}= E \Psi_{n_x,n_y},
\end{eqnarray}
where the potential $V_{n_x,n_y}$ takes different random values on different sites. Typically,
one chooses  $V_{n_x,n_y}$ randomly with a flat distribution $V_{n_x,n_y} \in [-W/2,W/2]$ where the parameter $W$ is the strength of the disorder. In a seminal paper, the ``gang of four''
\cite{abrahams1979} had predicted that the conductance $g(W,L)$  (where $L$ is the size of the sample) followed a simple scaling law from which one could predict the insulating or metallic behavior of the sample. The scaling hypothesis, which was supported by theoretical arguments both in the strong and weak disorder regimes, was that the function $\beta \equiv
\partial\log g/\partial\log L$ was a function of $g$ only
(i.e., $\exists f, \ \beta(L,W)=f[g(W,L)]$ ). One of the most striking conclusions of the scaling theory was that in three dimensions there was a critical value of disorder associated to a metal-insulator transition, but in one and two dimensions the system was always an insulator in the sense that the conductance would become vanishingly small in the $L\rightarrow\infty$ limit.

\begin{figure}
  \centering
  \includegraphics[width=0.45\linewidth]{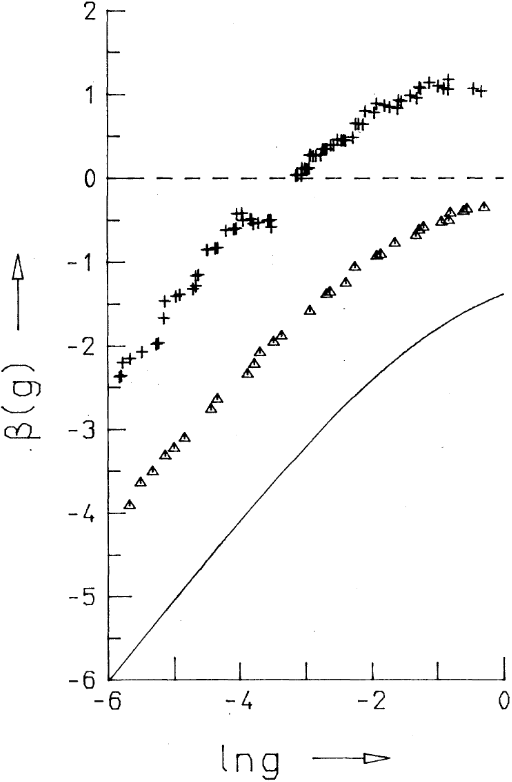}
  \caption{Scaling function $\beta(g)$ computed numerically for a 2D (triangles) and 3D (pluses) tight-binding model, confirming the scaling hypothesis.
  Reproduced with permission from \cite{mackinnon1981}. Copyright 1981 by the American Physical Society.}
  \label{fig:mackinnon1981}
\end{figure}

While the scaling hypothesis was plausible, it was merely a hypothesis. It was put on firm grounds in a series of works combining numerical calculations with finite-size scaling analysis. These calculations were pioneered by \cite{pichard1981,pichard1981b} who, however, wrongly concluded the existence of a metal-insulator transition in two dimensions. Subsequent more precise numerics \cite{mackinnon1981} could indeed confirm the finite-size scaling hypothesis, as shown in Fig.~\ref{fig:mackinnon1981}. A large activity followed in the following decade with the study of the effect of various parameters such as spin-orbit coupling, magnetic field, type of disorder, as reviewed in \cite{kramer1993}. The field has now matured into very precise calculations of the phase diagram and critical exponents \cite{slevin2014}.

Related to the Anderson localization regime, numerics have also been used to study the quantum corrections to the classical limit in the diffusive regime. An early work in this direction is the study of the universal conductance fluctuations \cite{stone1985}. More recent applications
include simulations of diffusive metal-superconductor junctions \cite{marmokos1993} or disproving of the conjecture of the existence of a two-parameter scaling regime in a system with partially broken time reversal symmetry \cite{schomerus2000}.

Here we showcased the role that numerical quantum transport played in understanding disorder physics. Extensive reviews on localization and disorder phenomena can be found, e.g., in \cite{kramer1993,beenakker1997,evers2008}.

\subsubsection{Ballistic systems and Quantum billiards}
\label{sec:quantum-billards}
\begin{figure}
  \centering
  \includegraphics[width=0.9\linewidth]{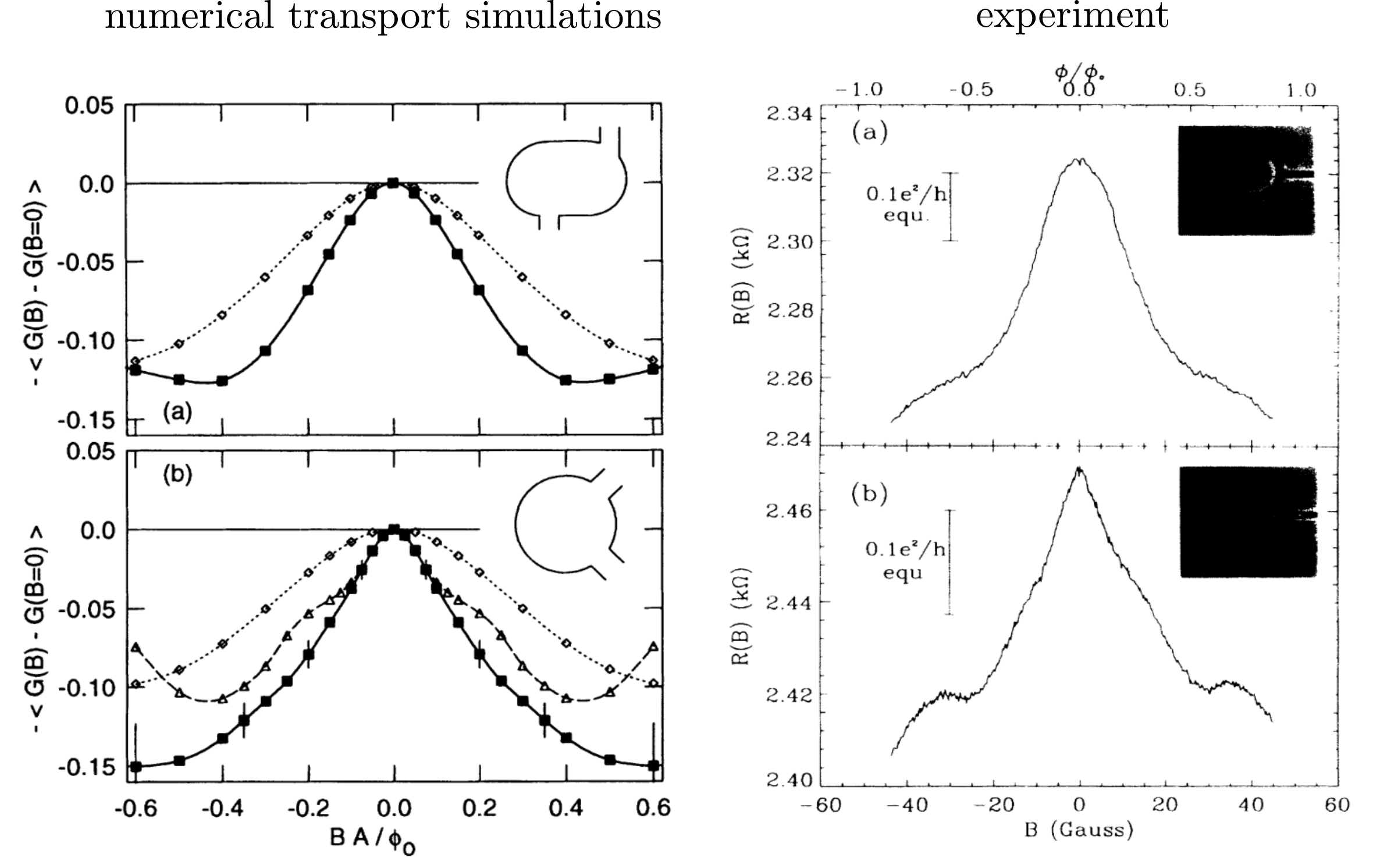}
  \caption{Left: Numerically computed weak localization correction to the conductance for a chaotic stadium billiard (top) and an integrable circular billiard (bottom). Solid lines correspond to specular scattering at the boundary, other lines to a situation with disorder at the boundary. Right: Corresponding experimental data obtained by averaging over 48 billiards defined in a high-quality GaAs electron gas. The lineshape in the chaotic case is Lorentzian and for the integrable billiard triangular, in agreements with simulation. Reproduced with permission from \cite{chang1994}. Copyright 1994 by the
American Physical Society.}\label{fig:chang1994}
\end{figure}

In the early nineties, the opposite, ballistic regime (no disorder) attracted wide interest in the context of ``quantum billiards''.
Quantum billiards consist of a high mobility two-dimensional electron gas (typically GaAs/GaAlAs heterostructures) where a billiard is patterned using electrostatic gates deposited on top of the heterostructure.
A study of such billiards (also known as quantum dots in the regimes where they are almost closed and charging energy plays an important role) was performed in \cite{jalabert1990} using the RGF algorithm. 4-terminal (X-shaped) and 3-terminal (T-shape) ballistic junctions were studied in \cite{baranger1991,sols1989}.
One of the successes of these ballistic calculations was the finding that the statistics of the height of the resonances (in the almost closed regime) matched the Porter--Thomas distribution prediction of random matrix theory \cite{jalabert1992}.
Numerics also clarified \emph{weak localization}, i.e., the interference contribution to the magneto-conductance \cite{baranger1993}.
Numerics could in particular clearly distinguish the difference in behavior between the magneto-conductance of \emph{integrable} and \emph{chaotic} billiards: a chaotic billiard gives rise to a Lorentzian lineshape for the magnetoconductance, whereas an integrable billiard features a triangular lineshape, as shown in the left panels of Fig.~\ref{fig:chang1994}.
This prediction was soon confirmed experimentally \cite{chang1994}, shown in the right panels of Fig.~\ref{fig:chang1994}. In billiards where there exists a short straight path between the entrance and the exit, the observation of Fano resonances \cite{gores2000} was also well reproduced in numerics \cite{clerk2001}.

Beyond billiards, recent ballistic devices have, e.g., implemented electronic interferometers analogous to those known in optics \cite{bauerle2018}. Those systems can also be simulated accurately both with \cite{kazymyrenko2008} (quantum Hall regime) and without \cite{bautze2014} magnetic field.

More extended reviews on quantum billiards can be found in, e.g., \cite{beenakker1997, alhassid2000}.

\subsection{Quantum transport in graphene}
\label{sec:graphene}

\subsubsection{Numerical tight-binding simulations}

In numerical transport studies, graphene is often well-described by a simple nearest-neighbor tight-binding model,
\begin{equation}
\label{eq:grapheneTBM}
t \sum_{j=N(i)} \Psi_j = E\Psi_i,
\end{equation}
where the site $i$ sits on the hexagonal lattice of graphene and the sum over $j=N(i)$ stands for the three neighbors $j$ of site $i$.  The value $t=2.7$ eV is the hopping integral between two nearest neighboring $p_z$ orbitals.

Early on, the RGF algorithm was adapted from the square lattice to the hexagonal case of graphene \cite{lewenkopf2013}.
A notable use was confirming the prediction that the conductivity of pristine graphene at the Dirac point was $4e^2/(\pi h)$ \cite{tworzydlo2006} and predicting that the Fano factor has a value of $1/3$. The Fano factor predicted from these quantum transport simulations was later indeed observed experimentally \cite{dicarlo2008,danneau2008}.
Going beyond pristine graphene, RGF calculations were also used extensively to study the effect of disorder
which makes the conductivity dependent on the disorder strength and correlation length \cite{lewenkopf2008} or
to assess the regime of validity of, e.g., semi-classical Boltzmann approaches \cite{klos2010}.

Another notable example of influential quantum transport simulations is the study of quantum Hall effect in graphene induced by a light field \cite{takashi2009}.
In fact, this work presented both analytical and numerical results, but the analytical results later proved to be incorrect \cite{takashi2009erratum}.
The numerical results based on the tight-binding model however upheld the conclusion.
Reference~\cite{takashi2009} was the starting point of the subsequent theories of Floquet topological insulators \cite{kitagawa2011,lindner2011}, which has since then grown into an active field of research (see, e.g., \cite{cayssol2013,rudner2020} for reviews).
Additionally, this numerical study spurred experimental efforts implementing this effect using photonic quantum simulation \cite{rechtsman2013}, as well as the observation of an anomalous Hall effect in illuminated graphene \cite{mciver2020}.

Besides these selected examples, there are numerous numerical transport studies of graphene, far beyond the scope of the present review.
We refer the reader to reviews such as \cite{perez2010, mucciolo2010} for a more comprehensive overview.

\subsubsection{Massless Dirac equation}
\label{sec:massless_dirac}
At low energy, pristine graphene is described by 2 independent relativistic Weyl (massless Dirac) equations
\begin{equation}
\label{eq:dirac}
H_\mathrm{D} = v_F
\begin{pmatrix}
0 & p_x-ip_y \\ p_x + ip_y & 0
\end{pmatrix},
\end{equation}
corresponding to two different valleys K and K',
where the $2\times 2$ structure arises from graphene having two atoms per unit cell, and $v_F$ is the Fermi velocity.
Much of the interesting physics of graphene are due to the low-energy massless Dirac equation (see, e.g., \cite{castroneto2009} for a review).
Beyond graphene, relativistic equations of this kind nowadays play an important role in modern condensed matter physics as they arise in a number of topological materials \cite{wehling2014}.

Physics related to the Dirac equation only shows if the scattering between valleys is small.
In the tight-binding model of the previous section this requires the use of a long-range scattering potential, making the simulated systems larger and thus more costly.
It would be advantageous to only simulate one of the two valleys, however  any discretization scheme faces the so-called ``fermion doubling'' problem: in any tight-binding model, Dirac points always come in pairs \cite{susskind1977,nielsen1981}.

This problem can be circumvented by borrowing solutions from the lattice gauge theory \cite{stacey1982}, such as by evaluating finite differences on a lattice that is displaced symmetrically from the original lattice \cite{tworzydlo2008} or by adding a small term which breaks the time-reversal symmetry \cite{hong2012,habib2015} or $k$-space discretization \cite{bardarson2007}.
For example, the simulation in Ref.~\cite{bardarson2007} was critical to determining the correct scaling function $\beta(g)$ of graphene.
For a more extended review of avoiding the Fermion doubling problem in transport, we refer to, e.g., \cite{pacholski2021, beenakker2023}.

\subsection{Quantum spin Hall effect}
\label{sec:topol-insul}

The seminal work of Kane and Mele introduced the concept of two-dimensional quantum spin Hall insulators \cite{kane2005,kane2005b}, and thus one of the earliest examples of symmetry-protected phases beyond the quantum Hall effect.
To this end, they demonstrated the effect in what is now know as Kane--Mele model, corresponding to a single orbital model for graphene Eq.~\eqref{eq:grapheneTBM} to which
one adds a spin-orbit coupling term of strength $\lambda$. The tight-binding equation reads
\begin{eqnarray}\label{eq:kanemele}
t \sum_{j=N(i)} \Psi_j + i\lambda \sum_{j=N''(i)} \nu_{ij} \sigma_z \Psi_j = E \Psi_i,
\end{eqnarray}
where $\Psi_i = (\Psi_{i\uparrow},\Psi_{i\downarrow})$ is a two component spinor, $\sigma_z$ a Pauli matrix,
$N(i)$ [resp. $N''(i)$] stands for the [resp. second] nearest neighbors of $i$, and $\nu_{ij}=-1$ (resp. $\nu_{ij}=1$) for clockwise (resp. counter-clockwise) hoppings. The spin-orbit term opens an energy gap in the spectrum. Inside this gap are topologically protected helical states that appear at the edge the system.
The prediction of a quantized conductance even in the presence of disorder was then confirmed explicitly using a numerical quantum transport calculation \cite{sheng2005}.

The quantum spin Hall effect was also predicted in a model describing type III semiconductor quantum wells, in what is now known as Bernevig--Hughes--Zhang (BHZ) model \cite{bernevig2006}. The BHZ Hamiltonian is a four band model
that reads
\begin{eqnarray}
\label{eq:bhz}
H = C + M \sigma_0 \tau_0 - B (p_x^2+p_y^2) \sigma_0 \tau_z \nonumber \\
- D (p_x^2+p_y^2) \sigma_0 \tau_0
+ A (p_x \sigma_z \tau_x - p_y \sigma_z \tau_y),
\end{eqnarray}
where $\sigma_a$ and $\tau_a$, $a \in \{0,x,y,z\}$ are Pauli matrices (or the identity matrix for $a=0$)
that act, respectively, on the spin degree of freedom or an extra orbital degree of freedom.
The interest for the BHZ model stems from the fact that it arises naturally as an effective 2D model for a quantum well made in inverted semi-conductor heterostructures such as HgTe/HgCdTe. The
parameters $A,B,C,D$ and $M$ are material-specific and depend on heterostructure geometry
parameters such as well thickness.

Discretized versions of the BHZ model are a natural playground for numerical transport calculations in devices.
As a particularly striking example, numerical quantum transport calculations were crucial in identifying  a ``topological Anderson insulator'' phase, a topological phase induced by disorder \cite{li2009}.
This numerical work opened a new field with many theoretical follow-up studies, explaining the effect in an effective model \cite{groth2009}, and expanding the concept to different dimensions \cite{guo2010,altland2014} as well as higher-order topological systems \cite{yang2021}.
It also motivated experimental efforts that observed this effect implementing the Hamiltonian using photonic systems \cite{stutzer2018,liu2020}, cold atoms \cite{meier2018} and classical electric circuits \cite{zhang2021}.

For a more comprehensive overview beyond these specific examples, we refer to the reviews of Refs.~\cite{hasan2010,qi2011,bansil2016}.

\subsection{Mesoscopic superconductivity}
\label{sec:mesosc-superc}

It is natural to apply computational quantum transport methods to mesoscopic superconductivity problems, and this has been done in this field since the early days.
A natural framework to do so is the mean field Bogoliubov De Gennes (BdG) method \cite{degennes1996} that extends an initial normal (non-superconducting) Hamiltonian $H$ to a larger Hilbert space with a $2\times 2$ Nambu block structure with an ``electron'' and a ``hole'' sector:
\begin{equation}
\label{eq:bdg}
H_\mathrm{BdG} =
\begin{pmatrix}
H & \Delta \\ \Delta^\dagger & -H^*
\end{pmatrix}.
\end{equation}
The underlying normal Hamiltonian $H$ can be either in the continuum or already discretized into some tight-binding model Hamiltonian \cite{zhu2016}, making this technique applicable to a wide class of devices where some part is superconducting.
The anti-symmetric matrix $\Delta$ accounts for the superconducting pairing.
For conventional $s$-wave superconductors, the restriction of $\Delta$ to one spin sector (say electron-spin up and hole-spin down) is a diagonal matrix $\Delta_{nm} = \Delta_n \delta_{nm}$ where $|\Delta_n|$ is the local superconducting gap on site $n$.
Using various forms of off-diagonal $\Delta$, it is also possible to account for more exotic pairings such as $p$-wave or $d$-wave superconductivity \cite{asano2006}.
In principle $\Delta$ must be calculated self-consistently from the mean-field treatment of the effective (attractive) electron-electron interaction.
However, in some situations (such as a bulk superconductor in contact with a small mesoscopic region) it is sufficient to consider $\Delta_n$ to be equal to its bulk value inside the superconductor and zero in the normal part. This approximation is often used in practice.

\subsubsection{Andreev reflection}
\label{sec:andreev-bound-states}

\begin{figure}
  \centering
  \includegraphics[width=0.6\linewidth]{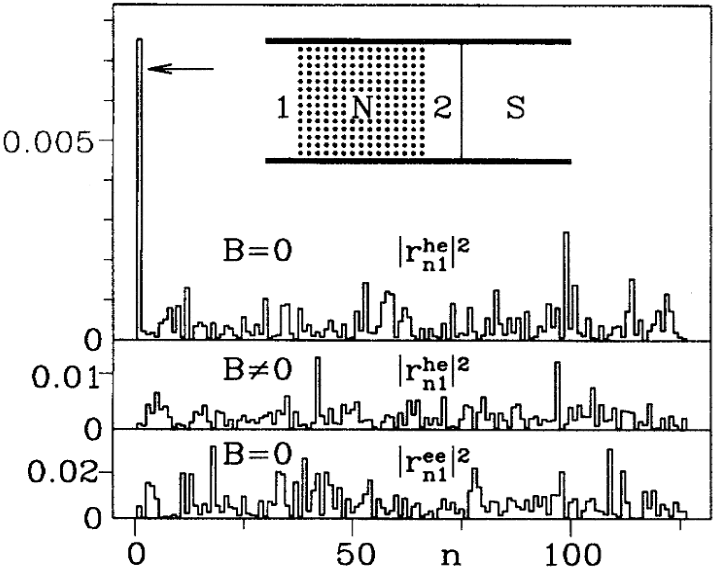}
  \caption{Numerical simulation of a $300\times 300$ tight-binding
    model for a disordered normal metal, in series with a
    superconductor (inset). The histograms give the modal distribution
    for reflection of an electron, with mode number
    1 indicating normal incidence. The top two panels give the distribution of reflected holes
    (in the absence and presence of a magnetic field $B$), the bottom
    panel of reflected electrons (for $B=0$). The arrow indicates the
    ensemble-averaged height of the giant backscattering peak for Andreev
    reflection, predicted from theory. Reproduced with permission from
    \cite{beenakker1995}. Copyright 1995 by the American Physical Society.}
  \label{fig:beenakker1995}
\end{figure}

In the presence of a single superconductor (or several at the same electrochemical potential) and normal
electrodes, the superconductor acts as an ``Andreev mirror'' that reflects electrons into holes.
The conductance of such a system can be obtained through a modified form of the Landauer formula Eq.~\eqref{eq:current_BdG} \cite{lambert1991}.
In the simple case where the superconducting gap is much smaller than the Fermi energy and one considers voltage biases smaller than the gap, it is sufficient to study solely the scattering matrix of the normal part of the system and infer the properties of the full normal-superconducting system from it \cite{marmorkos1993,beenakker1995}.
Numerical quantum simulation was particularly influential in understanding the physics of Andreev reflection in mesoscopic devices.
For example, Beenakker et al.\ \cite{beenakker1995} showed that Andreev reflection gives rise to an enhancement of Andreev back-scattering even in disordered systems, as shown in Fig.~\ref{fig:beenakker1995}.
Another example are numerical studies of Andreev billiards, the superconducting version of the quantum billiards described in Sec.~\ref{sec:quantum-billards}, such as in Refs.~\cite{melsen1997, kormanyos2004}.
For more extended coverage of Andreev reflection in mesoscopic systems, we refer to the reviews of \cite{beenakker1992, beenakker1995, beenakker2005}.

\subsubsection{Search for Majorana bound states in nanowires}

Majorana bound states (MBSs) are the simplest example of non-Abelian anyons \cite{nayak2008}.
They have been predicted to be realized, for instance, in one-dimensional hybrid superconducting/semiconducting nanowires.
The simplest model describing such a system is the Bogoliubov-de Gennes Hamiltonian
\begin{equation}
\label{eq:bdg_majorana}
  H = \left(\frac{p_x^2}{2m} - \mu + V(x) +\frac{\alpha_R}{\hbar} p_x \sigma_y\right) \tau_z + E_Z \sigma_x \tau_z + \Delta \sigma_y\tau_y,
\end{equation}
where $\sigma_i$ ($\tau_i$) are Pauli matrices in spin (electron-hole) space, $i=0,x,y,z$. $p_x$ is the momentum along the nanowire, $m$ the effective mass of the semiconductor, $\mu$ the global
chemical potential, and $V(x)$ a potential induced for example by external gates, allowing to
change the electron density in different parts of the wire and defining a tunnel barrier.
$\alpha_R$ is the strength of Rashba spin-orbit coupling in the semiconductor, $E_Z$ the Zeeman splitting induced by an external magnetic field along the nanowire, and $\Delta$ the induced superconducting gap. This model exhibits MBSs when $E_Z^2 > (\mu - V_\text{wire})^2 + \Delta^2$, in a region with potential $V_\text{wire}$ \cite{lutchyn2010, oreg2010}.
Arising through topological phase transition, the appearance of an MBS is always accompanied by a gap closing at the transition point.

Many transport simulations rely on a discretized version of this simple model, but it has also been extended to higher spatial dimensions, including additional physical effects such as the orbital effects of a magnetic field \cite{nijholt2016}, a more realistic modeling of the proximity effect \cite{antipov2018, reeg2018}, and including all these effects together with electrostatic simulations \cite{woods2018, winkler2019}.

Experimental and theoretical work in this field is numerous, with experiments being heavily scrutinized.
Here, we highlight selected numerical transport simulations that crucially shaped the understanding in this field, in particular how to interpret conductance measurements.

Early work focused on studying the appearance of the zero-bias peak in different device geometries \cite{prada2012, rainis2013}. It was also shown that the gap closing accompanying the appearance of MBS may not be visible in transport measurements \cite{stanescu2012, pientka2012}.

\begin{figure}
  \centering
  \includegraphics[width=0.7\linewidth]{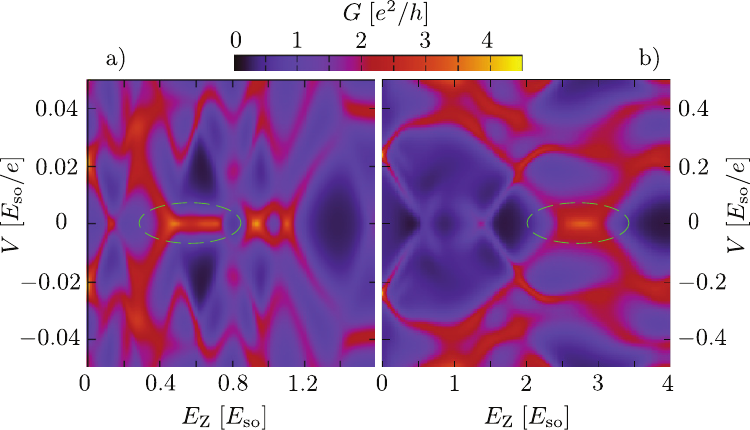}
  \caption{Numerical simulation of a two-dimensional version of
    Eq.~\eqref{eq:bdg_majorana} for a nanowire including a tunnel barrier and disorder. The differential conductance is shown for a single disorder realization as a function of bias voltage $V$ and Zeeman splitting $E_Z$ due to a parallel magnetic field. The system is in the trivial regime, but still exhibits a conductance peak pinned to zero voltage (green circles) that could be mistaken for a Majorana bound state. The two panels correspond to two parameter regimes. Reproduced from \cite{pikulin2012} under a CC BY-NC-SA 3.0 licence.}
  \label{fig:pikulin2012}
\end{figure}

Numerical transport simulations have been essential in shaping the understanding that zero-bias peaks are not a smoking-gun signature of MBSs. Early work showed that disorder can lead to trivial low-lying Andreev bound states giving rise to zero-bias peaks in the conductance \cite{potter2012,pikulin2012,rainis2013}. Figure~\ref{fig:pikulin2012} shows one example of such zero-bias peaks sticking to zero energy even in a trivial system. Later, it was shown that even in the absence of disorder particular Andreev bound states can appear very similar to MBS \cite{liu2017}, mimicking for example quantized conductance signatures \cite{moore2018}, that previously were considered a smoking gun. Even more, it was shown that any local measurement signal associated with MBSs can be mimicked by trivial states \cite{vuik2019}. For this reason, new measurement paradigms have been proposed, in particular non-local measurements to detect the topological phase transition \cite{rosdahl2018, danon2020}.

For a broader overview of the physics of Majorana bound states, we refer to the reviews of \cite{leijnse2012, prada2020}.

\subsubsection{Supercurrents}
\label{sec:supercurrents}

Another important question that can be considered is the current-phase relation for the supercurrent that can flow in-between two superconductors \cite{golubov2004}.
Though seemingly related, this is not strictly speaking quantum transport since the supercurrent actually flows {\it at} equilibrium.
Given the importance of this topic, we however briefly comment on how to compute supercurrents numerically.

Numerical calculations of supercurrents are particularly useful for the self-consistent problem, which is usually difficult to address analytically. There are many examples of such calculations including \cite{levyyeyati1995,nikolic2001,black-schaffer2008}.

Interestingly, when different superconducting electrodes do not share the same electrochemical potential, the problem becomes effectively time-dependent even though the initial formulation is not. For instance, a DC voltage $V_b$ across a superconducting-normal-superconducting junction gives rise to an AC current at frequency $2eV_b/h$---this is the AC Josephson effect. To simulate such problems, one needs to leave the DC framework of this article and use time-dependent techniques \cite{cuevas1996,perfetto2009,stefanucci2010,weston2016}.

\subsection{Self-consistent quantum-electrostatic simulations of nanoelectronics}
\label{sec:SCQE_1}

The model calculations discussed previously assume a known Hamiltonian, i.e., that the electric potential seen by the conducting electrons is an input of the model. This level of modeling is sufficient in many situations but suffers from a number of limitations: the chosen potential may be unrealistic, the relations between the electric potential and, e.g., the voltage applied to electrostatic gate may be unknown, or they may ignore important effects such as edge state reconstruction or non-linear corrections to the conductance. A second level in modeling thus consists in {\it calculating} the electric potential self-consistently by solving the quantum problem (the focus of this review) together with the Poisson equation. Below, we review the main aspects of the self-consistent problem.

\subsubsection{Formulation of the self-consistent problem}
The self-consistent problem is formulated as a set of three equations that account, respectively, for quantum mechanics, statistical physics, and electrostatics. We denote the electric
potential as $U$ and for simplicity suppose that the tight-binding model is formulated in real space (otherwise $U$ has to be replaced by the corresponding matrix elements in the quantum problem).
We write $H_0$ for the tight-binding Hamiltonian without electric potential and $\Psi_{\alpha E}$ for the scattering state in mode $\alpha$ and energy $E$ (ignoring for simplicity possible bound states present in the system). The problem reads
\begin{eqnarray}
\left[ H_0 -e U \right] \Psi_{\alpha E} = E \Psi_{\alpha E}, \label{eq:scs_1} \\
n = \int \frac{dE}{2\pi} \sum_\alpha  |\Psi_{\alpha E}|^2 f_\alpha (E), \label{eq:scs_2} \\
\vec\nabla  \cdot (\epsilon \vec\nabla U) = e n + e n_0\label{eq:scs_3},
\end{eqnarray}
where Eq.~\eqref{eq:scs_1} is the scattering problem formulated at the beginning of this review.
Equation~\eqref{eq:scs_2} relates the electronic density $n$ to the injectivities/partial local density of states $|\Psi_{\alpha E}|^2/(2\pi)$ in a non-equilibrium situation where different leads may have different chemical potentials or temperatures. Equation~\eqref{eq:scs_3} is the Poisson equation in a (possibly spatially dependent) dielectric constant $\epsilon$, $e$ is the electron charge, and $n_0$ an additional electronic density due to, e.g., dopants.

The self-consistent quantum-electrostatic problem in the presence of electrodes is a non-linear
integro-differential problem. Its solution is usually obtained by some sort of iterative scheme.
One starts with an initial electronic density, calculates the potential from the Poisson equation, then solves the quantum problem to recalculate the density. This sequence is iterated until convergence.
Many techniques have been developed to obtain better convergence properties for such schemes. We refer to \cite{armagnat2019,lacerda2025,lacerda2025b} for an entry point to the corresponding literature. Besides the convergence of the self-consistent loop, the calculation of the density itself, Eq.~\eqref{eq:scs_2}
can be computationally intensive.  To accelerate the computation, it is useful to separate the density into the sum of an equilibrium contribution and an out-of-equilibrium one. The calculation of the equilibrium density can be accelerated using integration in the complex plane \cite{brandbyge2002,ozaki2007,karrasch2010,papior2017} while the integration of the out-of-equilibrium contribution spans a small range of energies close to the Fermi energy \cite{sanvito2011}.

The importance of taking into account electrostatics can be illustrated by the non-linear
part of the current-voltage characteristics $I(V)$ of a device \cite{christen1996}. Let us write $g_2$ for the second-order contribution such that $I(V) = g V + g_2 V^2 + \mathcal{O}(V^3)$.  The non-interacting Landauer theory then gives $g_2 = \frac{1}{2}\frac{\partial g}{\partial E_F}$ with $g_2(B)=g_2(-B)$ in the presence of a magnetic field. When taking electrostatics into account, one needs to calculate the
\emph{emissivity} $\frac{\partial n_i}{\partial V}$ (change of density due to the bias voltage, also known as Landauer dipole), then compute the change of potential $\frac{\partial U_i}{\partial n_j}$ due to the change of density $n_j$ (through Poisson equation), then finally compute the \emph{injectivity}, i.e., the change of conductance
$\frac{\partial g}{\partial U_i}$ due to the change of potential. One arrives at
\begin{equation}
\label{eq:landauer_dipole}
g_2 = \frac{1}{2}\frac{\partial g}{\partial E_F} + \sum_{ij} \frac{\partial g}{\partial U_i}
\frac{\partial U_i}{\partial n_j} \frac{\partial n_j}{\partial V},
\end{equation}
where the second term is entirely due to electron-electron interactions \cite{angers2007,polianski2007,hernandez2013}. In particular
the antisymmetric part $g_2(B)-g_2(-B)$ vanishes in the absence of this term.

\subsubsection{Application to edge state reconstruction in the quantum Hall effect}
\begin{figure}
  \centering
  \includegraphics[width=0.7\linewidth]{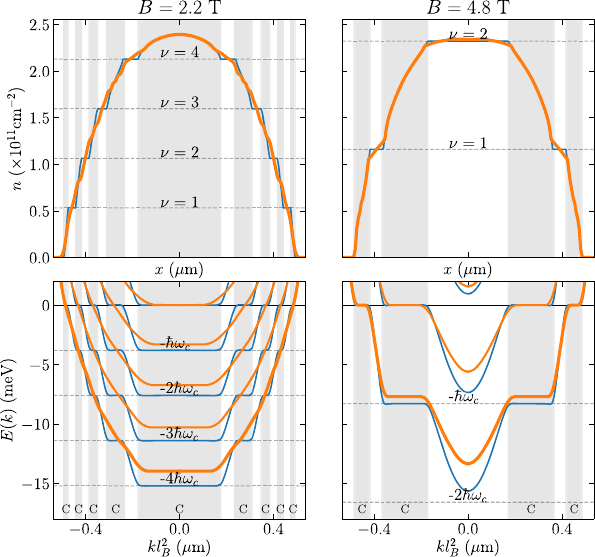}
  \caption{Numerical calculation of the electron density (top) and band structure (bottom) for a top gate-defined wire in a two-dimensional electron gas in the quantum Hall regime. The device is translationally invariant along $y$; the electrons are confined by two gates along $x$, with the magnetic field applied along $z$. Blue lines correspond to a self-consistent Thomas--Fermi calculation, orange lines to a full solution of the Poisson--Schrödinger problem. Gray regions indicate the compressible stripes while the white regions are incompressible. Reproduced from \cite{armagnat2020} under a CC BY 4.0 licence; the middle column has been omitted.}\label{fig:armagnat2020}
\end{figure}

Quantum transport within the self-consistent quantum electrostatic model has been studied in many situations. (Self-consistent calculations within the density functional theory framework will be discussed in the next section.) They are particularly important for the quantum Hall effect as the electrostatic interaction leads to a reconstruction of the edge states into compressible and incompressible stripes \cite{chklovskii1192} and the
non-interacting picture can be qualitatively wrong \cite{sahasrabudhe2018,armagnat2019,armagnat2020}. An example is shown in Fig.~\ref{fig:armagnat2020}, where the bottom panels show the dispersion relation and the upper panels the electronic density. One should focus on the orange curves; the blue curves correspond to the Thomas--Fermi approximation used in \cite{chklovskii1192}. On the left, the magnetic field is not very high and one recovers the usual Landauer--Büttiker picture of edge states. However, on the right, at a higher magnetic field, the reconstruction takes place and one observes a very different dispersion relation with almost flat plateaux at the Fermi level.

\subsubsection{Application to semiconductor device simulations}
\label{sec:semidev}

The simulation of semiconductor devices typically also requires self-consistent calculations to account for voltage drops and band bending.
Frequently, semiconductors can be described in terms of a single, parabolic band in the effective mass approximation, similar to the models considered in Sec.~\ref{sec:early-applications}.
However, in order to take into account, for instance, the $p$-character of the valence band, multi-band $\mathbf{k}\cdot\mathbf{p}$-models are employed. For example, the $4 \times 4$ Luttinger Hamiltonian \cite{luttinger1955},
\begin{equation}
\label{eq:luttinger}
H = \frac{1}{2m} \left[ \left(\gamma_1
+ \frac{5}{2}\gamma_2 \right) \hat{\mathbf{p}}^2  - 2\gamma_2 \left( \hat{\mathbf{p}} \cdot \hat{\mathbf{S}} \right)^2 \right],
\end{equation}
describes the heavy and light hole bands of semiconductors. Here $m$ is the electron mass and $\hat{\mathbf{S}}=(S_x,S_y,S_z)$ is a vector of spin-3/2 operators. The dimensionless parameters $\gamma_1$ and $\gamma_2$ are material specific.

Prior to quantum transport calculations, these models must be discretized using, e.g., finite differences.
When discretizing these models, particular care needs to be taken to avoid spurious solutions, e.g., in $8\times 8$ $\mathbf{k}\cdot\mathbf{p}$ semiconductor models \cite{foreman1997}.
Additionally, Dirac materials are not amenable to regular discretization due to the occurrence of the fermion doubling problem as explained in Sec.~\ref{sec:massless_dirac}.
As for other models, special care must be given to the treatment of boundaries and interfaces \cite{burt1992} so that they are not error-prone when used for confined structures such as quantum dots \cite{fu1998} or thin films \cite{nechaev2016}.

Self-consistent calculations using $\mathbf{k}\cdot \mathbf{p}$-models have been applied to a broad range of semiconductor devices.
For example, the technique has been used to study devices including quantum dots \cite{gao2014}, silicon nanowire field-effect transistors (FETs) \cite{shin2009, shin2009b}, tunneling FETs in III-V materials \cite{huang2016b}, as well as FinFET transistors \cite{khan2007} and quantum cascade lasers \cite{greck2015}.

For an overview of $\mathbf{k}\cdot\mathbf{p}$-models for semiconductors, we refer to \cite{lewyanvoon2009}, and for reviews on modelling quantum transport in semiconductor devices to \cite{anantram2008, ferry2022}. More generally, details of the self-consistent approach used in nonequilibrium situations (typically for calculating $I-V$ characteristics) can be found in \cite{evers2020,sanvito2011,smidstrup2020,papior2017,garcia-lekue2015}.

\subsection{Applications using atomistic tight-binding models}

We now turn to examples of quantum transport making use of ``more realistic'' modelling, i.e. taking into account the atomistic structure of devices.
To this end, we highlight two examples, namely semi-empirical and first-principles tight-binding models.

\subsubsection{Semi-empirical tight-binding models (SE-TBM)}\label{sec:setbh}

In semi-empirical tight-binding models (SE-TBM), a set of relevant orbitals and couplings is chosen {\it a priori}. The model is then fitted to ab-initio data (e.g., often to reproduce the DFT band structure) and/or to reproduce experimental data. SE-TBM models can have up to hundreds of parameters. The simplest SE-TBM models, with very few parameters, essentially match the model Hamiltonians discussed in the preceding sections.

The main difficulty in preparing SE-TBM for quantum transport calculations is a proper treatment of the boundaries of the device and the interface between different materials. For instance, the description of the interface between graphene and metallic contacts can require as many as a hundred parameters for a faithful description \cite{papaconstantopoulos2003,barraza-lopez2013}. The models are usually fitted with classical least-squares optimization methods, but recently machine-learning approaches have started to be used \cite{hegde2017}.

We cannot do justice to the breadth of SE-TBM literature in this review. They have been applied to a broad range of materials, ranging from, e.g., MoS$_2$ devices \cite{ridolfi2015}, adatoms on graphene \cite{weeks2011}, graphene-hBN devices \cite{marmolejotejada2018}, quantum dots \cite{luisier2014}, 3D topological insulators Bi$_2$Se$_3$, Bi$_2$Te$_3$, Sb$_2$Te$_3$ \cite{liu2010}, and many more. Instead, we will highlight two selected applications, namely semiconductor and carbon-based devices.

SE-TBMs for semiconductors go back to the eighties with minimum next-nearest-neighbor models accounting for the $s$ and $p$ orbitals \cite{vogl1983}. More accurate models also include the $d$ bands \cite{carlo2003,boykin2009}.
Assuming that the hopping matrix elements only depend on the distance between atoms \cite{papaconstantopoulos2003}, these models can be parametrized as a function of atomic distance and then used for an arbitrary atomic configuration. This is very useful for
semiconductor devices where stress can have a significant effect on transport properties \cite{niquet2014,esseni2017}. Another appealing aspect of these models is that they can correct typical deficiencies of DFT approaches, e.g., the inability to correctly predict the semiconductor gap, by fitting the model to experimental data. These models have been applied to quantum transport calculations in, e.g., semiconductor nanowires \cite{luisier2006, luisier2008} and nanowire transistors \cite{luisier2009}.

SE-TBMs are also extremely popular for modeling graphene nanoribbons or carbon nanotube devices. It was recognized early on that the simple nearest-neighbor TBM for a single $p_z$ orbital Eq.~\eqref{eq:grapheneTBM} is often inaccurate: it predicts perfectly flat bands in zigzag nanoribbons as well as a metallic (gapless) band structure for armchair nanoribbons that are $3N+2$ carbon atoms wide. A more accurate model uses
second- and third-nearest-neighbor hoppings \cite{reich2002}. Equivalently, one may use nearest-neighbor hoppings with more bands \cite{cresti2008,boykin2011}. These additional bands are also important when describing the spin-orbit interaction \cite{konschuh2010}. Again, the transport properties across these carbon-based devices are very sensitive to the accuracy of model and in particular the interfaces. For instance, SE-TBM modeling \cite{leonard2006} that did not properly address the interface between carbon nanotubes and the metallic Ohmic contact could not predict the experimental findings quantitatively \cite{franklin2012}. Another example is
the off-current of graphene nanoribbon transistors which varies by orders of magnitude upon
moving from the simple single-orbital $p_z$-SETBM model to a more elaborate
multi-orbital $p/d$-SETBM model \cite{boykin2011}. Similar conclusions were drawn for the validity of simulations of non-local resistances in graphene Hall bars \cite{marmolejotejada2018}.

For a broader overview of the properties of SE-TBM models we refer to the review \cite{papaconstantopoulos2003}, and for reviews on using SE-TBMs for quantum transport to \cite{luisier2014, klimeck2023}.

\subsubsection{First-principles tight-binding models (FP-TBM)}\label{sec:wtbh}

The first-principles tight-binding model (FP-TBM) approach projects a DFT Hamiltonian onto a smaller basis set, typically atomic or pseudo-atomic orbitals.
FP-TBM are designed to describe the physics in a finite energy window centered around the Fermi level.
This distinguishes the FP-TBM from the semi-empirical SE-TBM approach that relies on fitting band structures.
The systematic approach of FP-TBM methods is particularly important when several bands get hybridized at the Fermi level.

Perhaps the most well-known FP-TBM bases are the \emph{maximally localized Wannier functions}, which can be constructed out of Bloch wave functions \cite{marzari1997}.
These orbitals are rather compact, retain the original symmetries of the bands and reproduce well the original DFT results.
A popular approach uses a plane wave code such as VASP \cite{kresse1999} followed by a projection on Wannier orbitals using, e.g., the code Wannier90 \cite{mostofi2008}.

An illustrative example is the case of graphene.
Using a minimal basis of a single $p_z$-like orbital \cite{fang2016} per site,
Jung and MacDonald \cite{jung2013} find that including hoppings up to $15^{th}$ neighbor is required to accurately reproduce the $\pi$-bands (predicted by DFT) around the Fermi level across the entire Brillouin zone. This can be contrasted with the single-hopping model
Eq.~\eqref{eq:grapheneTBM} employed in the preceding section.
Examples of quantum transport calculations performed with Wannier functions include
the calculation of the conductance of disordered nanowires \cite{calzolari2004,shelley2011} and of single molecule junctions \cite{thygesen2005,strange2008}.

For a comprehensive review of maximally localized Wannier functions, we refer to \cite{marzari2012}, and for an overview of various applications including transport and existing software to \cite{marrazzo2024}.

There are several other FP-TBM methods beyond Wannier functions, and a comprehensive overview is beyond the scope of this review. Examples include linear-muffin-tin-orbitals (LMTO) and generalizations \cite{andersen2000}. Other FP-TBM include the quasiminimal basis orbitals \cite{wang2008} or predetermined basis of fixed localized wave functions \cite{agapito2016,agapito2016b}. All these techniques require particular care in the treatment of boundaries or interfaces if the model is to be transferred to other geometries or structures.

\subsection{DFT models}

First-principles density functional theory (DFT) can provide predictive calculations in the absence of any prior experimental data.
The combined approach is often referred to as DFT + NEGF, being typically based on the non-equilibrium Green's function technique reviewed in Sec.~\ref{sec:phys-obs-negf} as in Refs.~\cite{brandbyge2002, xue2002, palacios2002}.
For reviews and pedagogical materials with more detail about DFT + NEGF, we refer the interested reader to the articles that accompany corresponding software packages, a list of which can be found towards the end of Sec.~\ref{sec:short-hist-comp-quant}.

The accuracy of DFT+NEGF calculations depends on many details, such as the right choice of basis sets as discussed, e.g., in \cite{lejaeghere2016, driscoll2010, herrmann2010}.
These topics are beyond the scope of this review and we refer the reader to specialized literature.
Below, we highlight two particularly prominent use cases.

\subsubsection{Non-linear conductance in short junctions}
\label{sec:SCQE_2}

We start with a use case where density functional theory (DFT) is used to compute the non-linear
current-voltage characteristics of nanodevices.
As highlighted in Sec.~\ref{sec:SCQE_1}, it is crucial to treat the electron-electron interaction at least at the mean-field level to calculate non-linear corrections to the conductance.
DFT + NEGF calculations do this by design.

Early calculations and later benchmarks focused on short wires made of a few atoms, e.g., C, Al or Au \cite{lang1995,brandbyge2002,palacios2002,rocha2006}. The calculations soon extended to molecular nanoelectronics \cite{diventra2001,ke2004,faleev2005,varga2007,finch2009,saha2010,sergueev2010}. Despite the sensitivity of the conductance to the nature of the molecule-electrode contact, it was found in \cite{diventra2000} that the qualitative shape of the experimental $I-V$ characteristics could be recovered.
Another topic that attracted a strong interest is carbon nanotubes \cite{taylor2001,cook2011,cook2012} as well as graphene-based devices with either a transistor-like geometry \cite{ozaki2010,areshkin2010,papior2016} or multi-terminal crossbar geometry
 \cite{botello-Mendez2011,habib2012,saha2013}. Possible applications of these devices include DNA sensing using nanopores in graphene \cite{saha2012}. Recent works have extended the studies to other 2D materials such as MoS$_2$ \cite{garcia-lekue2015}.

For a comprehensive review of quantum transport in molecular junctions and the limitations of DFT methods, we refer to, e.g., \cite{sanvito2011, evers2020}.

\subsubsection{Applications to spintronics}
\label{sec:spintronics_realistic}

A very successful example of the DFT + NEGF approach is the theoretical prediction \cite{butler2001,mathon2001} of the existence of a large tunneling magnetoresistance of a Fe/MgO/Fe magnetic tunnel junction. There, the coherent spin-polarized tunneling and the symmetry of electron wave functions play an essential role that can only be captured by quantum transport calculations.
This prediction has motivated large experimental efforts \cite{yuasa2004} to fabricate such devices. These magnetic tunnel junctions now play a central role in both basic \cite{wang2011} and applied \cite{locatelli2014} spintronics research.

\begin{figure}
  \centering
  \includegraphics[width=0.55\linewidth]{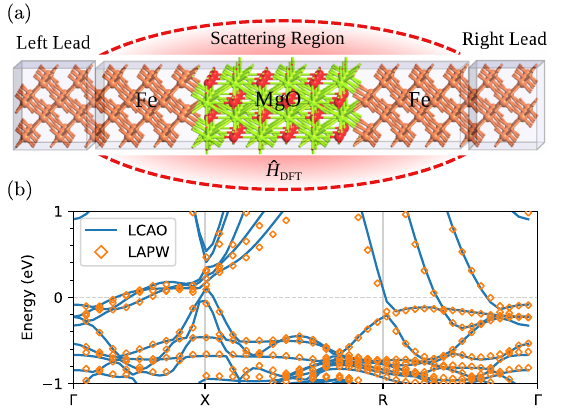}
  \caption{(a) Schematic view of a magnetic tunnel junction where 6 monolayers of MgO(001) insulating barrier are sandwiched between two semi-infinite Fe(001) leads. The junction is infinite in the transverse direction, so that the depicted supercell is periodically repeated in the direction transverse to current flow from the left to the right Fe lead. (b) Electronic band structure of a periodic   Fe/MgO/Fe/MgO\dots{}Fe/MgO lattice obtained by DFT calculations using LCAO basis (solid line) versus those using full potential
    linearized augmented plane-wave basis (circles). Panel (b) has been replotted from numerical coordinates extracted from Fig. 3(b) of \cite{waldron2007}.}
  \label{fig:fig1_bkn}
\end{figure}

Figure~\ref{fig:fig1_bkn}(a) shows a schematic of a Fe-MgO-Fe junction simulated in \cite{waldron2007}. The system is assumed to be translationally invariant along the direction perpendicular to the junction (no disorder) so that the transmission is calculated as a function of the transverse momentum, which must be integrated over at the end of the calculation.
The Hamiltonian of the scattering region was written in a linear combination of atomic orbitals (LCAO). An accurate calculation requires as many as 10 000 orbitals for this calculation while the supercell only contains a total of 24 atoms (Fe Mg and O). To control this accuracy, one
compares the LCAO calculation with a more accurate one done in a plane wave basis, as illustrated in Fig.~\ref{fig:fig1_bkn}(b). In addition to checking the convergence of the calculation with the number of LCAO orbitals, one must also obtain convergence of the momentum integration which requires $\sim 10^4$ momentum points at zero bias (only $\sim 10^2$ for the band structure, up to $\sim 10^5$ at finite bias voltage, up to $\sim 10^7$ for spin-torque \cite{wang2008b}).

Later works tried to optimize the magnetoresistance of the tunnel junction. It has been predicted, for instance, that a few layers of graphene with Ni electrodes act as a perfect spin filter \cite{karpan2007} with regimes showing a negative differential resistance \cite{saha2012b} at finite bias.
Another line of work used DFT+NEGF to address metallic magnetic systems \cite{schep1995,stiles2002}. In metals some effort has been devoted to
calculate effective material parameters such as Gilbert damping, spin-flip diffusion length,
spin Hall angle...that can later be incorporated into diffusion-like equations to address the macroscopic scale \cite{starikov2010,wang2016,gupta2020}. An important evolution of the field has been the discovery of the phenomenon of spin-transfer torque in non-collinear systems. Spin torque has enabled fully electric control of a magnetic configuration \cite{ralph2008}. DFT studies of spin transfer torque include \cite{stiles2002,haney2007,wang2008b,ellis2017}.
Recent studies also capture a related effect, the spin-orbit torque \cite{nikolic2018,belashchenko2019}.

For a more in-depth review on spintronics devices including MgO tunnel junctions and spin torques we refer to reviews such as Ref.~\cite{hirohata2020}.

\section{Final remarks}%

\subsection{Simulation workflow}
\label{sec:workflow}

In practice, numerical simulations of quantum transport devices are often carried out by approaching the complete problem step by step.
One starts by studying the bulk Hamiltonian(s) for the material(s) under consideration.
Written in $k$-space, these Hamiltonians are readily diagonalized and their spectra provide useful information such as the positions of gaps or Dirac points.

In a second step, one constructs electrodes of finite width and analyzes them in isolation.
The associated $k$-space Hamiltonians provide information such as the number of propagating modes in each electrode.
Packages such as KwantSpectrum \cite{kloss2021} provide tools to facilitate this analysis.

It is only in a third step that one proceeds to treat scattering problems that consist of a scattering region with attached electrodes.
It is often useful to start out with subparts of the final device.
For instance, in an electronic interferometer that contains several quantum point contacts,
one could study the transport properties of a system that contains a single quantum point contact in order to calibrate the corresponding model.
Another example is the interface between two materials that could be studied separately before being integrated into a larger device.
When the final device is very large, it can actually be computationally advantageous to stop at this level and concatenate the scattering matrices of the different pieces numerically (following, e.g., \cite{rotter2000}).
In most cases, however, one proceeds, at last, to simulate the entire device directly.

While in former times research groups relied on in-house codes,
nowadays in most cases one will want to take advantage of already existing solutions, at least as a starting point or foundation.
We refer to the list of codes at the end of Sec.~\ref{sec:short-hist-comp-quant}
as possible entrance points, either for standalone quantum transport calculations (model systems) or for a combined DFT/self-consistent quantum electrostatic + quantum transport (realistic materials and devices).

\subsection{Outlook}
\label{sec:outlook}

Numerical calculations of quantum transport have become a ubiquitous tool in many fields, from electrical engineering to materials science and quantum nanoelectronics.
In this review, we focused on providing a comprehensive presentation of the two equivalent formalisms used in this field,
the scattering matrix approach and the Green's function approach,
with an emphasis on explicitly providing the links between the two.
We also provided an in-depth review of the different algorithms that have been developed to perform actual calculations.

Many topics fall beyond the scope of this review.
One is the simulation of time-dependent quantum transport \cite{stefanucci2004, krueckl2012, kloss2021},
see \cite{gaury2014} for a review,
and its special case of finite-frequency transport or Floquet scattering \cite{moskalets2012}.
Another major aspect that we did not touch is the effect of electron-electron interactions
or the interaction of electrons with other degrees of freedom such as phonons or magnons. It is indeed a very frustrating fact that even the simplest and most common of these effects, for instance Coulomb blockade, cannot be easily incorporated into the type of approach presented in this review.
The effects of decoherence are also beyond the scope of this review.

In terms of applications, the range is so vast that we could only point to a few examples.
Our choice of topics was necessarily subjective and biased by our own interests.

With regard to future work, much remains to be done besides addressing the above points. One important aspect is improving the accuracy of predictions compared to experiments \cite{chatzikyriakou2022}.
This requires better models, efficient tools to address electrostatics, solvers capable of handling large systems (in particular in three dimensions), and close collaboration with experimental teams with the goal of producing systematic data sets dedicated to the calibration of models.

\section*{Acknowledgements}
Many colleagues have given us important help in putting this review together. We are particularly grateful to Piet Brouwer, Rodolfo Jalabert, Colin Lambert, Caio Lewenkopf, Horacio Pastawski, Klaus Richter, Stefano Sanvito, Henning Schomerus, and Dietmar Weinmann.

\paragraph{Author contributions}
AA, MW, and XW developed the formalism.
AA and MW developed the stable algorithm for solving the scattering problem.
MI described the bound state problem.
TOR and DV implemented and described symmetry analysis of the modes.
BKN wrote the section on applications to realistic systems.
AA, MW, CG, and XW wrote the rest of the manuscript.
XW organized the writing.
All authors reviewed the text.
The authors are listed in decreasing order of contributions.

\paragraph{Funding information}
D.V. acknowledges support from the Dutch national science organization (NWO) under VIDI grant 680-47-53, and the Deutsche Forschungsgemeinschaft (DFG, German Research Foundation) under Germany’s Excellence Strategy through the W\"{u}rzburg--Dresden Cluster of Excellence on Complexity and Topology in Quantum Matter (ct.qmat, EXC 2147, project IDs 390858490 and 392019). X.W. acknowledges funding from the French ANR TKONDO and the PEPR EQUBITFLY

\appendix
\numberwithin{equation}{section}

\section{Derivation of Feynman--Hellmann equation for the velocity}
\label{sec:deriv-feynm-hellm}
The Feynman--Hellmann equation relates the velocity of a mode to the average of the velocity operator. Considering normalized modes,
\begin{equation}
\phi^\dagger \phi = 1,
\end{equation}
we have
\begin{equation}
\frac{\partial\phi^\dagger}{\partial k} \phi + \phi^\dagger \frac{\partial\phi}{\partial k} = 0.
\end{equation}
From $E(k)=\phi^\dagger H(k) \phi$, we get
\begin{eqnarray}
 v &\equiv&  \frac{\partial E(k)}{\partial k} \nonumber \\
&=&  \frac{\partial\phi^\dagger}{\partial k} H(k) \phi + \phi^\dagger H(k)\frac{\partial\phi}{\partial k} +
\phi^\dagger \frac{\partial H(k)}{\partial k}  \phi \nonumber \\
&=& E(k) \left[\frac{\partial\phi^\dagger}{\partial k}\phi + \phi^\dagger\frac{\partial\phi}{\partial k}\right]
+ \phi^\dagger \frac{\partial H(k)}{\partial k}  \phi \nonumber \\
&=& \phi^\dagger \frac{\partial H(k)}{\partial k}  \phi \nonumber \\
&=& i\phi^\dagger \left[ \lambda V^\dagger - \lambda^{-1}V \right]\phi,
\end{eqnarray}
which is the Feynman--Hellmann equation. Note that inside the matrices $\Phi$,
the modes $\phi$ are renormalized to carry a unit velocity $v = \pm 1$.

\section{Properties of the lead modes.}

\label{app:hermit}

In this appendix, we establish a few technical results on the lead propagating and evanescent modes that are needed in various proofs of the main text.
We begin by multiplying Eq.~\eqref{eq:lead_def} by $\Phi^{\dagger}$:
\begin{equation}
\label{eq:AAA}
\Phi^{\dagger}V \Phi + \Phi^{\dagger}(H - E) \Phi \Lambda +\Phi^{\dagger} V^{\dagger} \Phi (\Lambda)^2 = 0.
\end{equation}
The complex conjugate of the above equation reads
\begin{equation}
\label{eq:BBB}
(\Lambda^*)^2 \Phi^{\dagger}V \Phi +\Lambda^* \Phi^{\dagger}(H - E) \Phi  +\Phi^{\dagger} V^{\dagger} \Phi  = 0.
\end{equation}
Now, multiplying Eq.~\eqref{eq:AAA} by $\Lambda^*$ on the left and Eq.~\eqref{eq:BBB} by $\Lambda$ on the right, we arrive after subtracting one equation from the other at
\begin{equation}
\label{eq:CCC}
[\lambda_\alpha^* - (\lambda_\alpha^{*})^2 \lambda_\beta ] \left. \Phi^{\dagger}V \Phi\right|_{\alpha\beta} =
[ -\lambda_\alpha^{*} (\lambda_\beta)^2 +\lambda_\beta]   \left. \Phi^{\dagger} V^{\dagger} \Phi\right|_{\alpha\beta}.
\end{equation}
which is valid for all types of modes, evanescent and/or propagating.

\subsection{Application to evanescent modes}
If one of the two modes $\alpha$ or $\beta$ is evanescent, we can simplify Eq.~\eqref{eq:CCC}
 by $1 - \lambda_\alpha^{*} \lambda_\beta \ne 0$ and arrive at
\begin{equation}
\label{eq:DDD}
\lambda_\alpha^*    \left. \Phi^{\dagger}V \Phi\right|_{\alpha\beta} =
\lambda_\beta   \left. \Phi^{\dagger} V^{\dagger} \Phi\right|_{\alpha\beta},
\end{equation}
which can be written in matrix form as,
\begin{equation}
\label{eq:veloc_ee}
\Lambda_{\mathrm{e}+}^* \Phi_{\mathrm{e}+}^\dagger V \Phi_{\mathrm{e}+} =
\Phi_{\mathrm{e}+}^\dagger V^\dagger \Phi_{\mathrm{e}+} \Lambda_{\mathrm{e}+},
\end{equation}
Similarly,
\begin{equation}
\label{eq:veloc_pe}
\Lambda_{\mathrm{p}+}^* \Phi_{\mathrm{p}+}^\dagger V \Phi_{\mathrm{e}+} =
\Phi_{\mathrm{p}+}^\dagger V^\dagger \Phi_{\mathrm{e}+} \Lambda_{\mathrm{e}+},
\end{equation}
and
\begin{equation}
\label{eq:veloc_ep}
\Lambda_{\mathrm{e}+}^* \Phi_{\mathrm{e}+}^\dagger V \Phi_{\mathrm{p}+} =
\Phi_{\mathrm{e}+}^\dagger V^\dagger \Phi_{\mathrm{p}+} \Lambda_{\mathrm{p}+}.
\end{equation}
\subsection{Application to propagating modes}
For propagating modes $1 - \lambda_\alpha^{*} \lambda_\beta$ vanishes whenever $\lambda_\alpha = \lambda_\beta= \lambda$,
i.e., the two modes $\alpha$ and $\beta$ have the same momentum. In that situation, we can form the velocity matrix
\begin{equation}
i \Phi_{\mathrm{p}+}^\dagger \left[ \lambda V^\dagger -\lambda^* V \right]\Phi_{\mathrm{p}+} |_{\alpha\beta},
\end{equation}
where $\alpha$ and $\beta$ lie inside the degenerate space. This matrix is Hermitian, hence can be diagonalized.
The resulting linear combination of propagating modes (for which we abusively use the same notation), normalized to carry unit
velocities, remains a solution of the mode equation and further satisfies,
\begin{equation}
i \Phi_{\mathrm{p}+}^\dagger \left[ \lambda V^\dagger -\lambda^* V\right]\Phi_{\mathrm{p}+} |_{\alpha\beta} =\delta_{\alpha\beta},
\end{equation}
from which we finally get,
\begin{equation}
\label{eq:veloc_pp}
\Lambda_{\mathrm{p}+}^* \Phi_{\mathrm{p}+}^\dagger V \Phi_{\mathrm{p}+} =
\Phi_{\mathrm{p}+}^\dagger V^\dagger \Phi_{\mathrm{p}+} \Lambda_{\mathrm{p}+} + i.
\end{equation}
A corresponding expression can be derived for incoming modes. In that case, the sign of the current is opposite and one arrives at,
\begin{equation}
\label{eq:veloc_pp_incoming}
\Lambda_{\mathrm{p}-}^* \Phi_{\mathrm{p}-}^\dagger V \Phi_{\mathrm{p}-} =
\Phi_{\mathrm{p}-}^\dagger V^\dagger \Phi_{\mathrm{p}-} \Lambda_{\mathrm{p}-} - i.
\end{equation}

\section{Proof of invertibility of mode matrices}
\label{sec:inverse_exist}
In this section, we discuss the relation between the rank of $\Phi_{\bar t+}$ and the absence of bound states in the corresponding semi-infinite lead.

Let us suppose that $\Phi_{\bar t+}$ is not full rank. It follows that $V\Phi_{\bar t+}$ is not full rank neither, hence that there is a vector $c\neq 0$ such that $V\Phi_{\bar t+} c =0$. On the other hand, we know that the wave function
\begin{equation}
\psi(j) = \Phi_{\bar t+}\Lambda_{\bar t+}^j c,
\end{equation}
is an eigenstate of the \emph{infinite} lead,
i.e., it satisfies the infinite lead equation
\begin{equation}
V\psi(j-1) +(H-E) \psi(j) +V^\dagger \psi(j+1) =0.
\end{equation}
In addition, we now have $V \phi(0) = 0$, which means that $\psi(j)$  also fulfills the boundary condition,
\begin{equation}
(H-E) \psi(1) + V^\dagger \psi(2) = 0.
\end{equation}
In other words, $\psi(j)$ is an eigenstate of the \emph{semi-infinite} lead.

Let us now decompose the vector $c$ into its propagating and evanescent sector
$c = (c_\mathrm{p+}, c_\mathrm{\bar e+})$. Using Eq.~\eqref{eq:veloc_ee} and Eq.~\eqref{eq:veloc_pp}, we get,
\begin{equation}
c^\dagger \Lambda_{\bar t+}^* \Phi_{\bar t+}^\dagger V \Phi_{\bar t+}c
= c^\dagger \Phi_{\bar t+}^\dagger V^\dagger \Phi_{\bar t+}\Lambda_{\bar t+}c
+i c_\mathrm{p+}^\dagger c_\mathrm{p+}
\end{equation}
and since $V \Phi_{\bar t+}c = c^\dagger \Phi_{\bar t+}^\dagger V^\dagger = 0$,
it follows that $c_\mathrm{p+}^\dagger c_\mathrm{p+} =0$ hence,
\begin{equation}
c_\mathrm{p+} = 0.
\end{equation}
Hence, $V \Phi_{\mathrm{t}+}c=0$ reduces to $V \Phi_{\mathrm{e}+}c_{\bar e+} = 0$,
which in turn means that $\psi(j)$ is a \emph{bound state} solution of the semi-infinite lead. This last paragraph is a technical way to reflect the fact that
the probability current of an eigenstate of the semi-infinite lead must be zero everywhere---this is current conservation---hence the propagating contribution must vanish since we include only modes that propagate along one direction.

Conversely, if there are no bound states at energy $E$, then
$V\Phi_{\bar t+}$ is full rank. Since $V=AB^\dagger$, $V\Phi_{\bar t+}$ being full rank implies that $B^\dagger\Phi_{\bar t+}$ is full rank and since it is a square matrix, it must therefore be invertible.

\section{Self-energy from the stable solution of the lead eigenproblem}
\label{sec:stable_selfenergy}

We can rewrite Eq.~\eqref{eq:self_energy_out_modes} using the definition \eqref{eq:def_phi_ab} as
\begin{equation}
\Sigma = B \Phi_{A\mathrm{\bar t}+}\frac{1}{\Phi_{B\mathrm{\bar t}+}} B^\dagger\,.
\end{equation}
Furthermore, making use of the orthogonal basis computed from the QZ decomposition, Eq.~\eqref{eq:scatt_form_stable3}, we find
\begin{equation}
\begin{split}
\Sigma & =B\,(\Phi_{A\mathrm{p}+}|Z_{A\mathrm{\bar e}+}R)\frac{1}{(\Phi_{B\mathrm{p}+}|Z_{B\mathrm{\bar e}+}R)} \,B^\dagger \\
& = B\, (\Phi_{A\mathrm{p}+}|Z_{A\mathrm{\bar e}+})\frac{1}{(\Phi_{B\mathrm{p}+}|Z_{B\mathrm{\bar e}+})} \,B^\dagger\,.
\end{split}
\end{equation}
The last equation holds as $R$ is invertible as long as $\Phi_{B\mathrm{\bar t}+} = B^\dagger \Phi_{\mathrm{\bar t}+}$ is invertible.
As shown in Appendix~\ref{sec:inverse_exist}, this is the case unless the semi-infinite lead has a bound state at energy $E$.

\section{Derivation of symmetry relations of the scattering matrix}
\label{sec:deriv-symmetry}

\subsection{Time Reversal Symmetry}

In a time-reversal symmetric system, time reversal symmetry $\mathcal{T}$ relates incoming to outgoing modes. We can choose a basis of modes where
\begin{equation}
\Phi_{\mathrm{p}+} = \mathcal{T} \Phi_{\mathrm{p}-}, \quad \Lambda_{\mathrm{p}+} = \Lambda_{\mathrm{p}-}^*\,.
\end{equation}

We can thus write the scattering wave function in the lead as
\begin{equation}
\begin{split}
\Psi(j) & = \Phi_{\mathrm{p}-} \Lambda^j_{\mathrm{p}-} q_- + (\mathcal{T} \Phi_{\mathrm{p}-})(\Lambda^*_{\mathrm{p}-})^j q_+ + \Phi_{\mathrm{e}+} \Lambda_{\mathrm{e}+}^j q_\mathrm{e},
\end{split}
\end{equation}
with $q_+ = S_\mathrm{pp} q_-$.
The notation $(\mathcal{T} \dots)$ indicates that the anti-unitary symmetry operator $\mathcal{T}$ only acts inside the brackets, and not on the remaining parts of the expression.

When $\Psi$ is a solution of the Schrödinger equation, so is $\mathcal{T}\Psi$ and
\begin{equation}
\begin{split}
\mathcal{T}\Psi(j) & = (\mathcal{T}\Phi_{\mathrm{p}-}) (\Lambda^*_{\mathrm{p}-})^j q_-^* + \mathcal{T}^2 \Phi_{\mathrm{p}-} \Lambda^j_{\mathrm{p}-} q^*_+ \\
& \quad + \mathcal{T}( \Phi_{\mathrm{e}+} \Lambda_{\mathrm{e}+}^j q_\mathrm{e}).
\end{split}
\end{equation}
The action of $\mathcal{T}$ keeps the evanescent modes. Hence, $\mathcal{T}\Psi$ is also a valid scattering wave function, but now with incoming and outgoing modes interchanged such that $q^*_- = \mathcal{T}^2 S_\mathrm{pp} q_+^*$. Hence we find
\begin{equation}
q_- = \mathcal{T}^2 S^*_\mathrm{pp} S_\mathrm{pp} q_-,
\end{equation}
and since this equation holds for all possible vectors $q_-$, we find
\begin{equation}
S_\mathrm{pp} = \mathcal{T}^2 S_\mathrm{pp}^T.
\end{equation}

\subsection{Chiral Symmetry}

Chiral symmetry relates incoming modes to outgoing modes at $E=0$. We thus choose a basis of modes such that
\begin{equation}
\Phi_{\mathrm{p}+} = \mathcal{C} \Phi_{\mathrm{p}-}, \quad \Lambda_{\mathrm{p}+} = \Lambda_{\mathrm{p}-}.
\end{equation}

We can thus write the scattering wave function in the lead as
\begin{equation}
\Psi(j) = \Phi_{\mathrm{p}-} \Lambda^j_{\mathrm{p}-} q_- + (\mathcal{C} \Phi_{\mathrm{p}-}) \Lambda_{\mathrm{p}-}^j q_+ + \Phi_{\mathrm{e}+} \Lambda_{\mathrm{e}+}^j q_\mathrm{e},
\end{equation}
with $q_+ = S_\mathrm{pp} q_-$. Then $\mathcal{C}\Psi$ is also a valid scattering wave function with
\begin{equation}
\mathcal{C}\Psi(j) = (\mathcal{C}\Phi_{\mathrm{p}-}) \Lambda^j_{\mathrm{p}-} q_- + \Phi_{\mathrm{p}-} \Lambda_{\mathrm{p}-}^j q_+ + \mathcal{C}(\Phi_{\mathrm{e}+} \Lambda_{\mathrm{e}+}^j q_\mathrm{e}),
\end{equation}
since $\mathcal{C}^2=1$, and hence also $q_- = S_\mathrm{pp}q_+ = S_\mathrm{pp} S_\mathrm{pp} q_-$, and thus
\begin{equation}
S_\mathrm{pp}(E=0) = S_\mathrm{pp}^\dagger.
\end{equation}

\subsection{Particle-Hole Symmetry}

When there is a particle-hole symmetry $\mathcal{P}$ together with a conservation law $\tau_z$ in particle-hole space, we can classify modes as ``electron'' ($e$) and ``hole'' ($h$) and at $E=0$ use a basis of modes
\begin{equation}
\begin{split}
\Phi_{\mathrm{p}\pm} \Lambda_{\mathrm{p}\pm} & =
\left(\Phi_{\mathrm{p}\pm}^{(e)} \Lambda_{\mathrm{p}\pm}^{(e)} \,\middle|\, \Phi_{\mathrm{p}\pm}^{(h)} \Lambda_{\mathrm{p}\pm}^{(h)}\right)\\
& =
\left(\Phi_{\mathrm{p}\pm}^{(e)} \Lambda_{\mathrm{p}\pm}^{(e)} \,\middle|\, \left(\mathcal{P}\Phi_{\mathrm{p}\pm}^{(e)}\right)\Lambda_{\mathrm{p}\pm}^{(e)}{}^*\right).
\end{split}
\end{equation}

In this case, the particle-hole symmetry relates electron modes to hole modes, but does not switch between incoming and outgoing modes.

The scattering wave function then takes the form
\begin{equation}
\begin{split}
\Psi(j) = & \sum_{\sigma=\pm}\left(\Phi_{\mathrm{p}\sigma}^e \left(\Lambda_{\mathrm{p}\sigma}^{(e)}\right)^j \,\middle|\, \left(\mathcal{P}\Phi_{\mathrm{p}\sigma}^{(e)}\right) \left(\Lambda_{\mathrm{p}\sigma}^{(e)}{}^*\right)^j\right) \begin{pmatrix}q^{(e)}_\sigma\\q_\sigma^{(h)} \end{pmatrix}\\
& +
\Phi_{\mathrm{e}+}\Lambda_{\mathrm{e}+}^j q_\mathrm{e}\,.
\end{split}
\end{equation}

The terms corresponding to the incoming and outgoing modes have the same structure, so we write them for compactness as a sum over $\sigma=\pm$ (not to be confused with a sum over spins!). We then have
\begin{equation}
\begin{pmatrix}q^{(e)}_+\\q_+^{(h)}\end{pmatrix}=S_\mathrm{pp}\begin{pmatrix}q^{(e)}_-\\q_-^{(h)}\end{pmatrix},\quad \text{with} \quad S_\mathrm{pp}=\begin{pmatrix}
S_\mathrm{pp}^{(ee)} & S_\mathrm{pp}^{(eh)}\\
S_\mathrm{pp}^{(he)} & S_\mathrm{pp}^{(hh)}
\end{pmatrix},
\end{equation}
having a corresponding block structure.

$\mathcal{P}\Psi$ is then also a solution and
\begin{equation}
\begin{split}
&\mathcal{P}\Psi(j) = \\
& = \sum_{\sigma=\pm}\left(\left(\mathcal{P}\Phi_{\mathrm{p}\sigma}^{(e)}\right) \left(\Lambda_{\mathrm{p}\sigma}^{(e)*}\right)^j \,\middle|\, \mathcal{P}^2\Phi_{\mathrm{p}\sigma}^{(e)} \left(\Lambda_{\mathrm{p}\sigma}^{(e)}\right)^j\right) \begin{pmatrix}q^{(e)}_\sigma{}^*\\q_\sigma^{(h)}{}^* \end{pmatrix} \\
& \quad\quad + \mathcal{P}(\Phi_{\mathrm{e}+}\Lambda_{\mathrm{e}+}^j q_\mathrm{e}) \\
&  = \sum_{\sigma=\pm}\left(\Phi_{\mathrm{p}\sigma}^e \left(\Lambda_{\mathrm{p}\sigma}^{(e)}\right)^j \,\middle|\, \left(\mathcal{P}\Phi_{\mathrm{p}\sigma}^{(e)}\right)\left(\Lambda_{\mathrm{p}\sigma}^{(e)}{}^*\right)^j\right) \begin{pmatrix}\mathcal{P}^2 q^{(h)}_\sigma{}^*\\q_\sigma^{(e)}{}^* \end{pmatrix} \\
& \quad\quad + \mathcal{P}(\Phi_{\mathrm{e}+}\Lambda_{\mathrm{e}+}^j q_\mathrm{e})\,.
\end{split}
\end{equation}

Hence we find
\begin{equation}
\begin{pmatrix}\mathcal{P}^2 q^{(h)}_+{}^*\\q_+^{(e)}{}^* \end{pmatrix}
= S_\mathrm{pp}
\begin{pmatrix}\mathcal{P}^2 q^{(h)}_-{}^*\\q_-^{(e)}{}^* \end{pmatrix},
\end{equation}
which can be recast as
\begin{equation}
\begin{pmatrix}q^{(e)}_+\\q_+^{(h)}\end{pmatrix} = \begin{pmatrix} 0 & 1 \\ \mathcal{P}^2 & 0\end{pmatrix} S_\mathrm{pp}^* \begin{pmatrix} 0 & \mathcal{P}^2 \\ 1 & 0\end{pmatrix} \begin{pmatrix}q^{(e)}_-\\q_-^{(h)}\end{pmatrix},
\end{equation}
and thus
\begin{equation}
S_\mathrm{pp}(E=0) = \begin{pmatrix} 0 & 1 \\ \mathcal{P}^2 & 0\end{pmatrix} S_\mathrm{pp}^* \begin{pmatrix} 0 & \mathcal{P}^2 \\ 1 & 0\end{pmatrix}.
\end{equation}

Note that different choices for the mode basis will lead to different symmetry relations for the scattering matrix.
For example, for $\mathcal{P}^2=1$, it is also possible to define a ``Majorana basis'' such that $\mathcal{P}\Phi_{\mathrm{p}\pm} = \Phi_{\mathrm{p}\pm}$.
Then
\begin{equation}
S_\mathrm{pp}(E=0) = S_\mathrm{pp}^*.
\end{equation}
That basis would be a natural basis when one has, e.g., propagating Majorana edge modes.

\section{Scattering problem with non-orthogonal basis set}

This review has focused on tight-binding models with an orthogonal basis set. However, non-orthogonal basis sets appear naturally for basis sets used in Slater--Koster tight-binding methods or density functional theory, as, e.g., in \cite{junqera2001,ozaki2003,ozaki2004,papaconstantopoulos2003}. In this case, different basis states $\ket{n_i}$ have finite overlap, $\braket{n_i|n_j}=W_{ij}$. Note that the overlap matrix is usually denoted by the letter $S$, which is already taken by a central object of this review, the scattering matrix. Hence, we resort to $W$ to denote the overlap matrix.

In this appendix, we argue how the main formulas presented in this review can be generalized to tight-binding models with a non-orthogonal basis set with a simple replacement rule. More explicit and extensive derivations can be found in the literature, such as for the Green's function formalism in \cite{lohez1983} and its non-equilibrium version in \cite{lake2006}.

The Schrödinger equation for the scattering system with a non-orthogonal basis reads
\begin{equation}\label{eq:schroedinger_nonortho}
  \hat{H}_\text{sys} \hat{\psi} = E \hat{W}_\text{sys} \hat{\psi},
\end{equation}
with the overlap matrix
\begin{equation}
  W_\text{sys} = \begin{pmatrix}
    W_\text{sr}&P_\text{sr}^TW_V^\dagger&\\
    W_VP_\text{sr}& W_H & W_V^\dagger\\
    & W_V&W_H&W_V^\dagger\\
    & & W_V & W_H & W_V^\dagger\\
    & & & \ddots &\ddots &\ddots
  \end{pmatrix},
\end{equation}
where $W_\text{sr}^\dagger = W_\text{sr}$ and $W_H^\dagger=W_H$ by definition. Here we have also assumed that the overlap matrix $W_\text{sys}$ has the same sparsity structure as the Hamiltonian $H_\text{sys}$, which can always be achieved by choosing the unit cell of the leads appropriately.

Most of the results in this review are obtained by direct manipulation of the Schrödinger equation $\hat{H}_\text{sys} \hat{\psi} = E \hat{\psi}$. To generalize to the case of a non-orthogonal basis, i.e., \eqref{eq:schroedinger_nonortho}, it is thus sufficient to replace
\begin{equation}\label{eq:nonortho_rules}
  \begin{split}
    H_\text{sr}-E &\rightarrow H_\text{sr} - E\, W_\text{sr},\\
    H-E &\rightarrow H - E\, W_H\text{, and}\\
    V &\rightarrow V - E\, W_V,
  \end{split}
\end{equation}
in our main results such as Eq.~\eqref{eq:scatt_form}.

The Bloch equation for the lead modes now reads
\begin{equation}
H(k) \phi = E(k) W(k) \phi,
\end{equation}
with $W(k) = W_V e^{-ik} + W_H + W_V^\dagger e^{ik}$. In analogy to Appendix~\ref{sec:deriv-feynm-hellm}, the expression for the velocity of a lead mode is obtained by projecting the equation onto $\phi^\dagger$ and taking a derivative with respect to $k$. This yields
\begin{equation}
v = \frac{1}{\hbar} \frac{d E}{d k} = \frac{i}{w}
\phi^\dagger\left[ \lambda (V^\dagger - E W_V^\dagger) - \lambda^{-1} (V - EW_V) \right] \phi,
\end{equation}
where $w=\phi^\dagger(\lambda W_V^\dagger + W_H + \lambda^{-1} W_V)\phi$, and again $\lambda = e^{ik}$. In our derivations, we used the convention that propagating modes are normalized such that $v=1$. Hence, for the case of a non-orthogonal basis set, we need to absorb both $v$ and $w$ into the normalization of $\phi$. With this convention, the properties of the lead modes derived in Appendix~\ref{app:hermit} hold with the replacement rules \eqref{eq:nonortho_rules}, as they are again obtained by direct manipulation of the Bloch equation in the leads.

Hence, with the replacement rules \eqref{eq:nonortho_rules}, the results presented in this review can immediately be applied to systems with a non-orthogonal basis set.

\section{Stabilized algorithm for the Schur complement}
\label{sec:stab-schur}
The approach of stabilizing some part of a Hermitian linear system by considering linear combinations of equations applies to a broad class of problems.
We consider a linear system
\begin{equation}
\begin{pmatrix}
  A & B \\ C & D
\end{pmatrix}
\begin{pmatrix}
  x_1 \\ x_2
\end{pmatrix}=
\begin{pmatrix}
  b_1 \\ b_2
\end{pmatrix},
\end{equation}
where $A=A^\dagger$ is Hermitian and potentially not invertible.
This structure is common both to the scattering problem, Eq.~\eqref{eq:scatt_form_general}, and the computation of the Green's function at the lead interface, Eq.~\eqref{eq:dyson-lead-sr}.
In both cases, $A = H_{sr}$ is large and sparse, so that our goal is to solve for $x_2$ by eliminating the first block $x_1$.
Direct elimination of the first block requires the inversion of $A$, which may be ill-defined or numerically unstable.
To avoid this, we add the second equation multiplied by $iC^\dagger$ to the first equation, yielding
\begin{equation}
\begin{pmatrix}
  A + i C^\dagger C & B + i C^\dagger D \\ C & D
\end{pmatrix}
\begin{pmatrix}
  x'_1 \\ x_2
\end{pmatrix}=
\begin{pmatrix}
  b_1 + i C^\dagger b_2 \\ b_2
\end{pmatrix}.
\end{equation}
The new top-left block $A + i C^\dagger C$ is invertible unless $A$ and $C$ have a common right nullspace, or in other words, unless the entire system is not invertible.
This allows us to eliminate $x'_1$ safely, yielding
\begin{equation}
  \left(D - C (A + i C^\dagger C)^{-1} (B + i C^\dagger D)\right) x_2
  = b_2 - C (A + i C^\dagger C)^{-1} (b_1 + i C^\dagger b_2).
\end{equation}
The reduction to this Schur complement form is supported by the sparse linear algebra library MUMPS~\cite{amestoy1996,amestoy2001}.
Furthermore, in the case where $C$ is a coupling of the lead to the scattering region, the additional term only modifies onsite elements of a small fraction of the degrees of freedom of the scattering region and preserves its sparsity structure.

\newpage
\section{Notation}
\label{sec:notation}
\setlength{\LTpost}{0pt}
\begin{longtable}{@{}p{0.15\linewidth}p{\dimexpr0.85\linewidth-2\tabcolsep\relax}@{}}
  $\hat{H}_\text{sys}$ / $\hat{G}_\text{sys}$ / $\hat\psi$ & Hamiltonian / Green's function / Wave function of full (infinite) systems. The hat indicates an infinite system, while the subscript specifies the system (infinite lead alone $H_\text{lead}$ or semi-infinite lead + scattering region $H_\text{sys}$).\\
  $\hat H_\text{lead}$ / $\hat\phi$ & Hamiltonian / Wave function of the full infinite lead. \\
  $H_\text{sr}$ / $G_\text{sr}$ & Hamiltonian / Green's function of the scattering region\\
  $H$, $V$ & Hamiltonian and hopping matrix of the lead\\
  $\Sigma$ & Lead self-energy \\
  $H(k)$ & Bloch Hamiltonian of the lead\\
  $V_j$ & hopping in the lead between the $j$-th nearest-neighbor cell ($V= V_1 +V_2 +\dots$)\\
  $P_\text{sr}$ & projection from lead to scattering region\\
  $\Phi$ / $\phi$ & matrix of translation eigenvectors (transverse modes) / vector of a single translation eigenvector. The columns of $\Phi$ are made of the $\phi$ vectors.\\
  $\Lambda$ / $\lambda$ & matrix of translation eigenvalues / single eigenvalue\\
  $\tilde{\Lambda}$ & truncated $\Lambda$ matrix where the vanishing diagonal elements have been disregarded.\\
  $_{\mathrm{p},\pm}$ / $_{\mathrm{e},\pm}$ & subscript indicating a propagating ($\mathrm{p}$) / evanescent ($\mathrm{e}$) mode, with $\pm$ indicating whether it is incoming ($-$, going toward the scattering region) or outgoing ($+$, going away from the scattering region).\\
  $_{\mathrm{t} }$ & subscript ($\mathrm{t}$) indicating both propagating ($\mathrm{p}$) and evanescent ($\mathrm{e}$) modes.\\
  $_\mathrm{\bar t}$ & subscript ($\mathrm{\bar t}$) similar to $(\mathrm{t})$ but the modes $\lambda_\alpha=0$ are excluded. See Table~\ref{tab:whichmodes}.\\
  $_\mathrm{\bar e}$ & subscript ($\mathrm{\bar e}$) similar to $(\mathrm{e})$ but the modes $\lambda_\alpha=0$ are excluded. See Table~\ref{tab:whichmodes}.\\
  $v$ & velocity\\
  $J$ & current\\
  $N$ & number of modes or sites. $N_\mathrm{t} = N_\mathrm{p}+N_\mathrm{e}$ is the number of sites in each unit cell of the lead. $N_\text{sr}$ is the number of sites of the scattering region.\\
  $V = AB$ & decomposition of the lead hopping matrix\\
  $S$ & scattering matrix\\
  $S_{\text{tp}}$ & Generalized scattering matrix containing both outgoing propagating and evanescent modes.\\
  $\psi_\text{sr}$ / $\Psi_\text{sr}$ & wave function in scattering region / wave function matrix: the columns of $\Psi_\text{sr}$ are made of the vectors $\psi_\text{sr}$\\
  $E$, $\varepsilon$ & energy \\
  $M$ & transfer matrix\\
  $T$ & transmission\\
  $g$ & conductance\\
  $\Theta$ & temperature \\
  $\mu$ & chemical potential \\
  $E_F$ & Fermi energy \\
  $\lambda_F$ & Fermi wavelength \\
  $V_b$ & Bias voltage \\
  $f(E)$ & Fermi function \\
  $I_a$ & Electric current in lead $a$ (counted positive when flowing towards the scattering region) \\
  $g_{ab}(E)$ & Dimensionless conductance matrix between lead $a$ and $b$ \\
  $\conductance_{ab}$ & Conductance matrix between lead $a$ and $b$ \\
  $r,t$ & reflection and transmission matrices \\
  $\Sigma$ & self-energy\\
  $P$ & projector onto conservation law blocks\\
  $\mathcal{C}$, $\mathcal{P}$, $\mathcal{T}$ & chiral, particle-hole, and time-reversal symmetry\\
  $k$ & $k$-vector  \\
  $q$ & vector of amplitudes of the different (propagating/evanescent) modes. \\
  $\alpha,\beta, \gamma, \delta$ & Greek letters label the mode index. \\
  $n,m$ & Latin letters label the site index. \\
  $i, j$ & label the unit cells within a single lead\\
  $a, b, c, d$ & label different leads \\
  $c_n$ and $c^\dagger_n$ & are the fermionic annihilation and creation operators on site $n$.\\
  ${\cal H}$ & is the second-quantized Hamiltonian\\
  $\Theta(\tau)$ & is the Heaviside function\\
  $\tau$ & is time.\\
  $\eta$ & is a small imaginary energy\\
  $BS$ & subscript denoting solutions of Eq.~\eqref{eq:bs_problem_discarded}\\
  $\mathds{G}$, $\mathds{H}$ & are generic Green's function and Hamiltonian matrices
\end{longtable}

\newpage
\bibliography{computationalqtransport}
\end{document}